\documentstyle[epsf]{l-aa}

\def\nix{\discretionary{}{}{}}

\def\Eq#1{Eq.~(\protect\ref{#1})}

\def\ave#1{\langle #1 \rangle}
\def\eck#1{\left\lbrack #1 \right\rbrack}
\def\rund#1{\left( #1 \right)}
\def \la {\mathrel{\vcenter
     {\offinterlineskip \hbox{$<$}\hbox{$\sim$}}}}
\def \ga {\mathrel{\vcenter
     {\offinterlineskip \hbox{$>$}\hbox{$\sim$}}}}

\begin{document}

\thesaurus{06(02.08.1, 08.02.1, 02.14.1, 08.14.1, 13.07.1, 02.05.1)}

\title{Coalescing neutron stars -- a step towards physical models}
\subtitle{II. Neutrino emission, neutron tori, and gamma-ray bursts}

\author{M.~Ruffert\inst{1}\thanks{e-mail: {\tt mruffert@mpa-garching.mpg.de}}
\and H.-Th.~Janka\inst{1}\thanks{e-mail: {\tt thj@mpa-garching.mpg.de}}
\and K.~Takahashi\inst{1,2}\thanks{e-mail: {\tt kjt@mpa-garching.mpg.de}}
\and G.~Sch\"afer\inst{3}\thanks{e-mail: {\tt gos@gravi.physik.uni-jena.de}} }
\institute{Max-Planck-Institut f\"ur Astrophysik, Karl-Schwarzschild-Str.~1,
Postfach 1523, 85740 Garching, Germany
\and Physik-Department E12, Technische Universit\"at M\"unchen, 
James-Frank-Str., 85748 Garching, Germany
\and Max-Planck-Arbeitsgruppe "Gravitationstheorie",
Friedrich--Schiller--Universit\"at,
Max-Wien-Platz~1, 07743 Jena, Germany}
\offprints{M.~Ruffert}


\maketitle

\begin{abstract}

Three-dimensional hydrodynamical, Newtonian calculations of the 
coalescence of equal-mass binary neutron stars are performed with
the ``Piecewise Parabolic Method''. The properties of neutron
star matter are described by the equation of state of Lattimer
\& Swesty (1991) which allows us to include the emission of 
neutrinos and to evaluate our models for the $\nu\bar\nu$-annihilation
in the vicinity of the merging stars.
When the stars have merged into one rapidly spinning massive body,
a hot toroidal cloud of gas with a mass of about 0.1--$0.2\,M_{\odot}$
forms around the wobbling and pulsating central $\sim 3\,M_{\odot}$
object. At that time the total neutrino luminosity climbs to
a maximum value of 1--$1.5\cdot 10^{53}$~erg/s of which 
90--95\% originate from the toroidal gas cloud surrounding the
very dense core. The mean
energies of $\nu_e$, $\bar\nu_e$, and heavy-lepton neutrinos $\nu_x$
are around 12~MeV, 20~MeV, and 27~MeV, respectively. The characteristics
of the neutrino emission are very similar to the emission from type-II
supernovae, except for the $\bar\nu_e$ luminosity from the merged
neutron stars which is a factor 3--6 higher than the luminosities
of the other neutrino species.

When the neutrino luminosities are highest, $\nu\bar\nu$-annihilation
deposits about 0.2--0.3\% of the emitted neutrino energy in the
immediate neighborhood of the merger, and the maximum integral 
energy deposition rate is 3--$4\cdot 10^{50}$~erg/s.
Since the $3\,M_{\odot}$ core of the merged object will most likely
collapse into a black hole within milliseconds, the energy that
can be pumped into a pair-photon fireball is insufficient by a
factor of about 1000 to explain $\gamma$-ray bursts at cosmological 
distances with an energy of the order of $10^{51}/(4\pi)$~erg/steradian.
Analytical estimates show that the additional energy provided by the
annihilation of $\nu\bar\nu$ pairs emitted from a possible accretion
torus of $\sim 0.1\,M_{\odot}$ around the central black hole
is still more than a factor of 10 too small, unless focussing of the
fireball into a jet-like expansion plays an important role.
A few $10^{-4}\,M_{\odot}$ of very neutron-rich, low-entropy matter
may be dynamically ejected shortly after the neutron stars have
merged, and another $10^{-4}$ up to a few $10^{-2}\,M_{\odot}$ of 
strongly neutronized, high-entropy material could be carried
away from the accretion torus in a neutrino-driven wind. 
The contamination with this baryonic material is a severe
threat to a relativistic fireball. 
Aspects of a possible r-processing in these ejecta are discussed.

\end{abstract}

\section{Introduction}

During the first two and a half years of operation,
the Burst and Transient Source Experiment (BATSE)
on the Compton Gamma Ray Observatory (GRO)
has observed 1122 cosmic gamma-ray bursts (Meegan et al.~1995a).
The angular distribution of the bursts on the sky is amazingly well
compatible with isotropy (e.g., Tegmark et al.~1995). There
are recent claims that a tenuous indication of an association 
with host galaxies is present in the burst sample evaluated by 
Larson et al.~(1996) and that a correlation with Abell clusters
has been found on a 95\% confidence level in the BATSE 3B catalog
(Kolatt \& Piran 1996).
If confirmed, these would be the first hints of a correlation
of gamma-ray bursts with any other astronomical population.
Number counts show a deficiency of faint bursts relative to 
the $-3/2$ power law expected for the brightness distribution 
of a homogeneous spatial distribution
of standard candle sources (Fenimore 1996). 
This paucity of weak bursts could be caused
by a truncation of the distance to very far burst sources or might
be an effect due to the expansion of the universe or could be 
associated with an evolution of the spatial density of bursts
(Bloom et al.~1996). Both observational facts, isotropy and bend
in the brightness distribution, would be naturally explained
if the gamma-ray bursters were situated at cosmological distances
(e.g., Paczy\'nski 1995, Hartmann et al.~1996,
Kolatt \& Piran 1996). Nevertheless,
the assumption that they might populate an extended Galactic halo
cannot be ruled out yet (Lamb 1995), and Galactic halo models have
been constructed (Podsiadlowski et al.~1995) which are able to 
fulfill the stringent limits set by the isotropy of the detected 
bursts and their inhomogeneous spatial distribution. 

The distribution of measured gamma-ray burst durations exhibits 
a bimodal structure with peaks at about 0.5~s and about 30~s 
(Meegan et al.~1995a). The bursts can be as short as $\sim 1$~ms
but can also last for several 100~s with variabilities and 
fluctuations on a millisecond time scale (Fishman et al.~1994).
The extremely short sub-ms rise times of the gamma-ray luminosities
suggest that the energy sources for the bursts must be
connected with very compact astrophysical objects which have a typical
size of the order of 100~km. This favors neutron stars or black 
holes as most likely candidates for the enigmatic origin of the 
cosmic gamma-ray bursts. 

Despite of more than 25~years of gamma-ray
burst observations, there is neither a convincing identification
of counterparts in any other energy range of the electromagnetic
spectrum (Greiner 1995a, 1995b; Vrba 1996), nor has a
generally accepted, satisfactory theoretical model been developed
yet (Nemiroff 1994a, Hartmann \& Woosley 1995, Woosley 1996).
About 120 gamma-ray burst models have been published in the refereed
literature until 1992 (Nemiroff 1994a), until 1994 there were
135 (Nemiroff 1994b), and maybe another one or two dozens have
been added since. 

Cosmological explanations have become increasingly popular in the more
recent publications, a fair fraction of which suggests collisions of 
two neutron 
stars or mergers of binaries consisting of either two neutron stars
(NS-NS) or a black hole and a neutron star (BH-NS) as possible
sources of the bursts (e.g., Paczy\'nski 1986;
Goodman 1986; Eichler et al.~1989; Piran 1990; Paczy\'nski 1991;
Narayan et al.~1991, 1992; Piran et al.~1992; M\'esz\'aros \&
Rees 1992a,b, 1993; Woosley 1993a; Mochkovitch et al.~1993, 1995a;
Hernanz et al.~1994, Katz \& Canel 1995a). One of the reasons for 
the attractivity of these scenarios is the desired compactness of
the objects, another reason is the knowledge that these events
should happen and should release large amounts of energy
(Dermer \& Weiler 1995).
The frequency of NS-NS and BH-NS mergers was estimated to be between
$10^{-6}$ and $10^{-4}$ per year per galaxy (Narayan et al.~1991,
Phinney 1991, Tutukov et al.~1992, Tutukov \& Yungelson 1993, 
Lipunov et al.~1995) and is therefore 
sufficient to explain the observed burst rate which requires an event 
rate of about $10^{-6}$~yr$^{-1}$ per galaxy (Narayan et al.~1992).
These rates per galaxy are so low that burst repetition in the same
region of the sky is practically excluded which is in agreement 
with the observations (Lamb 1996, Meegan et al.~1995b,
Brainerd et al.~1995, Efron \& Petrosian 1995).
If the merger rate is near the high end of the estimated range, some
beaming of the gamma-ray emission might be involved, or the majority 
of the bursts has to be very dim and escapes detection. 
Beaming would also lower the energy that must be converted into 
gamma rays at the source in order to cause the observed fluences.

Since the detected gamma-ray bursts appear to be isotropically
distributed and do not visibly trace the
large-scale structure of luminous matter in the universe, in
particular, are not concentrated towards the supergalactic plane
like nearby galaxies, constraints on the distance scale to 
cosmological bursts can be placed. From the BATSE 3B catalog 
Quashnock (1996) infers that the comoving distance to the ``edge'' of 
the burst distribution is greater than 630$\,h^{-1}$~Mpc and the
nearest bursts are farther than 40$\,h^{-1}$~Mpc (at the 95\% 
confidence level), the median distance to the nearest burst being
170$\,h^{-1}$~Mpc ($h$ is the Hubble constant in units of 
100~km/s/Mpc). From the absence of anisotropies in supergalactic
coordinates, Hartmann et al.~(1996) conclude that the minimum
sampling distance is 200$\,h^{-1}$~Mpc, and Kolatt \& Piran (1996)
find for their accurate position sub-sample members locations 
within 600$\,h^{-1}$~Mpc. In case of isotropic emission, 
standard candle non-evolving burst sources
at these cosmological distances must release $\gamma$-ray energies
of the order of $(3...4)\cdot 10^{51}h^{-2}$~erg 
(Woods \& Loeb 1994, Quashnock 1996).

This energy is about 0.1--0.2\% of the rest-mass energy of one
solar mass or roughly 1\% of the gravitational binding energy set free
when two 1.5~$M_{\odot}$ neutron stars merge. A large part of the
energy released during the merging, i.e., up to more than 10\% of 
$M_{\odot}c^2$, 
is carried away by gravitational waves, the exact value depending
on the nuclear equation of state and thus on the compactness of the 
neutron stars and of the merged object (see Ruffert et al.~1996 
and references therein). A similar amount of energy
could be radiated away in neutrinos which are abundantly produced
when the matter of the coalescing and merging stars is heated up 
to very high temperatures by tidal forces, friction, and viscous
dissipation of kinetic energy in shocks (Eichler et al.~1989, 
Narayan et al.~1992, Harding 1994). The duration of the neutrino
emission, the neutrino luminosity, and the total energy radiated
in neutrinos will be determined by the structure and dynamical 
evolution of the merger, by the thermodynamical conditions in the
merging stars, and by the lifetime of the merged object 
before it collapses into a black hole or before the surrounding
material is swallowed by the black hole. Gravitational waves as well
as neutrinos from these very distant sources cannot be measured with
current experiments, but detectors are planned or built for 
gravitational waves (LIGO, VIRGO, GEO600; see, e.g., Thorne 1992) 
and are envisioned for neutrinos (a kilometer-scale neutrino telescope
with 1~km$^3$ of instrumented volume; see Weiler et al.~1994,
Halzen \& Jaczko 1996, Halzen 1996), which might be lucky to catch
the death throes of massive binaries in the not too distant future.

Due to the very high opacity
of neutron star matter, high energy photons cannot
be emitted by the merging object directly, unless the very outer 
layers are heated to sufficiently high temperatures. Even if they
were radiating with luminosities substantially above the Eddington
limit (Duncan et al.~1986), NS-NS or NS-BH mergers at
cosmological distances would still be far too faint to be visible from 
Earth. However, if only a tiny fraction (less than 1\%) of the 
potentially
emitted neutrinos and antineutrinos annihilate in the vicinity of
the merger (Goodman et al.~1987, Cooperstein et al.~1987)
and create a fireball of electron-positron pairs
and photons (Cavallo \& Rees 1978), an intense outburst of 
gamma-radiation can be produced with an overall power output 
exceeding the Eddington luminosity by up to 15 orders of magnitude
and energetic enough to account for a cosmological gamma-ray 
burst (Eichler et al.~1989). 

Relativistic expansion of the pair-photon
plasma and the final escape of high energy gamma-radiation 
with the observed non-thermal spectrum (e.g., Band 1993) from an optically
thin fireball can only occur if the baryon load of the radiation-pressure
ejected shells is sufficiently small, i.e., the contamination with baryons
must be below a certain limit (Goodman 1986, Paczy\'nski 1986). If the 
admixture of baryons is too large, highly relativistic outflow
velocities will be impeded through conversion of radiation energy to
kinetic energy and the flow will remain optically thick, leading to
a degradation of the photons to the UV range (Paczy\'nski 1990,
Shemi \& Piran 1990). In order to suppress pair production 
$\gamma + \gamma \to e^+ + e^-$ and to make the fireball optically
thin to its own photons, the Lorentz factor $\Gamma$ of the emitting 
region must obey $\Gamma \ga 100$ (Fenimore et al.~1993, M\'esz\'aros
et al.~1993). The bulk Lorentz factor is related with the 
ratio of total energy to (baryon) rest mass energy:
$\Gamma = E_{\gamma}/(\xi Mc^2)$ where $\xi$ is the
fraction of the total energy in the fireball that ends up as observed
energy in $\gamma$-rays, $E_{\gamma}$. Therefore a lower limit for 
$\Gamma$ places an upper bound on the amount of baryonic mass $M$
released in the explosive event or being present as ambient gas near
the burst source, $M \la 0.6\cdot 10^{-5}\xi^{-1}
(E_{\gamma}/10^{51}\,{\rm erg})\,M_{\odot}$. Based on the empirical
data from the first BATSE catalog (169 bursts), Woods \& Loeb (1994)
deduce a mean minimum Lorentz factor $\Gamma$ of about 500, 
corresponding to an average maximum baryon load of 
$10^{-5}\xi^{-1}\,M_\odot$ for gamma-ray burst events at cosmological 
distances. Recent introductory reviews and summaries of
the properties of fireballs in the context of cosmological models
of gamma-ray bursts were given by Dermer \& Weiler (1995),
M\'esz\'aros (1995), and Piran (1995, 1996).

The limits for the baryon load in the fireball set by
observational requirements are very stringent and
the baryonic pollution of the pair-photon plasma is a major concern
for gamma-ray burst models based on mergers of massive binary stars.
In order to obtain $\nu\bar\nu$-annihilation in a baryon-poor
region around the merger, anisotropies of the merger geometry
are considered to be essential (Narayan et al.~1992; 
M\'esz\'aros \& Rees 1992a,b; Woosley 1993a; Mochkovitch 1993, 1995a;
Hernanz 1994). The hope is that centrifugal forces
can keep a region near the rotation axis of the merging NS-NS
or NS-BH system relatively baryon free and that the neutrino-driven 
mass loss from an accretion disk or torus formed after the merging
of the binary will leave a ``clean'' funnel along the axis of symmetry 
where $\nu\bar\nu$-annihilation can create collimated $e^+e^-$ jets
expanding relativistically in both axis directions. Also, general
relativistic bending of the $\nu$ and $\bar\nu$ trajectories and
$\nu\bar\nu$-annihilation inside the innermost stable orbit for
massive particles around the accreting black hole have been 
suggested to be potentially helpful. Secondary processes could
lead to the reconversion of baryonic kinetic energy, e.g.~by 
external shock interactions of the expanding fireball 
in the ambient interstellar medium or pre-ejected gas 
(Rees \& M\'esz\'aros 1992; M\'esz\'aros \& Rees 1992a, 1993; 
Sari \& Piran 1995) or by internal shock interactions of shells
having different speeds within the expanding fireball 
(Narayan et al.~1992, Paczy\'nski \& Xu 1994,
Rees \& M\'esz\'aros 1994, Mochkovitch et al.~1995b). 
Even more complex physics like the upscattering of interstellar
photons of a local thermal radiation field by collisions with
electrons in the ultrarelativistic wind
(Shemi 1994) or highly amplified magnetic
fields (M\'esz\'aros \& Rees 1992a, Usov 1992, Smolsky \& Usov 1996)
might be important in determining
the temporal and spectral structure of the observable gamma-ray burst.
In fact, such secondary processes may be crucial to produce the 
detected very high energy photons with energies of up to more than
1~GeV (Hurley et al.~1994, Teegarden 1995).

In this work we shall not deal with the possibly very complex
and complicated processes that are involved in the formation of
the finally
observable gamma-ray signature. Instead, our interest will be
concentrated on the hydrodynamical modelling of the last stages
of the coalescence of binary neutron stars employing an elaborate
equation of state for neutron star matter with the aim to compute
the neutrino emission from the merger. The results of our models
will allow us to calculate the 
efficiency of $\nu\bar\nu$-annihilation and the energy deposition
in the vicinity of the merging stars. After dozens of papers that
suggest and refer to the annihilation of $\nu\bar\nu$ pairs from
merging compact binaries as the energy source of cosmological 
gamma-ray bursts, we shall try to put this hypothesis to a 
quantitative test. We shall investigate the questions whether
the neutrino emission from the tidally heated neutron stars 
just prior to merging (M\'esz\'aros \& Rees 1992b), during the 
dynamical phase of the merging or collision (Narayan et al.~1992, 
M\'esz\'aros \& Rees 1992b, Dermer \& Weiler 1995, Katz \& Canel
1995a) or after the merging when a hot accretion disk 
or torus has possibly formed around a central black hole
(Woosley 1993a, Mochkovitch et al.~1993), is sufficiently luminous
and lasting to yield
the energy required for gamma-ray bursts at cosmological distances.
Also, our simulations will provide information about how much mass 
might remain in an accretion disk or torus and about the 
thermodynamcial conditions in this disk matter. 
These aspects might have
interesting implications for the possible contributions of NS-NS
and NS-BH mergers to the nucleosynthesis of heavy elements.

The paper is organized as follows. In Sect.~\ref{sec:numer} the 
computational
method and initial conditions for our simulations are described
and the hydrodynamical evolution of the merger is shortly 
summarized from the results given in detail by Ruffert et al.~1996
(Paper~I). In Sect.~\ref{sec:neuttherm} the results for the 
neutrino emission and thermodynamical evolution of the merger
will be presented. Section~\ref{sec:annihil} deals with the 
neutrino-antineutrino annihilation and contains information about
the numerical evaluation (Sect.~\ref{sec:anninum}), about the numerical
results (Sect.~\ref{sec:annires}), and about an analytical model
which was developed to estimate the neutrino emission and 
annihilation for an accretion torus around a black hole 
(Sect.~\ref{sec:simod}). In Sect.~\ref{sec:end} the results
will be discussed concerning their implications for 
heavy element nucleosynthesis and for
gamma-ray bursts, and Sect.~\ref{sec:summary} contains a
summary and conclusions.

\section{Computational procedure, initial conditions, and hydrodynamical
evolution\label{sec:numer}}

In this section we summarize the numerical methods and the
treatment of the input physics used for the presented
simulations. In addition, we specify the initial conditions by which
our different models are distinguished. Also, the results for the 
dynamical evolution as described in detail in Paper~I 
are shortly reviewed.

\subsection{Numerical treatment}

The hydrodynamical simulations were done with a code based on the
Piecewise Parabolic Method (PPM) developed by Colella \& Woodward (1984).
The code is basically Newtonian, but contains the terms 
necessary to describe gravitational wave emission and the corresponding
back-reaction on the hydrodynamical flow (Blanchet et al.~1990).
The modifications that follow from the gravitational
potential are implemented as source terms in the PPM algorithm.
The necessary spatial derivatives are evaluated as standard centered
differences on the grid. 

In order to describe the thermodynamics of the neutron star matter, 
we use the equation of state (EOS) of Lattimer \& Swesty (1991)
in a tabular form. The inversion for the temperature is done with
a bisection iteration. Energy loss and changes of the electron
abundance due to the emission of neutrinos and antineutrinos are 
taken into account by an elaborate ``neutrino leakage scheme''.
The energy source terms contain the production of all types of 
neutrino pairs by thermal processes and of electron neutrinos and
antineutrinos also by lepton captures onto baryons. The latter 
reactions act as sources or sinks of lepton number, too, and are
included as source term in a continuity equation for the electron
lepton number. When the neutron star
matter is optically thin to neutrinos, the neutrino source 
terms are directly calculated from the reaction rates, while 
in case of optically thick conditions lepton number and energy
are released on the corresponding neutrino diffusion time scales.
The transition between both regimes is done by a smooth interpolation. 
Matter is rendered optically thick to neutrinos due to the main 
opacity producing reactions which are  
neutrino-nucleon scattering and absorption of neutrinos onto baryons.

More detailed information about the employed numerical procedures
can be found in Paper~I, in particular about
the implementation of the gravitational wave radiation
and back-reaction terms and 
the treatment of the neutrino lepton number and energy loss terms
in the hydrodynamical code.

\subsection{Initial conditions\label{sec:initial}}

We start our simulations with two identical Newtonian
neutron stars with a baryonic mass of about 1.63~$M_\odot$
and a radius of 15~km which are placed at a center-to-center 
distance of 42~km on a grid of 82~km side length. With the
employed EOS of Lattimer \& Swesty (1991), this baryonic mass 
corresponds to a (general relativistic) gravitational mass of 
approximately 1.5~$M_\odot$ for a cool star with a 
radius as obtained from the general 
relativistic stellar structure equations of 11.2~km. Having a 
compressibility modulus of bulk nuclear matter of $K = 180$~MeV
(which is the ``softest'' of the three available cases), the 
Lattimer \& Swesty EOS may overestimate the stiffness of 
supranuclear matter, in particular, since in its present form 
it neglects the possible occurrence of new hadronic
states besides the neutron and proton at very high densities.
For a softer supranuclear EOS,
neutron stars would become more compact and their gravitational
binding energy larger so that a baryonic mass of 1.63 $M_\odot$
would more likely correspond to a gravitational mass between 1.4 
and 1.45 $M_\odot$.

The distributions of density $\rho$ and electron fraction $Y_e$ are
taken from a one-dimensional model of a cold, deleptonized neutron
star in hydrostatic equilibrium. For numerical reasons the 
surroundings of the neutron stars cannot be assumed to be 
evacuated. The density of the ambient medium was set to 
$10^9$~g/cm$^3$, more than five orders of magnitude smaller than 
the central densities of the stars. The internal energy density and 
electron fraction of this gas were taken to be equal to the
values in the neutron stars at a density of $10^9$~g/cm$^3$.
In order to ensure sufficiently good numerical resolution, we
artificially softened the extremely steep density decline towards
the neutron star surfaces by not allowing for a density change of
more than two orders of magnitude from zone to zone. From this
prescription we obtain a thickness of the neutron star surface 
layers of about 3 zones.

\begin{table}
\caption[]{
Parameters and some computed quantities for all models.
$N$ is the number of grid zones per dimension in the orbital plane,
$S$ defines the direction of the spins of the neutron stars
relative to the direction of the orbital angular momentum, and
$k_{\rm B}T_{\rm ex}$ gives the maximum temperature (in energy units)
reached on the grid during the simulation of a model.
$L_{\nu_e}$ is the electron neutrino luminosity after approaching
a saturation level at about 8~ms, $L_{\bar{\nu}_e}$ is the
corresponding electron antineutrino luminosity, and
$L_{\nu_x}$ the luminosity of each individual species of $\nu_\mu$,
$\bar{\nu}_\mu$, $\nu_\tau$, $\bar{\nu}_\tau$. $L_\nu$
gives the total neutrino luminosity after a quasi-stationary
state has been reached at $t\ga 6$--8~ms and $\dot{E}_{\nu\bar{\nu}}$
denotes the integral rate of energy deposition by
neutrino-antineutrino annihilation at that time
}
\begin{flushleft}

\tabcolsep=0.7mm
\begin{tabular}{lccccccccc}
\hline\\[-3mm]
Model & $\,N\,$ & $\,S\,$ & $k_{\rm B}T_{\rm ex}$ & $L_{\nu_e}$ & $L_{\bar{\nu}_e}$ &
           $L_{\nu_x}$ & $L_\nu$ & $\dot{E}_{\nu\bar{\nu}}$ \\
     &  &  &{\scriptsize MeV}
     &{\scriptsize$10^{53}{\rm erg/s}$}
     &{\scriptsize$10^{53}{\rm erg/s}$}
     &{\scriptsize$10^{53}{\rm erg/s}$}
     &{\scriptsize$10^{53}{\rm erg/s}$}
     &{\scriptsize$10^{50}{\rm erg/s}$}
\\[0.3ex] \hline\\[-3mm] 
A64 & $\phantom{1}$64 &$\phantom{-}$0 &$\phantom{>}$40 
                      & 0.13 & 0.58 & 0.11 & 1.15 & 2.1 \\
B64 & $\phantom{1}$64 &        $\,+$1 &$\phantom{>}$30 
                      & 0.18 & 0.55 & 0.09 & 1.09 & 3.0 \\
C64 & $\phantom{1}$64 &          $-$1 &          $>$50 
                      & 0.15 & 0.67 & 0.10 & 1.22 & 3.6 \\
A128&             128 &$\phantom{-}$0 &$\phantom{>}$39 
                      & 0.16 & 0.43 & 0.06 & 0.83 & --- \\
\hline
\end{tabular}
\end{flushleft}
\label{tab:models}
\end{table}

The orbital velocities of the coalescing neutron stars were prescribed
according to the motion of point masses spiralling towards each other
due to the emission of gravitational waves. The tangential velocities
of the neutron star centers were set equal to the Kepler velocities on 
circular orbits and radial velocity components were attributed as 
calculated from the quadrupole formula.
In addition to the orbital angular momentum, spins around their 
respective centers were added to the neutron stars. The assumed
spins and spin directions were varied between the calculated models. 
Table~\ref{tab:models} lists the distinguishing model parameters,
the number $N$ of grid zones per spatial dimension (in the orbital 
plane) and the spin parameter $S$.
The neutron stars in models~A64 and~A128 did not have any additional
spins ($S = 0$), in model B64 the neutron star spins were parallel to the 
orbital angular momentum vector ($S = +1$), in model C64 the spins were 
in the opposite direction ($S = -1$).
In both models B64 and C64, the angular velocities
of the rigid neutron star rotation and of the orbital motion were 
chosen to be equal. Model A128 had the same initial setup as model
A64 but had twice the number of grid zones per spatial dimension and
thus served as a check for the sufficiency of the numerical resolution
of the computations with $64^3$ zones.

The rotational state of the neutron stars is determined by
the action of viscosity. If the dynamic viscosity of neutron star 
matter were large enough, tidal forces could lead to tidal
locking of the two stars and thus spin-up during inspiral.
Kochanek (1992), Bildsten \& Cutler (1992),
and Lai (1994), however, showed that microscopic shear and bulk 
viscosities are probably orders of magnitude too small to achieve 
corotation. Moreover, they argued that for the same reason it is 
extremely unlikely that the stars are heated up to more than about
$10^8$~K by tidal interaction prior to merging. In this sense,
models A64 and A128 can be considered as reference case for two 
non-corotating neutron stars, while model B64 represents the
case of rigid-body like rotation of the two stars, and model C64 
the case where the spin directions of both stars were inverted.

Since in the case of degenerate matter the temperature is extremely
sensitive to small variations of the internal energy, e.g.~caused
by small numerical errors, we did not start our simulations with cold
($T = 0$) or ``cool'' ($T \la 10^8$~K) neutron stars as suggested
by the investigations of Kochanek (1992), Bildsten \& Cutler (1992),
and Lai (1994). Instead, we constructed initial temperature distributions
inside the neutron stars by assuming thermal energy densities of 
about 3\% of the degeneracy energy density for a given density $\rho$
and electron fraction $Y_e$. The corresponding central temperature 
was around 7~MeV and the average temperature was a few MeV and thus
of the order of the estimates obtained by M\'esz\'aros \& Rees (1992b)
for the phase just prior to the merging.
Locally, these initial temperatures were much smaller
than the temperatures produced by the compression and dissipative 
heating during coalescence (see Table~\ref{tab:models} for the maximum
temperatures). However, they are orders of magnitude larger than can
be achieved by tidal dissipation with plausible values for the microscopic
viscosity of neutron star matter. Even under the most extreme 
assumptions for viscous shear heating, the viscosity of neutron star 
matter turns out to be at least four orders of magnitude too small
(see Janka \& Ruffert 1996).

Models A64, B64, and C64 were computed on a Cray-YMP~4/64 where they
needed about 16~MWords of main memory and took approximately 40~CPU-hours
each. For the better resolved model A128 we employed a grid with  
$128\times 128\times 64$ zones instead of the $64\times 64\times 32$ grid 
of models A64, B64, and C64.
Note that in the direction orthogonal to the orbital
plane only half the number of grid zones was used but the spatial
resolution was the same as in the orbital plane. 
Model A128 was run on a Cray-EL98~4/256 and required about 22~MWords of 
memory and 1700~CPU-hours.

\subsection{Hydrodynamical evolution\label{sec:hydevol}}

The hydrodynamical evolution and corresponding gravitational wave 
emission were detailed in Paper~I. Again, we only summarize the most
essential aspects here. 

The three-dimensional hydrodynamical simulations were started at
a center-to-center distance of the two neutron stars of 2.8 neutron
star radii. This was only slightly larger than the separation where the
configurations become dynamically unstable which is at a distance of
approximately 2.6 neutron star radii. Gravitational wave emission leads
to the decay of the binary orbit, and already after about one quarter of
a revolution, approximately 0.6~ms after the start of the computations,
the neutron star surfaces touch because of the tidal deformation and
stretching of the stars. 

When the two stars begin to plunge into each other, compression 
and the shearing motion of the touching surfaces 
cause dissipation of kinetic energy and lead to
a strong increase of the temperature. From initial values of a
few MeV, the gas heats up to peak temperatures of several ten MeV. In 
case of model C64, a temperature of nearly 50~MeV is reached about 1~ms 
after the stars have started to merge. After one compact, massive body
has formed from most of the mass of the neutron stars about 3~ms later,
more than 50~MeV are found in two distinct, extremely hot off-center regions.

At the time of coalescence and shortly afterwards, spiral-arm or wing-like 
extensions are formed from material spun off the outward directed sides
of the neutron stars by tidal and centrifugal forces. Due to the
retained angular momentum, the central body performs large-amplitude
swinging motions and violent oscillations.
This wobbling of the central body of the merger
creates strong pressure waves and small shocks that heat the exterior
layers and lead to the dispersion of the spiral arms into a vertically
extended ``ring'' or thick toroidal disk of gas 
that surrounds the massive central
object. While the mass of the compact body is larger than 3~$M_\odot$
and its mean density above $10^{14}\,{\rm g/cm}^3$, the surrounding
cloud contains only 0.1--0.2~$M_\odot$ and is more dilute with an average
density of about $10^{12}\,{\rm g/cm}^3$. 

The central body is too massive to
be stabilized by gas pressure for essentially all currently discussed
equations of state of nuclear and supranuclear matter. Moreover, 
it can be argued (see Paper~I) that its angular momentum is not large
enough to provide rotational support. Therefore, we expect that in 
a fully general relativistic simulation, the central object will 
collapse into a black hole on a time scale of only a few milliseconds
after the neutron stars have merged. Dependent on the total angular
momentum corresponding to the initial setup of the neutron star spins, 
some of the material in the outer regions of the disk obtains enough
angular momentum (e.g., by momentum transfer via pressure waves) 
to flow across the boundary of the computational grid at a distance of 
about 40~km from the center. In model B64 which has the largest 
total angular momentum because of the assumed solid-body type rotation, 
0.1--0.15~$M_\odot$ are able to leave the grid between 2~ms and
4~ms after the start of the simulation. However, at most a few
times $10^{-4}$~$M_\odot$ of this material have a total energy that
might allow them to escape from the gravitational field of the merger.
In model C64 the anti-spin setup leads to violent oscillations of
the merged body which create a very extended and periodically 
contracting and reexpanding disk. In this model nearly 0.1~$M_\odot$ 
are lost across the outer grid boundaries even during the later stages
of the simulation ($t \ga 5$~ms).

While by far the major fraction of the merger mass ($\ga 3$~$M_\odot$)
will be swallowed up by the forming black hole almost immediately on a 
dynamical time scale, some matter with sufficiently high specific angular
momentum might be able to remain in a disk around the slowly rotating
black hole. Estimates show (Paper~I) that with the typical rotation
velocities obtained in our models, this matter must orbit around the 
central body at radii of at least between about 40~km and 100~km.
This means that
only matter that has been able to leave the computational grid used
in our simulations has a chance to end up in a toroidal disk around the 
black hole. From these arguments we conclude that such a possible disk 
might contain a mass of at most 0.1--0.15~$M_\odot$.

%
%





\begin{figure*}
 \begin{tabular}{cc}
  \epsfxsize=8.8cm \epsfclipon \epsffile{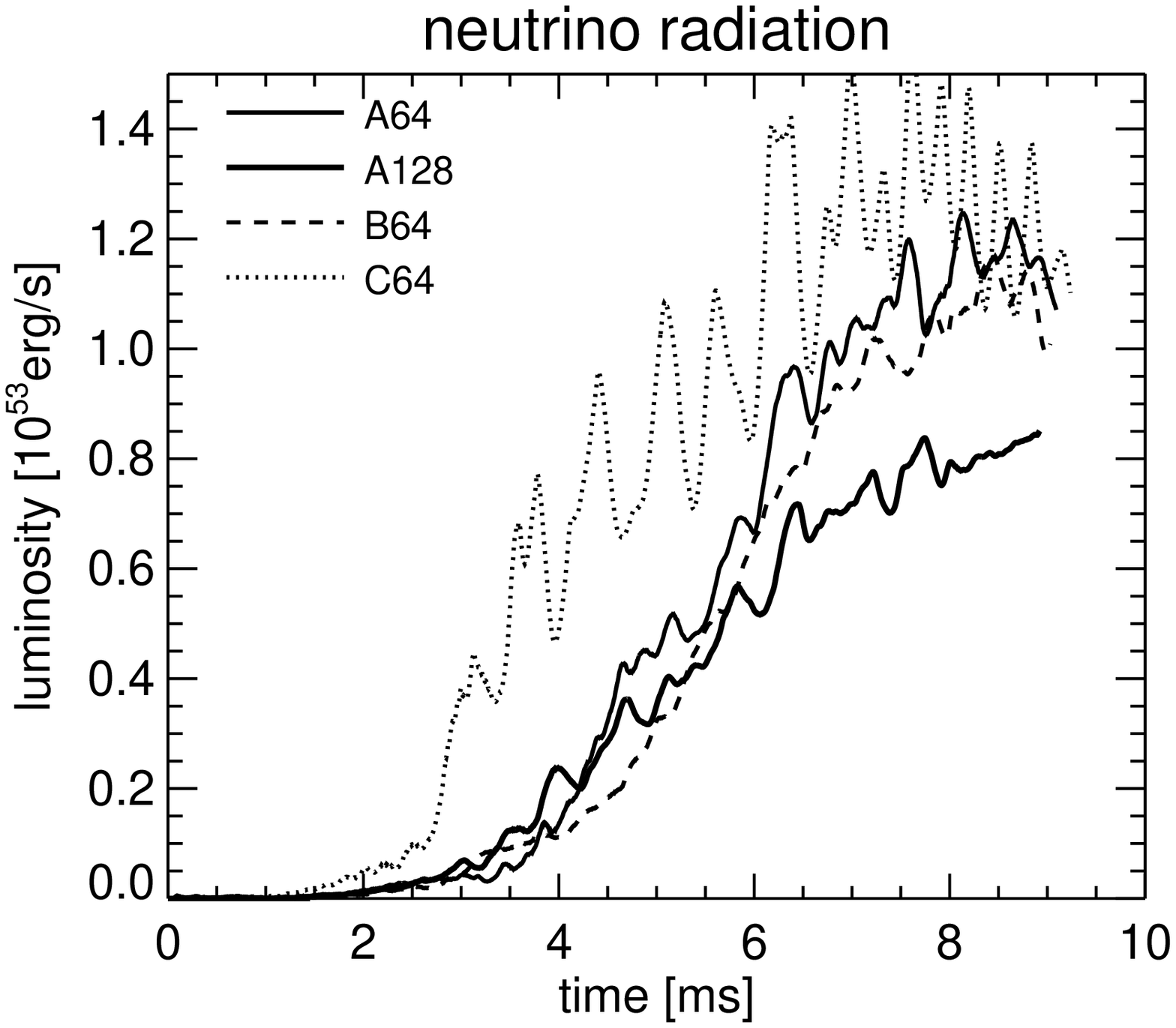} &
  \epsfxsize=8.8cm \epsfclipon \epsffile{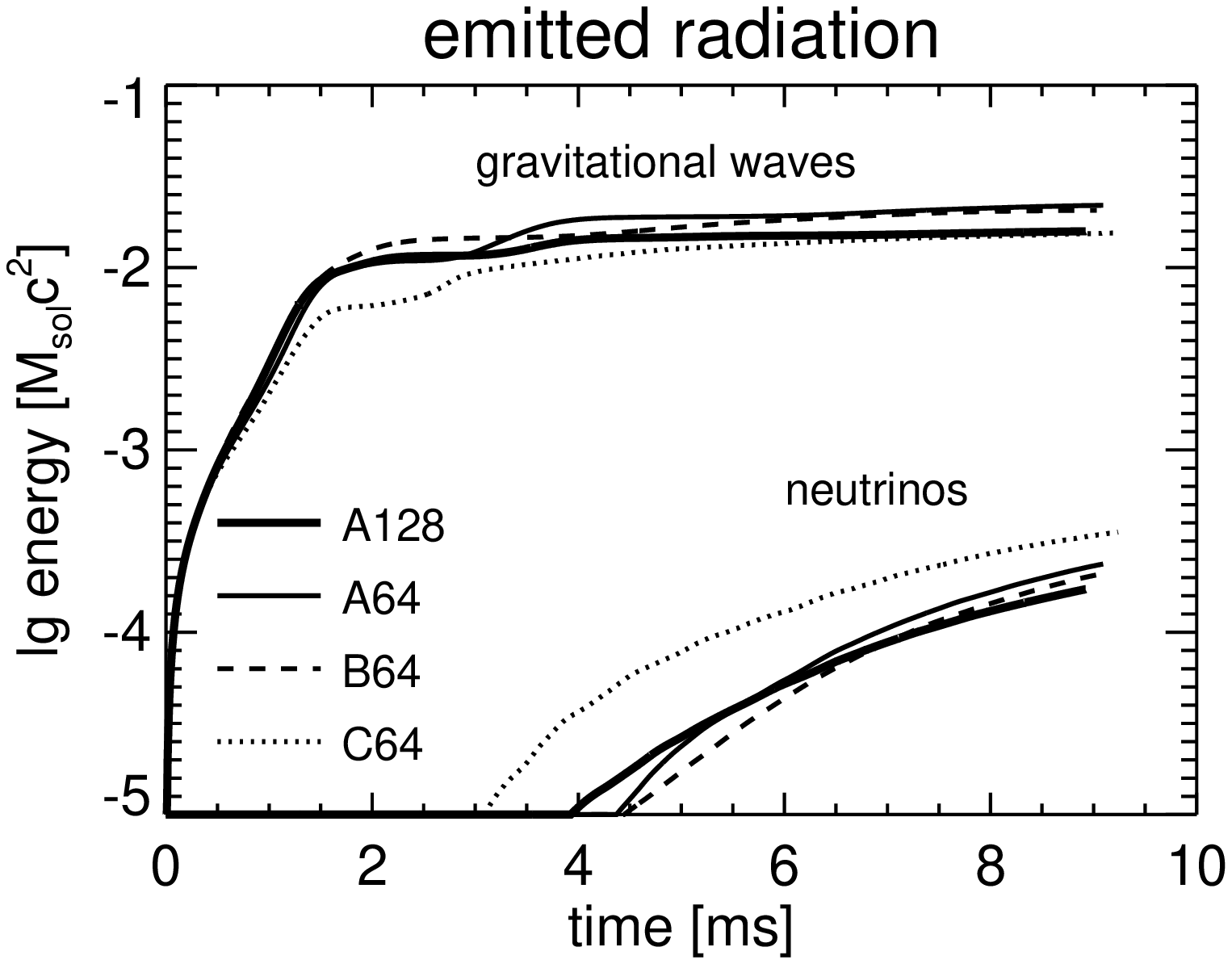} \\
  \parbox[t]{8.8cm}{\caption[]{Total neutrino luminosities as
	 functions of time for all four models}\label{fig:neutrad}} &
  \parbox[t]{8.8cm}{\caption[]{Comparison of the cumulative energies
	 emitted in gravitational waves and in neutrinos as functions
         of time for all four models}\label{fig:neutot}} \\[15ex] 
  \epsfxsize=8.8cm \epsfclipon \epsffile{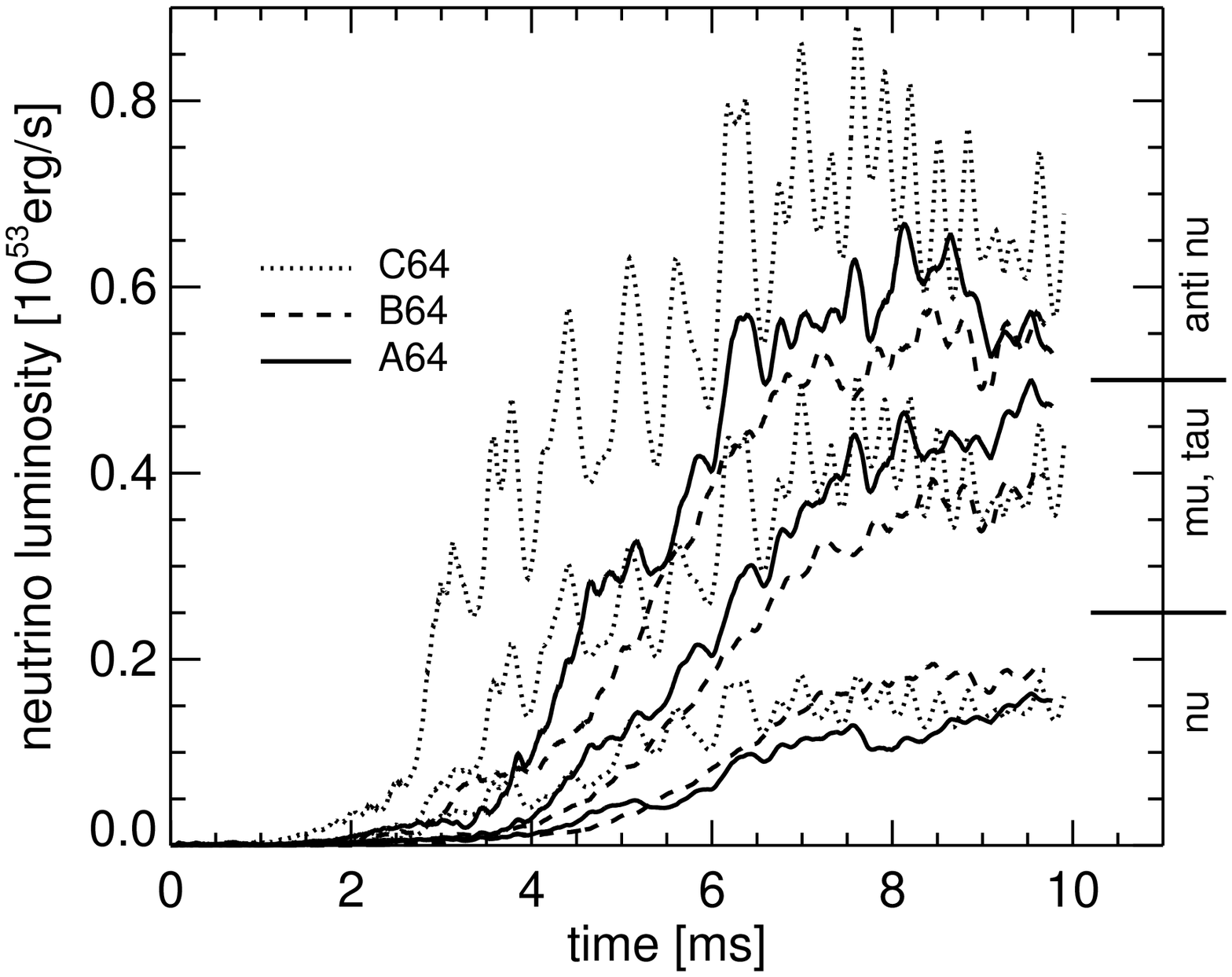} &
  \epsfxsize=8.8cm \epsfclipon \epsffile{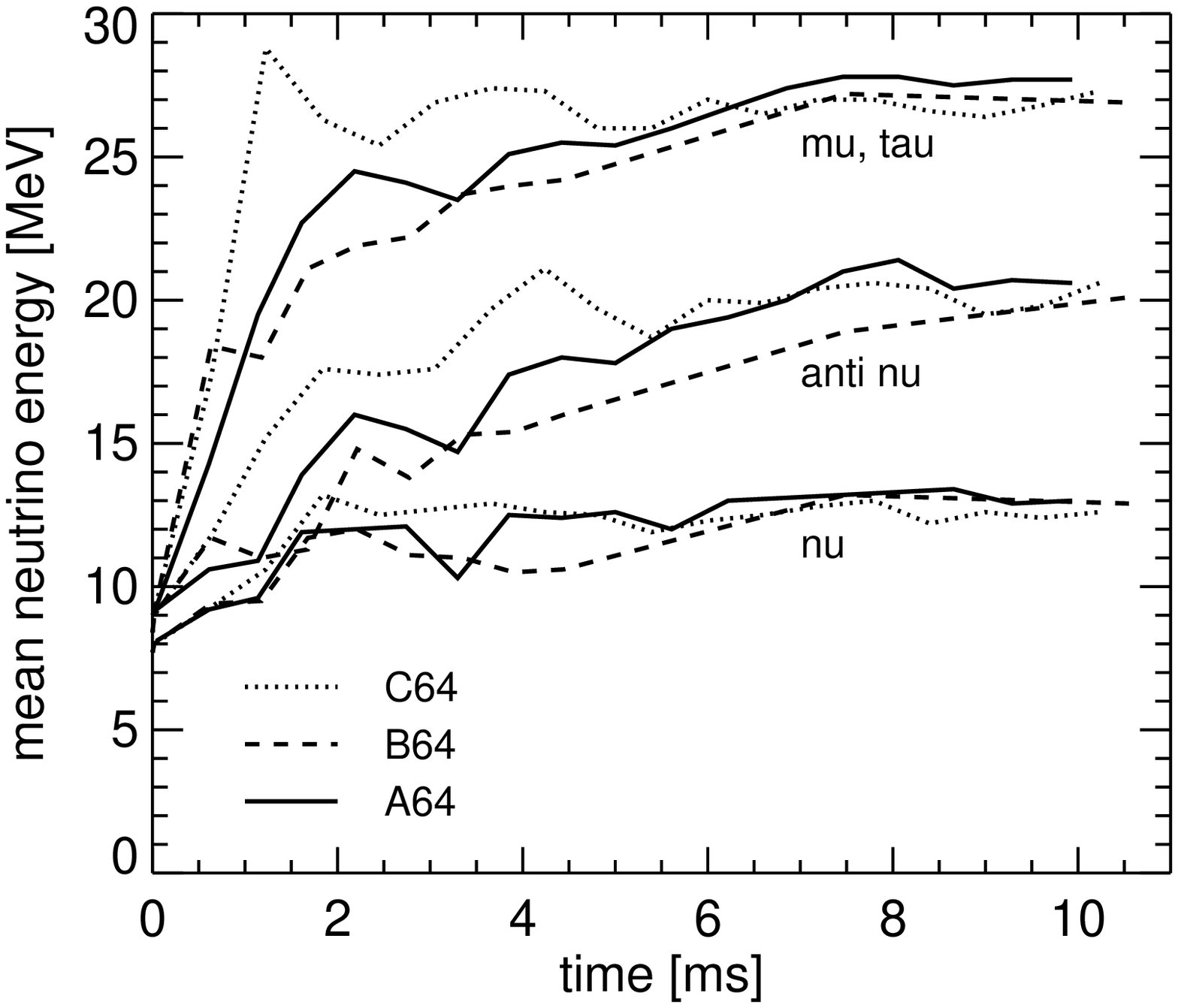} \\ 
  \parbox[t]{8.8cm}{\caption[]{Luminosities of individual
	 neutrino types ($\nu_e$, $\bar\nu_e$, and the sum of all 
	 $\nu_x$) as functions of time for the three models A64, B64,
	 and C64}\label{fig:ergne}} &
  \parbox[t]{8.8cm}{\caption[]{Average energies of emitted neutrinos 
	 $\nu_e$, $\bar\nu_e$, and $\nu_x$ as functions of time for
	 the three models A64, B64, and C64}\label{fig:mene}} \\
 \end{tabular}
\end{figure*}

\section{Neutrino emission and thermodynamical 
evolution\label{sec:neuttherm}}

\subsection{Neutrino emission\label{sec:nuem}}

The local energy and lepton number losses due to neutrino emission
are included via source terms in our code as described in the 
appendix of Paper~I. We treated the neutrino effects in terms of an
elaborate leakage scheme that was calibrated by comparison with 
results from diffusion calculations in one-dimensional situations. 
By adding up the local source terms over the whole computational 
grid, one obtains the neutrino luminosities of all individual neutrino
types, the sum of which gives the total neutrino luminosity. In the 
same way number fluxes of electron neutrinos ($\nu_e$), electron 
antineutrinos ($\bar\nu_e$), and heavy-lepton neutrinos ($\nu_{\mu}$,
$\bar\nu_{\mu}$, $\nu_{\tau}$, and $\bar\nu_{\tau}$, which
will be referred to as $\nu_x$ in the following)
can be calculated. The mean energies of the 
emitted neutrinos result from the ratios of neutrino luminosities to
neutrino number fluxes.

\begin{figure*}
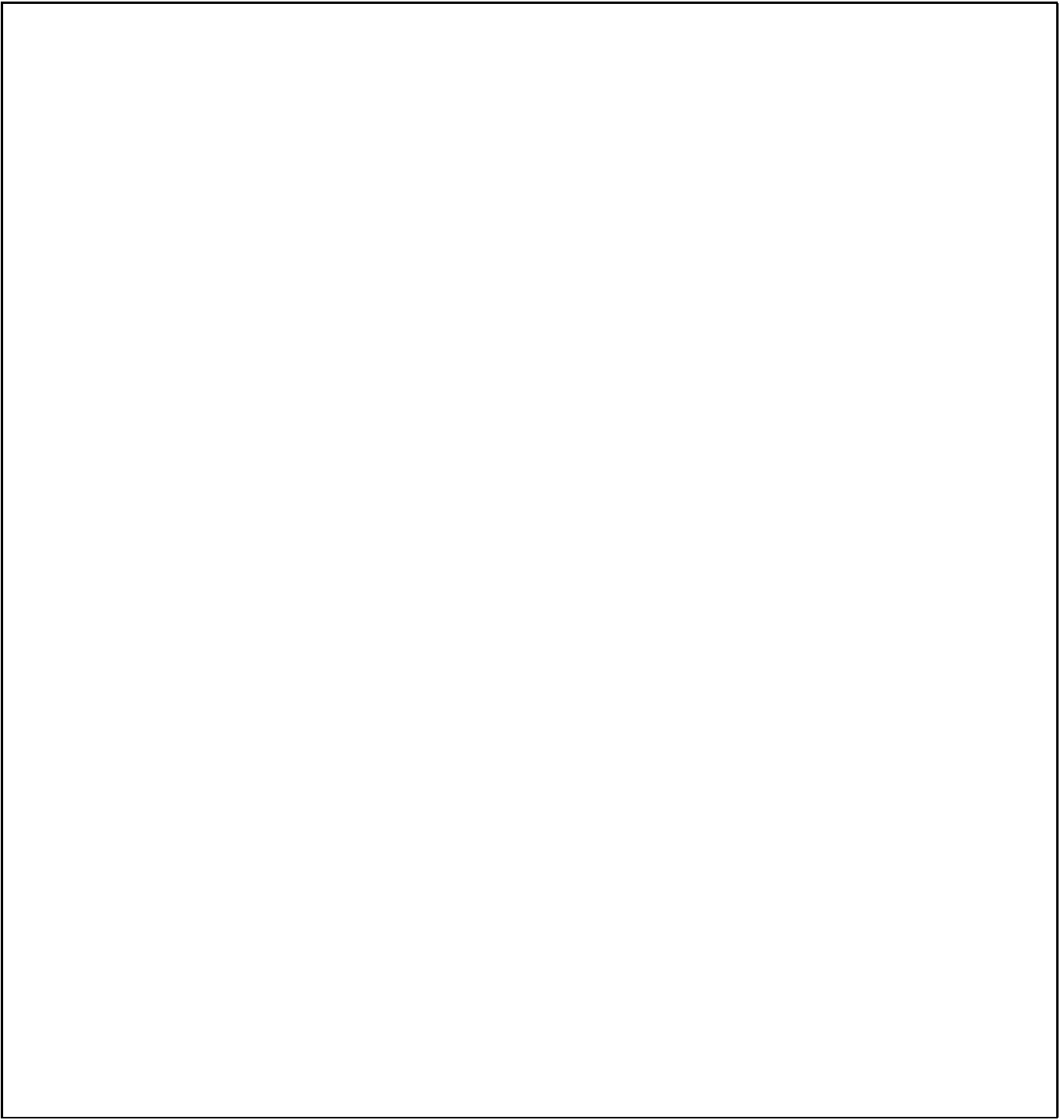

\picplace{19cm}
\caption[]{
Energy emission rates (in erg/cm$^3$/s)
of electron neutrinos (panel {\bf a}),
electron antineutrinos (panel {\bf b}), the sum of all heavy-lepton
neutrinos (panel {\bf c}), and the total neutrino energy loss rate
in the orbital plane of model A64 at the end of the
simulation (time in the top right corner of the panels).
The contours are logarithmically spaced in intervals of
0.5 dex, bold contours are labeled with their respective values.
The grey shading emphasizes the emission levels, dark grey 
corresponding to the strongest energy loss by neutrino emission
}
\label{fig:neutA64}
\end{figure*}


\begin{figure*}
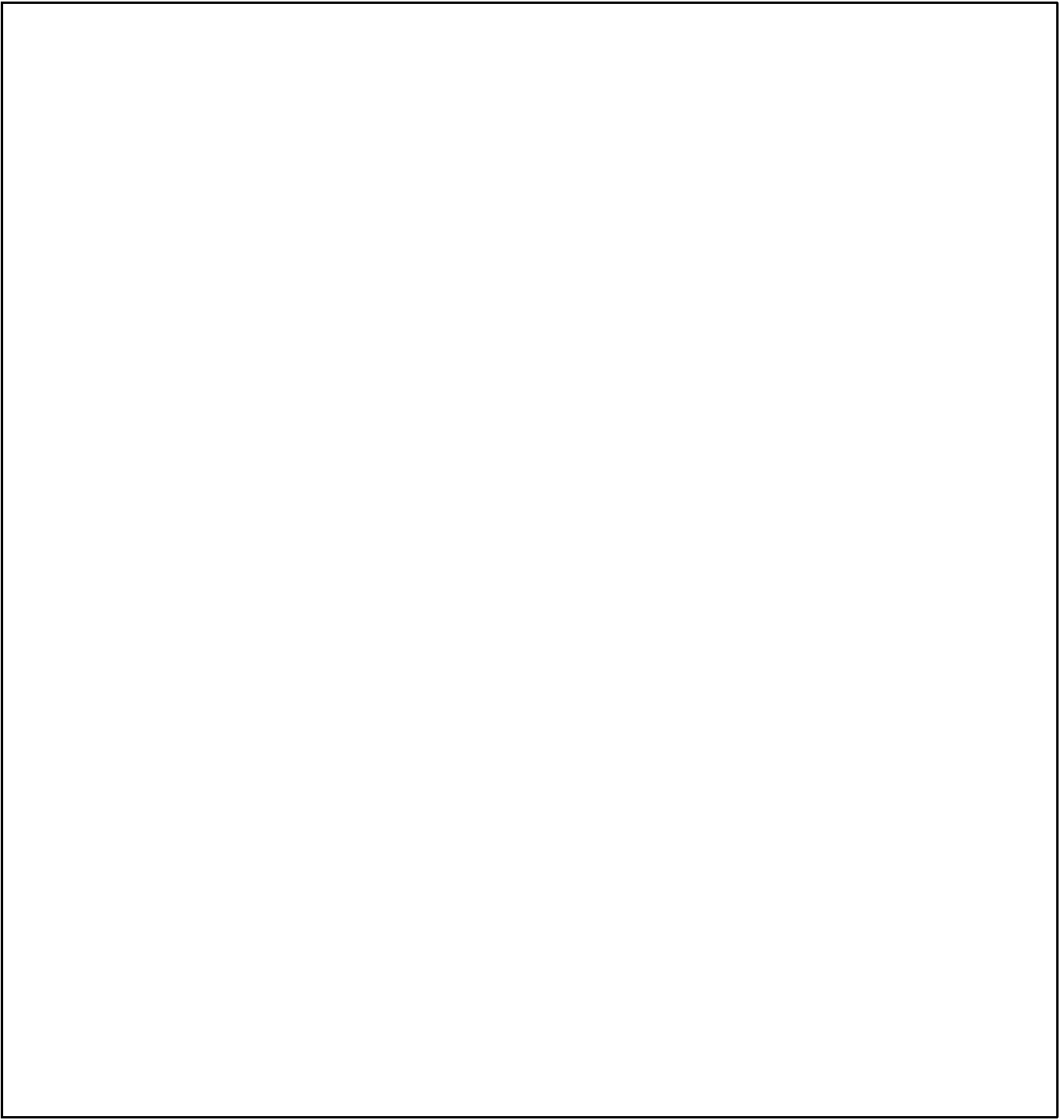

\picplace{19cm}
\caption[]{
Same as Fig.~\protect\ref{fig:neutA64} but for model A128 at
time $t = 8.80$~ms. The higher resolution of this simulation
allows more fine structure to be visible
}
\label{fig:neutA128}
\end{figure*}

\begin{figure*}
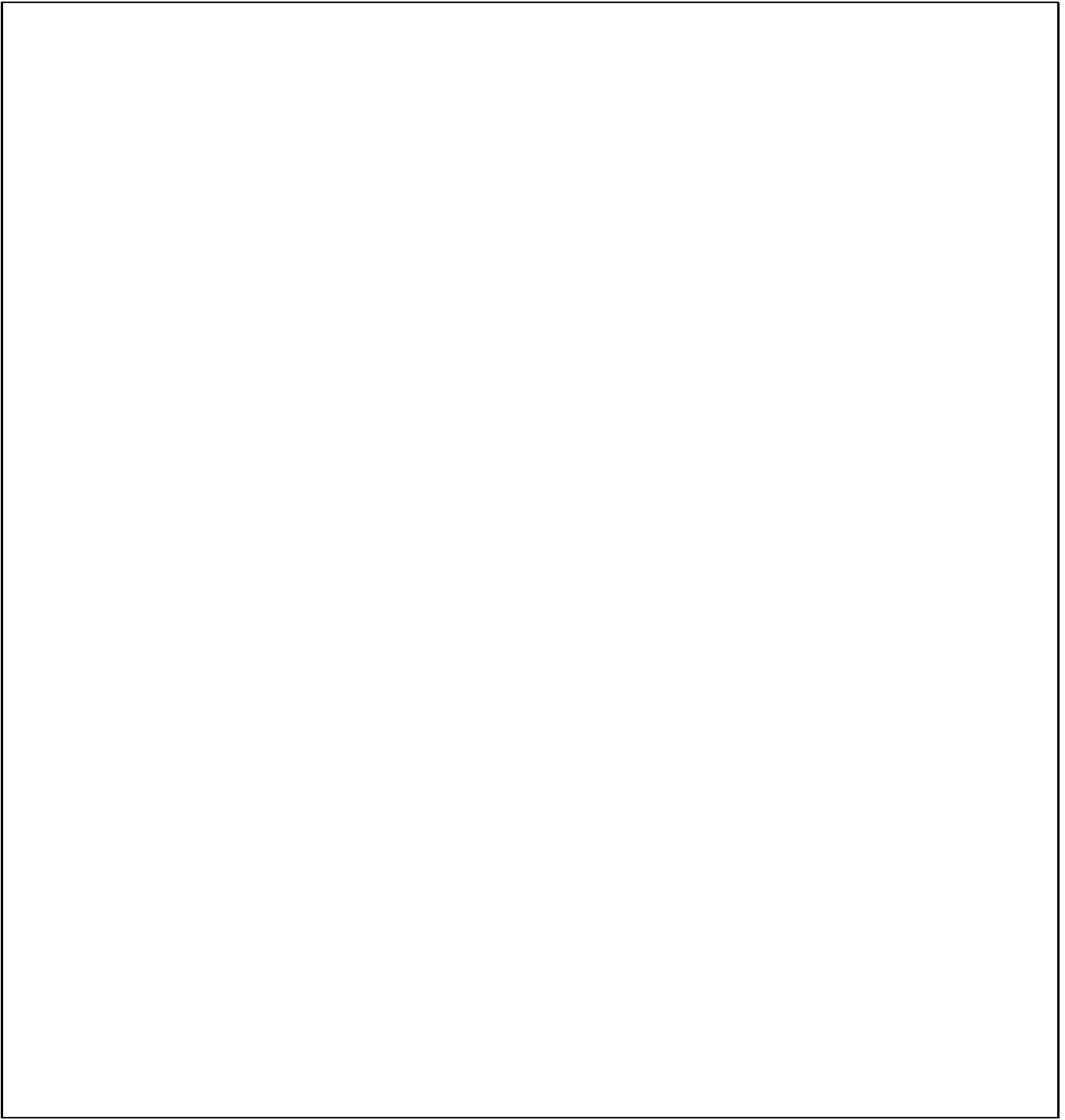

\picplace{19cm}
\caption[]{
Energy loss rates by neutrino emission in two orthogonal planes
vertical to the orbital plane for model A128 at time
$t = 8.80$~ms. The displayed information is the same as in 
Fig.~\protect\ref{fig:neutA64}
}
\label{fig:nepeA128}
\end{figure*}

\begin{figure*}
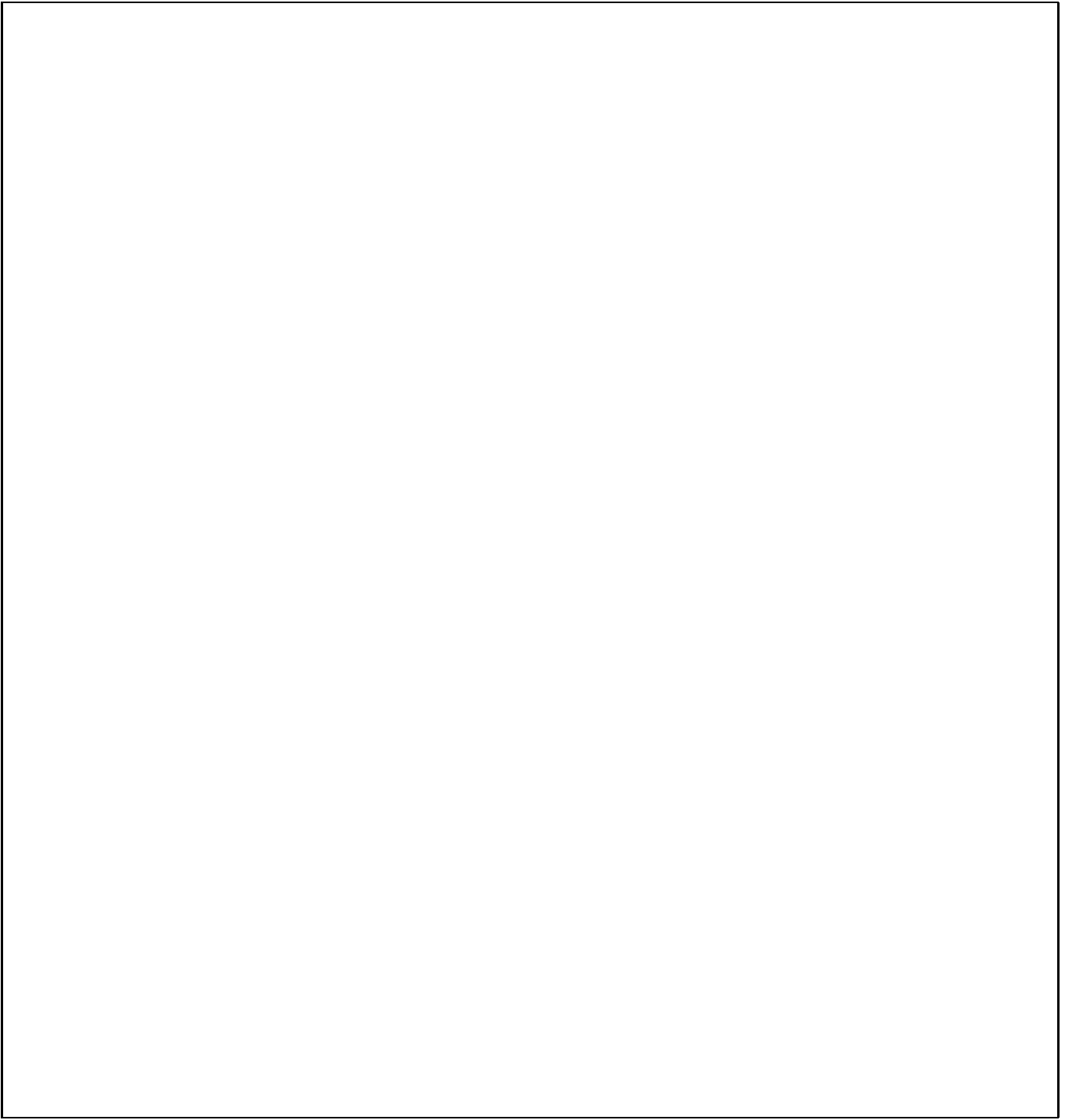

\picplace{19cm}
\caption[]{
Same as Fig.~\protect\ref{fig:neutA64} but for model B64 at time
$t = 10.55$~ms
}
\label{fig:neutB64}
\end{figure*}


\begin{figure*}
\picplace{19cm}
\caption[]{
Same as Fig.~\protect\ref{fig:neutA64} but for model C64 at time
$t = 10.86$~ms
}
\label{fig:neutC64}
\end{figure*}

\begin{figure*}
\picplace{19cm}
\caption[]{
Same as Fig.~\protect\ref{fig:nepeA128} but for model C64 at 
time $t = 10.86$~ms
}
\label{fig:nepeC64}
\end{figure*}

Figures~\ref{fig:neutrad}--\ref{fig:mene} 
display the results for the neutrino fluxes and mean 
energies of the emitted neutrinos for our models A64, B64, C64,
and A128. In Fig.~\ref{fig:neutrad} one can see that the total 
neutrino luminosities start to increase above 
$\sim 10^{52}\,{\rm erg/s}$ at about 2.5--3.5~ms after the
start of the simulations. This is about the time when the spiral
arm structures become dispersed into a spread-out cloud of material 
that surrounds the merger and is heated by the interaction with
compression waves. At the same time the temperatures in the 
interior of the massive central body reach their peak values. 
Since the neutrino luminosities are by far dominated by the 
contributions from the disk emission (see below), the heating of 
the cloud and torus material is reflected in a continuous increase
of the neutrino fluxes. Also, the dynamical expansions and 
contractions caused by the oscillations and wobbling of the 
central body impose fluctuations on the light curves. The
periodic expansions and contractions have particularly large 
amplitudes in model C64 because of the anti-spin setup of the
neutron star rotations which leads to very strong internal shearing
and turbulent motions after merging as well as to higher 
temperatures than in the other models (see Table~\ref{tab:models}). 
The light curve fluctuations in model C64 are as large
as 25--30\% of the average luminosity and proceed with a 
period of 0.5--1~ms which is about the dynamical time scale of
the merger. When the oscillating central
body enters an expansion phase, 
very hot matter that is located in a shell around the
central core of the merger, is swept to larger radii. In course of 
the expansion the neutrino optical depth decreases and the very hot 
material releases enhanced neutrino fluxes. The neutrino outburst 
is terminated, when the surface-near matter has cooled by adiabatic
expansion or when re-contraction sets in and the neutrino optical 
depth increases again. At times later than about 6--8~ms
quasi-stationary values of the fluxes are reached which are 
between about $8\cdot 10^{52}\,{\rm erg/s}$ in case of model
A128 and about $1.3\cdot 10^{53}\,{\rm erg/s}$ for C64.

Figure~\ref{fig:neutot} shows that the energy lost in gravitational
waves during the merging is about two orders of magnitude larger
than the energy radiated away in neutrinos during the 
simulated period of approximately 10~ms.
While the gravitational wave luminosity peaks
around the time when the dynamical instability of the orbit sets 
in and the two neutron stars start to interact dynamically
and fuse into a single object ($t$ between 0.5~ms and 1.5~ms),
the energy emitted in neutrinos becomes sizable only after the
extended toroidal cloud of matter has formed around the merged
stars. 

The heated torus or ``disk'' consists of decompressed 
neutron star matter with
an initially very low electron number fraction $Y_e$ between
about 0.01 and 0.04. The neutrino emission of the disk
is therefore clearly dominated by the loss of electron
antineutrinos which are primarily produced in the process 
$e^+ + n \to p + \bar\nu_e$, because positrons are rather 
abundant in the hot and only moderately degenerate, neutron-rich 
matter (see Sect.~\ref{sec:therm}). 
In Fig.~\ref{fig:ergne} one sees that the $\bar\nu_e$
luminosity $L_{\bar\nu_e}$ is a factor 3--4.5 larger than 
$L_{\nu_e}$ and between $5.5\cdot 10^{52}\,{\rm erg/s}$
for model B64 and about $6.7\cdot 10^{52}\,{\rm erg/s}$
for C64 when quasi-stationary conditions have been established 
after about 6~ms from the start of the simulations 
(Table~\ref{tab:models}). The sum
of all heavy-lepton neutrino fluxes which is four times the individual
luminosities of $\nu_{\mu}$, $\bar\nu_{\mu}$, $\nu_{\tau}$ or
$\bar\nu_{\tau}$, is around $4\cdot 10^{52}\,{\rm erg/s}$.
The better resolved model A128 yields slightly smaller values
for all luminosities than A64 (Table~\ref{tab:models}).

The mean energies of the emitted neutrinos are displayed in
Fig.~\ref{fig:mene} for models A64, B64, and C64. The differences
between the different models are smaller than in case of the
neutrino luminosities. This means that the effective temperatures
in the neutrinospheric regions are very similar in all models.
Electron neutrinos are emitted with 
an average energy of 12--13~MeV, electron antineutrinos have
19--21~MeV, and heavy lepton neutrinos leave the merger with mean
energies of 26--28~MeV during the stationary phase. 
Except for the dominance of $L_{\bar\nu_e}$ relative to 
$L_{\nu_e}$, both the mean neutrino energies and the neutrino
luminosities are in good overall
agreement with typical numbers obtained in stellar core 
collapse and supernova simulations at a stage some time after
the prompt $\nu_e$ burst has been emitted and after the collapsed
stellar core has heated up along with the post-shock settling
(see, e.g., Mayle et al.~1987, Myra \& Burrows 1990, Bruenn 1993, 
Bruenn et al.~1995). 
 
Just like the neutrino emission from the less opaque, hot mantle region      
of the protoneutron star dominates the supernova neutrino fluxes for
several ten to some hundred milliseconds after core bounce, more
than 90\% of the neutrino emission of the merger comes from
the extended, hot torus and only a minor fraction of less than 
10\% originates from the very opaque and dense central core of the
merged object. Figures~\ref{fig:neutA64}, \ref{fig:neutA128},
\ref{fig:neutB64}, and \ref{fig:neutC64} display the local 
emission rates of $\nu_e$, $\bar\nu_e$, the sum of $\nu_x$, 
and the sum of all neutrino types in the orbital plane
for models A64, A128, B64, and C64, respectively, at a time when
quasi-stationary conditions have been established.
Figures~\ref{fig:nepeA128} 
and \ref{fig:nepeC64} give the corresponding information in two
orthogonal cut planes vertical to the equatorial plane for the
representative models A128 and C64. One can
clearly see that the surface-near regions, in particular the 
toroidal cloud of matter surrounding the central, compact body,
emits neutrinos at much higher rates. 

Using the values from these figures, one can estimate the
relative importance of core and disk emission.
The dense core regions ($\rho \ga 10^{13}\,{\rm g/cm}^3$) inside
a radius of 
$R_{\rm c}\approx 20$~km lose neutrino energy with a typical 
rate of 
$Q_{\varepsilon}\approx 10^{32}\,{\rm erg/cm}^3/{\rm s}$ and
account for a luminosity of about 
$L_{\nu,{\rm c}}\approx (4\pi/3)R_{\rm c}^3 Q_{\varepsilon} \approx 
3.4\cdot 10^{51}\,{\rm erg/s}$, while the emission from the
surrounding torus-shaped disk region with outer radius 
$R_{\rm d}\approx 40$~km, height $2h \approx 40$~km,
and typical total neutrino energy loss rate 
$Q_{\varepsilon}\approx 5\cdot 10^{32}\,{\rm erg/cm}^3/{\rm s}$
emits a luminosity of roughly
$L_{\nu,{\rm d}}\approx (2\pi R_{\rm d}^2h - 4\pi R_{\rm c}^3/3)
Q_{\varepsilon}\approx 8.4\cdot 10^{52}\,{\rm erg/s}$. The sum
of $L_{\nu,{\rm c}}$ and $L_{\nu,{\rm d}}$ is of the size of 
the results displayed in Fig.~\ref{fig:neutrad}.

Figures~\ref{fig:neutA64}--\ref{fig:nepeC64} also demonstrate that
the disk emits $\bar\nu_e$ fluxes that are larger than the 
$\nu_e$ fluxes. Electron neutrinos are most strongly emitted 
from surface-near regions where the optical depth to $\nu_e$
by absorption on neutrons ($\nu_e + n \to p + e^-$) 
in the neutron-rich matter is smallest. 
In contrast, the production of electron antineutrinos by positron
captures on neutrons and of heavy-lepton neutrinos via 
electron-positron pair annihilation ($e^- + e^+ \to \nu + \bar\nu$)
requires the presence of large numbers of positrons and therefore
occurs predominantly in those parts of the disk which have been
heated most strongly and have a low electron degeneracy. 
Since $\bar\nu_e$ are absorbed on the less abundant protons 
in the inverse $\beta^+$ process and
the opacity to $\nu_x$ is primarily caused by neutrino-nucleon
scatterings only, electron antineutrinos and heavy-lepton 
neutrinos can escape on average from deeper and hotter
layers than electron neutrinos 
(see Figs.~\ref{fig:neutA64}--\ref{fig:nepeC64}). 
This explains the higher mean 
energies of the emitted $\bar\nu_e$ and $\nu_x$.

As expected from the very similar neutrino luminosities and nearly
equal mean energies of emitted neutrinos, the neutrino emission maps
do not reveal major differences between the models A64, B64, and C64.
Most neutrino emission comes from the disk region where the neutrino 
optical depths are lower, and in Figs.~\ref{fig:neutA64}, 
\ref{fig:neutB64}, and \ref{fig:neutC64} one can clearly recognize
the high-density, very opaque inner part of the core of the merger 
from its roughly two orders of magnitude smaller energy loss rates. The
cuts perpendicular to the orbital plane (see Fig.~\ref{fig:nepeC64}) 
show the ring-like main emission region that surrounds the central
core and has a banana- or dumb-bell shaped cross section.
Model A128 has significantly more fine structure 
but the overall features and characteristics of the neutrino emission
do not change with the much better numerical resolution of this 
simulation.

\begin{figure*}
 \begin{tabular}{cc}
 \epsfxsize=8.8cm \epsfclipon \epsffile{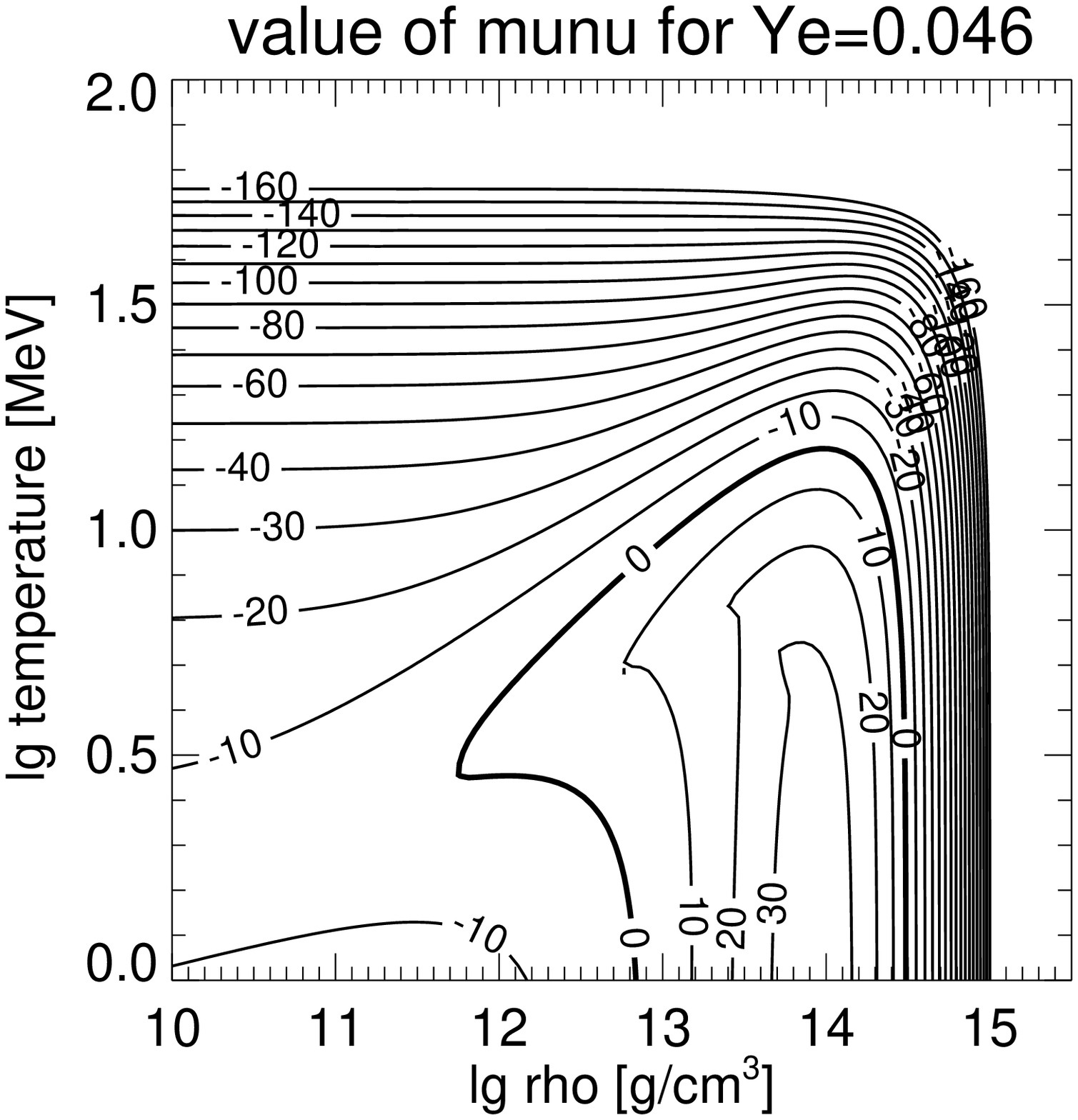} &
 \epsfxsize=8.8cm \epsfclipon \epsffile{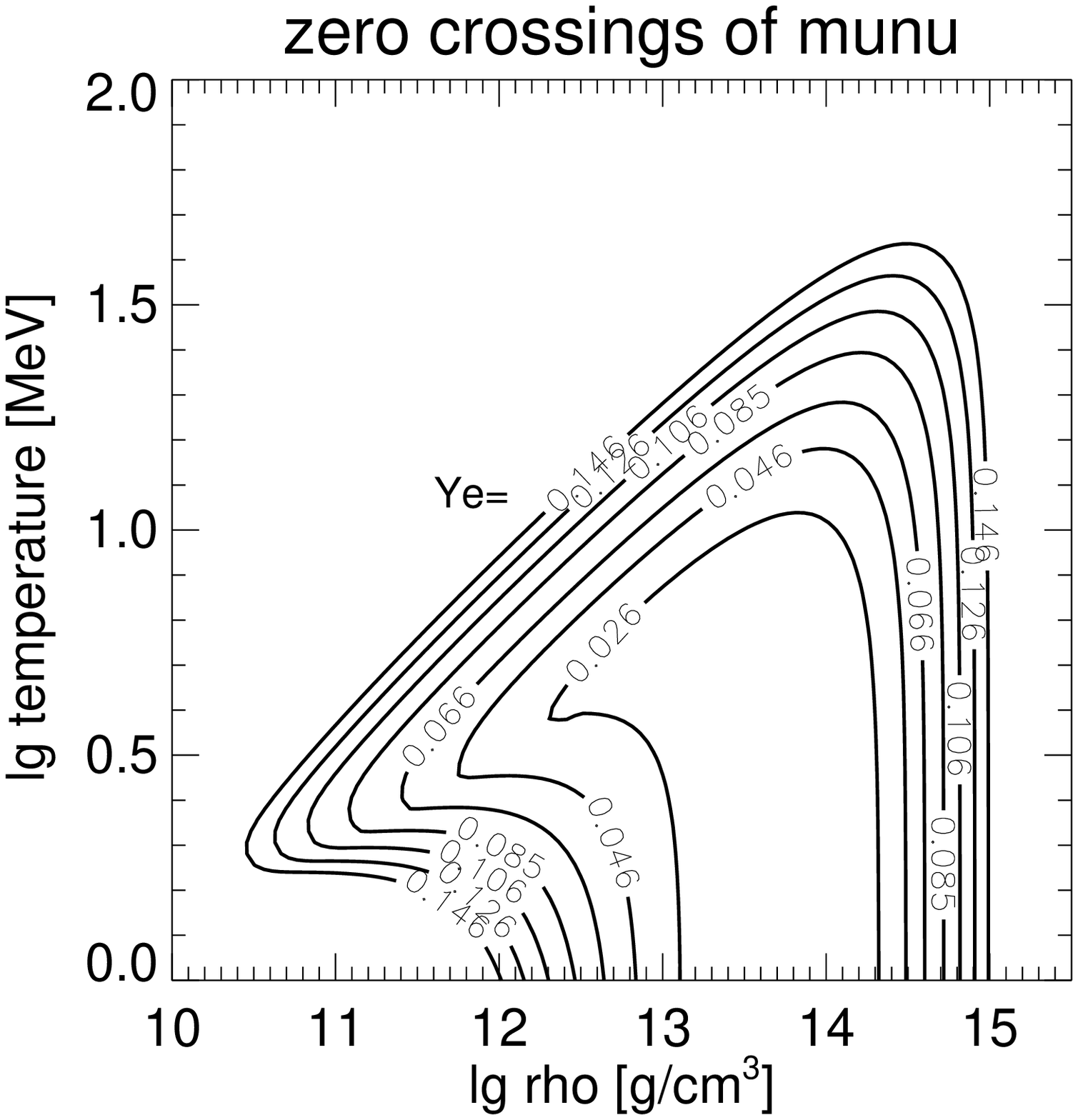} \\
 \parbox[t]{8.8cm}{\caption[]{Contours of the neutrino chemical potential
	 (measured in MeV)
	 in the temperature-density plane for fixed electron fraction $Y_e$.
	 The contours are spaced with increments of 10 and are labeled with
	 their respective values}\label{fig:munuYe}} &
  \parbox[t]{8.8cm}{\caption[]{Contours of vanishing neutrino chemical
	 potential ($\mu_\nu=0$) in the temperature-density plane for
	 different values of the electron fraction $Y_e$ as indicated 
	 by the line labels}\label{fig:munu}} \\
 \end{tabular}
\end{figure*}

\begin{figure*}
 \begin{tabular}{cc}
  \epsfxsize=8.8cm \epsfclipon \epsffile{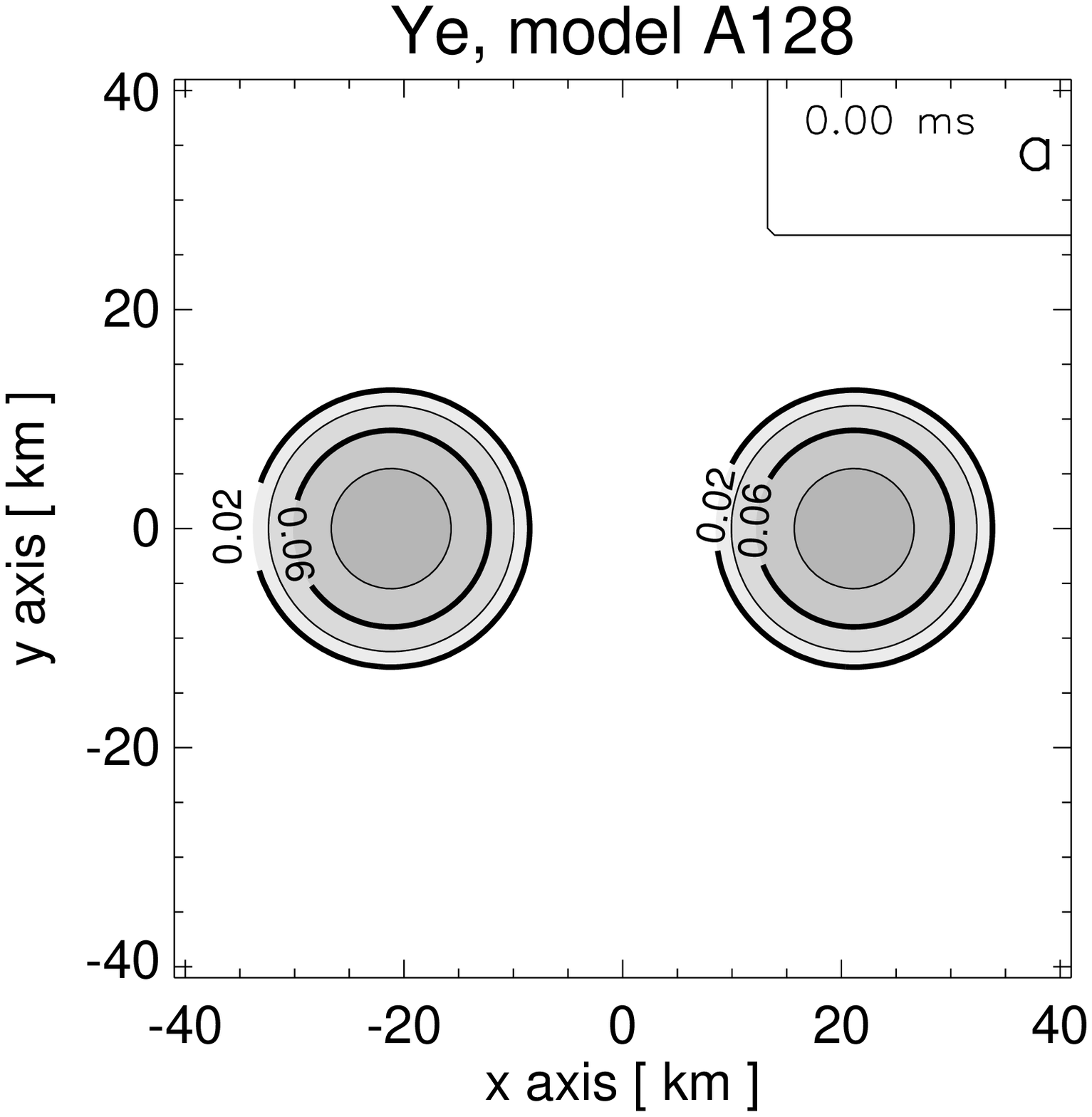} &
  \epsfxsize=8.8cm \epsfclipon \epsffile{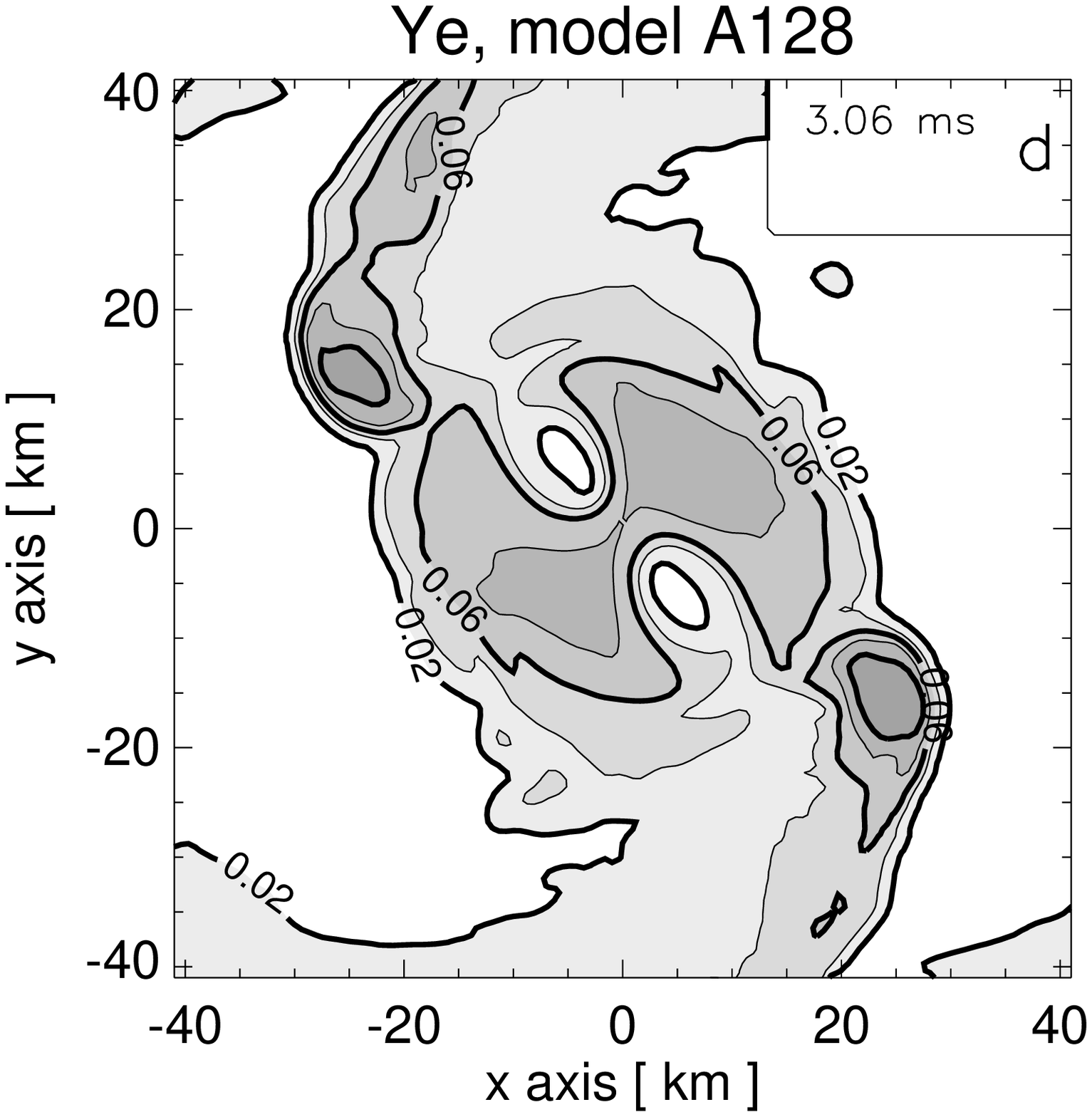} \\
  \epsfxsize=8.8cm \epsfclipon \epsffile{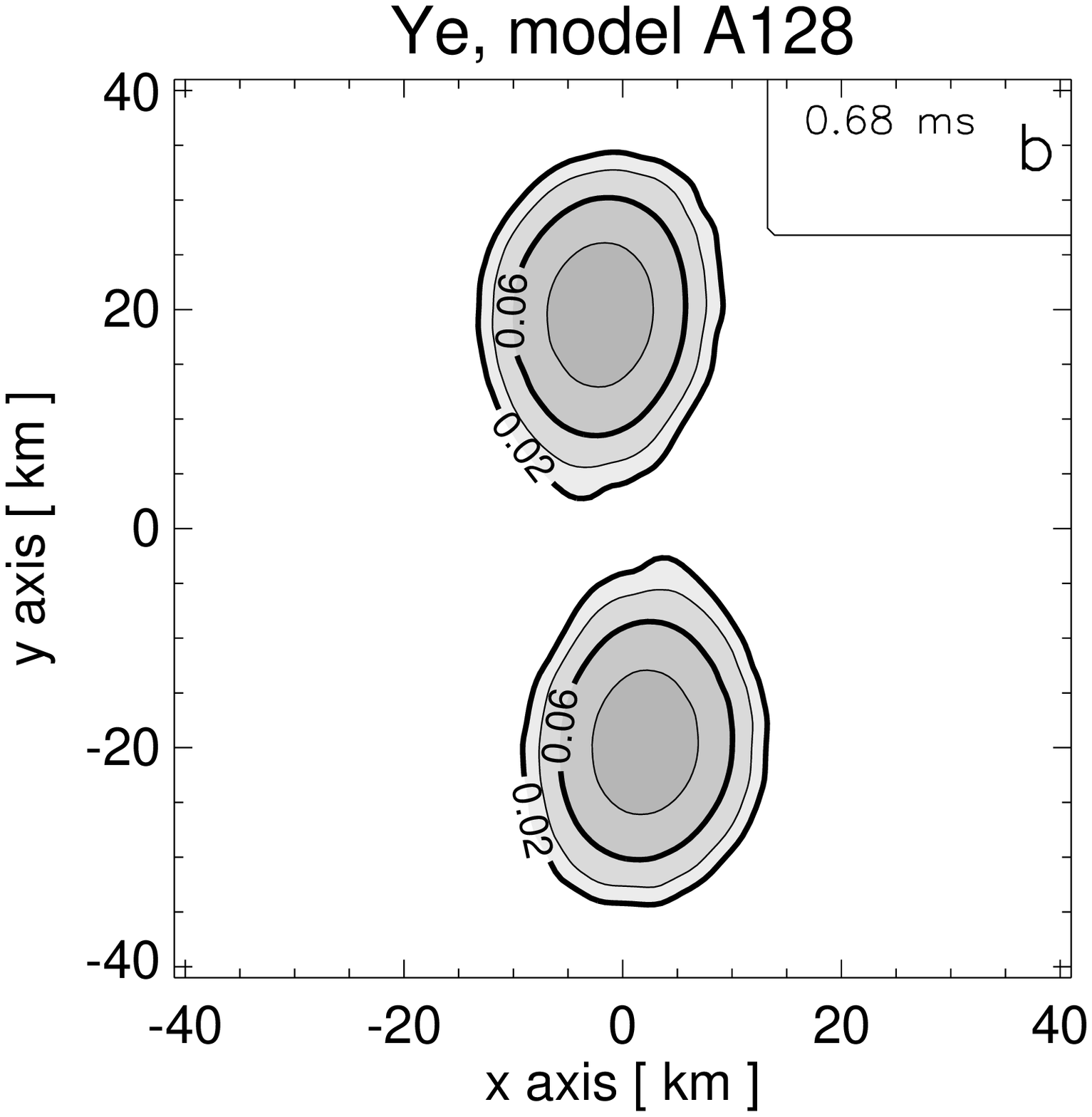} &
  \epsfxsize=8.8cm \epsfclipon \epsffile{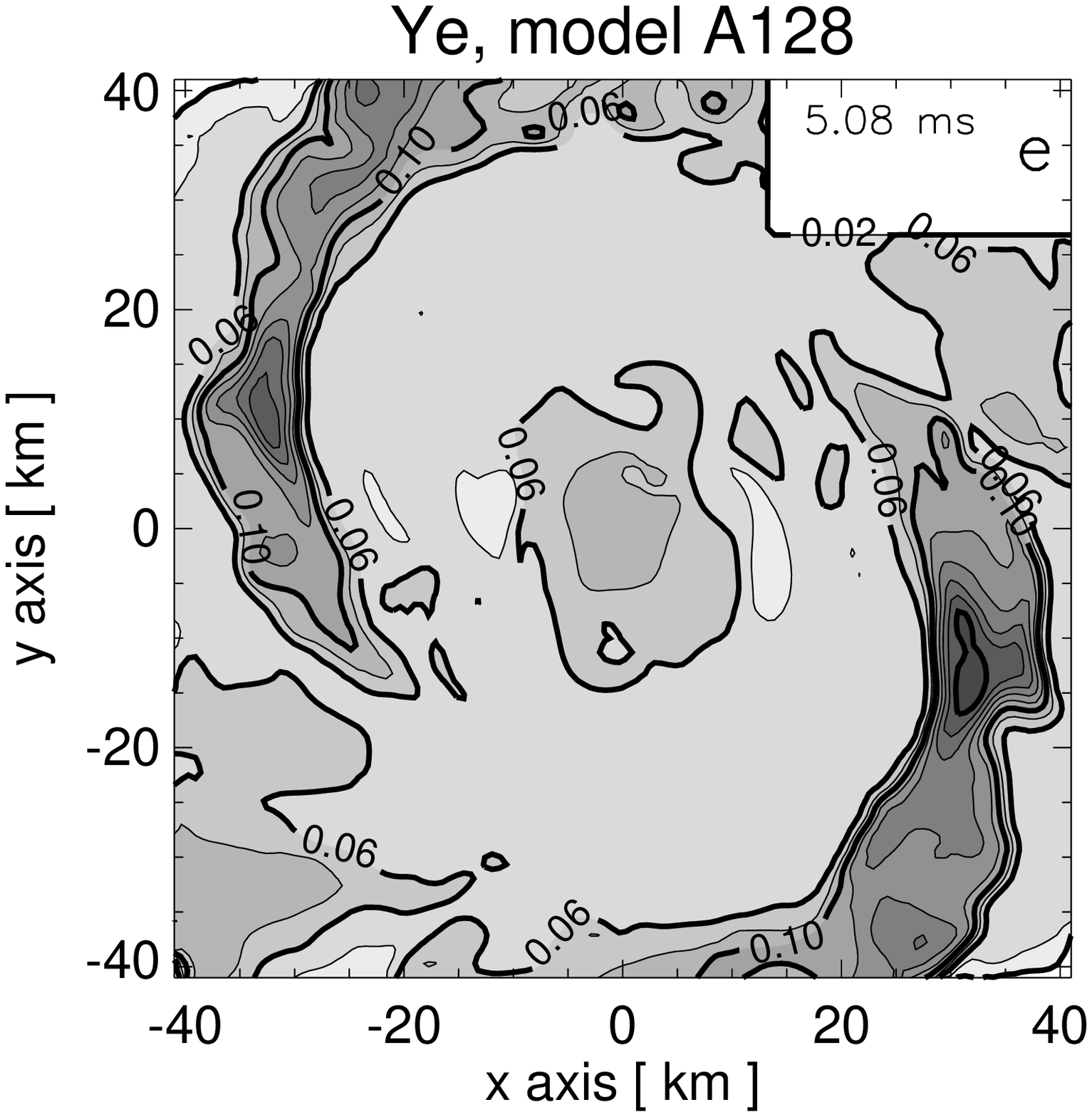} \\
  \epsfxsize=8.8cm \epsfclipon \epsffile{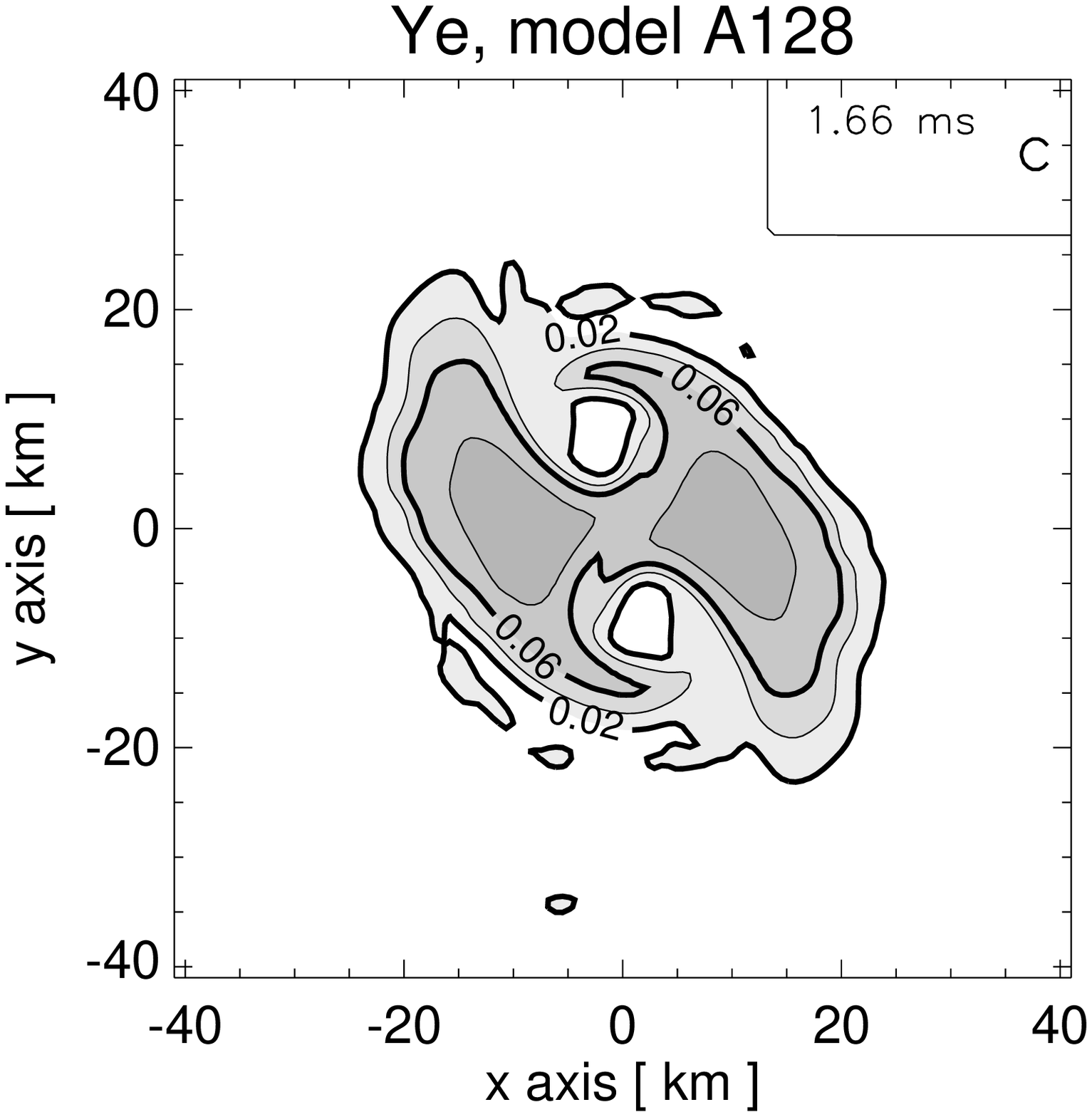} &
  \epsfxsize=8.8cm \epsfclipon \epsffile{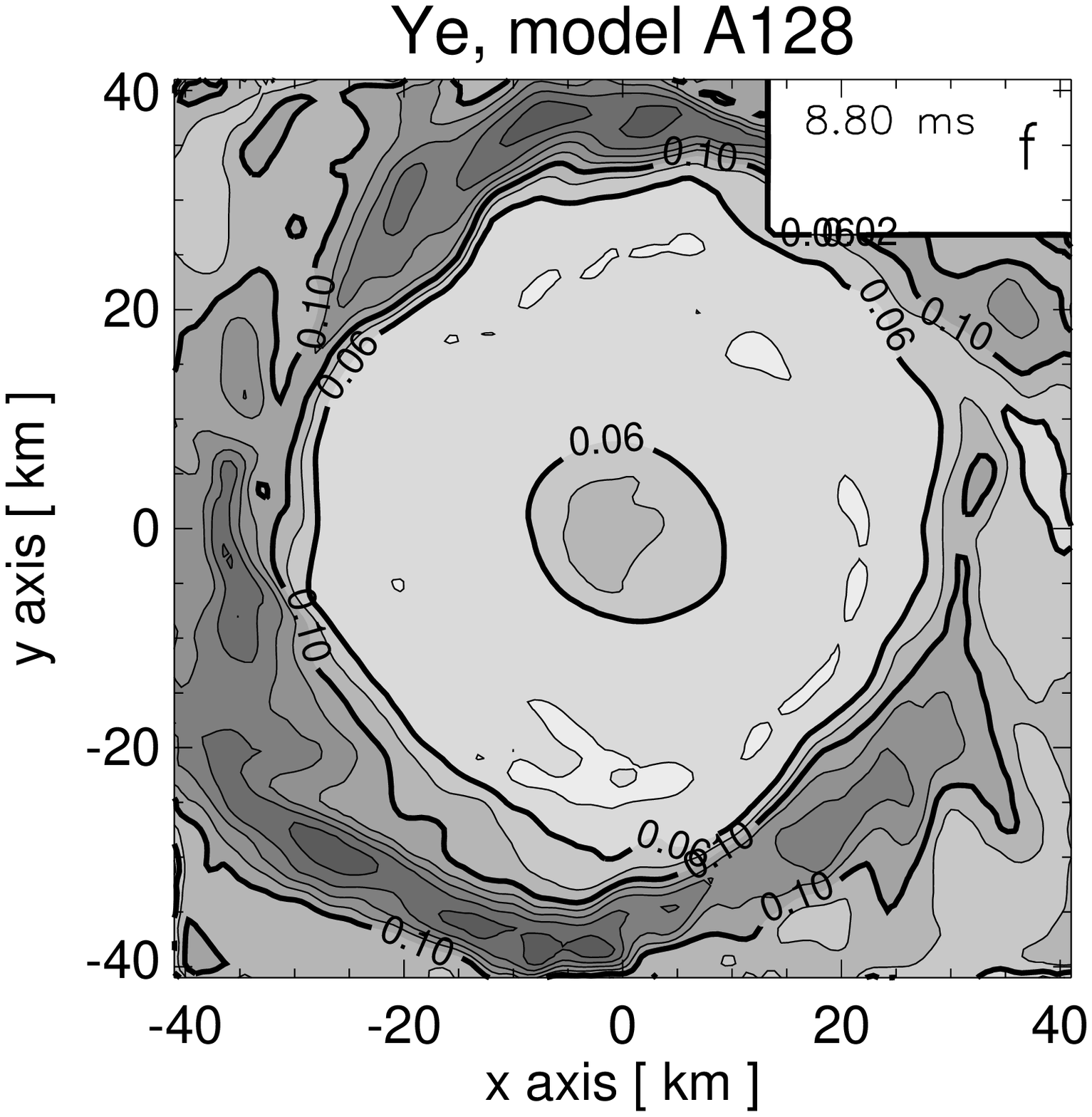}
 \end{tabular}
\caption[]{Time evolution of the spatial distribution of the
electron fraction $Y_e$ in the orbital plane of model A128. The
times of the snapshots are given in the upper right corners of 
the panels. The
contours are linearly spaced with intervalls of 0.02, bold lines
are labeled with their respective values, and the grey shading 
emphasizes the contrasts, higher values of $Y_e$ being associated
with darker grey 
}
\label{fig:A128Ye}
\end{figure*}

\begin{figure*}
 \begin{tabular}{cc}
  \epsfxsize=8.8cm \epsfclipon \epsffile{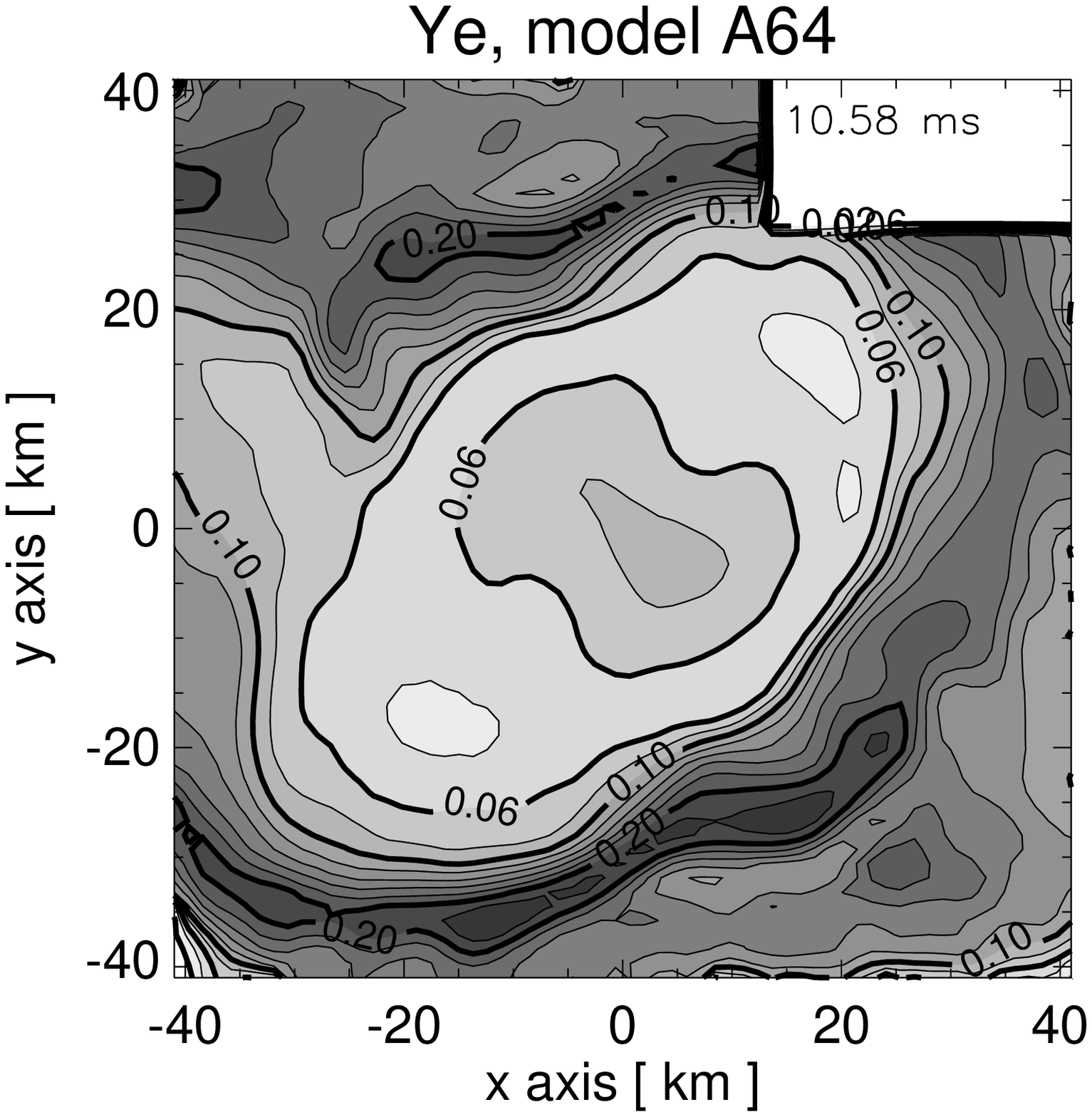}
  \put(-1.0,6.8){{\Large \bf a}} &
  \epsfxsize=8.8cm \epsfclipon \epsffile{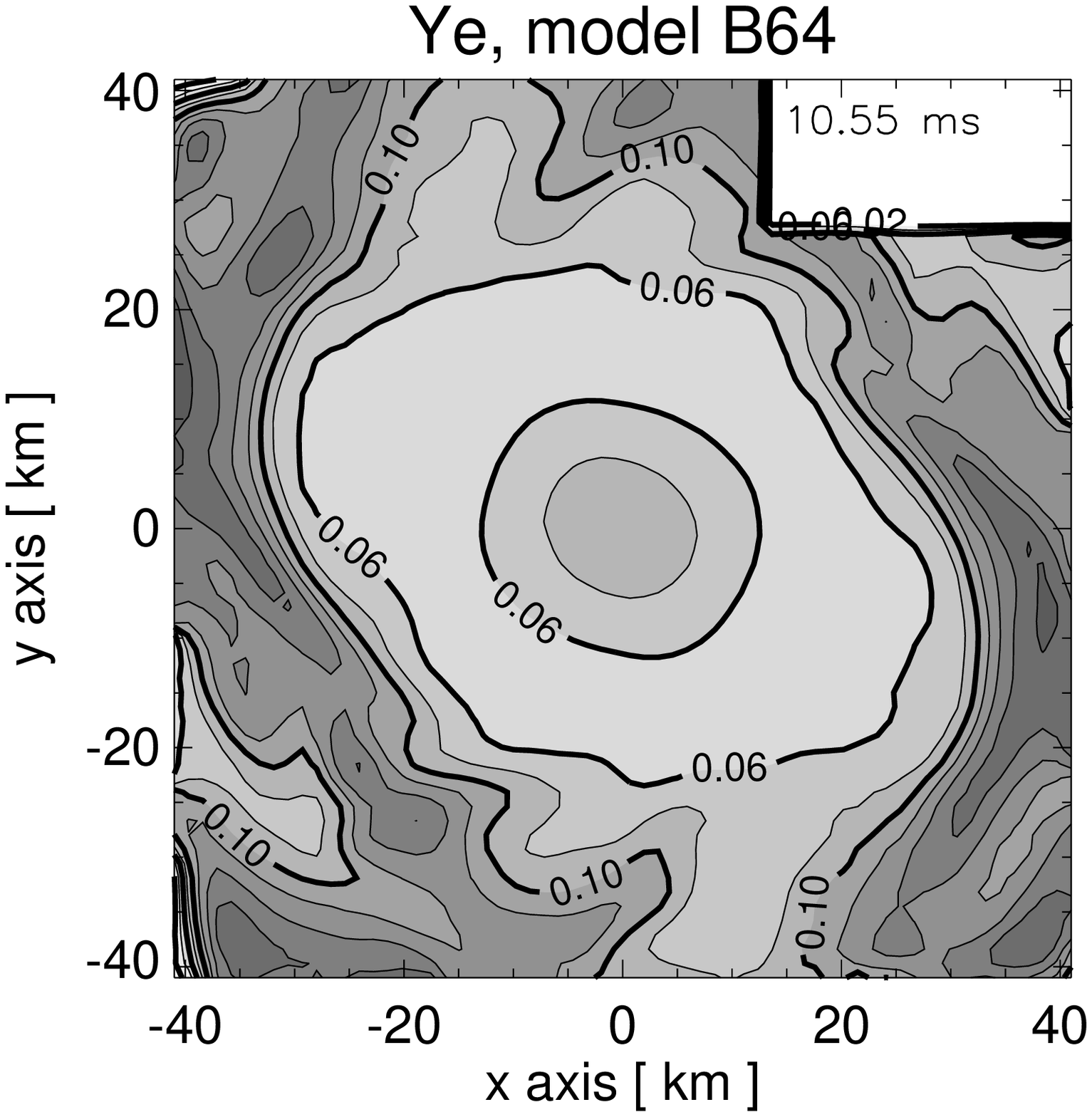}
  \put(-1.0,6.8){{\Large \bf b}} \\
  \epsfxsize=8.8cm \epsfclipon \epsffile{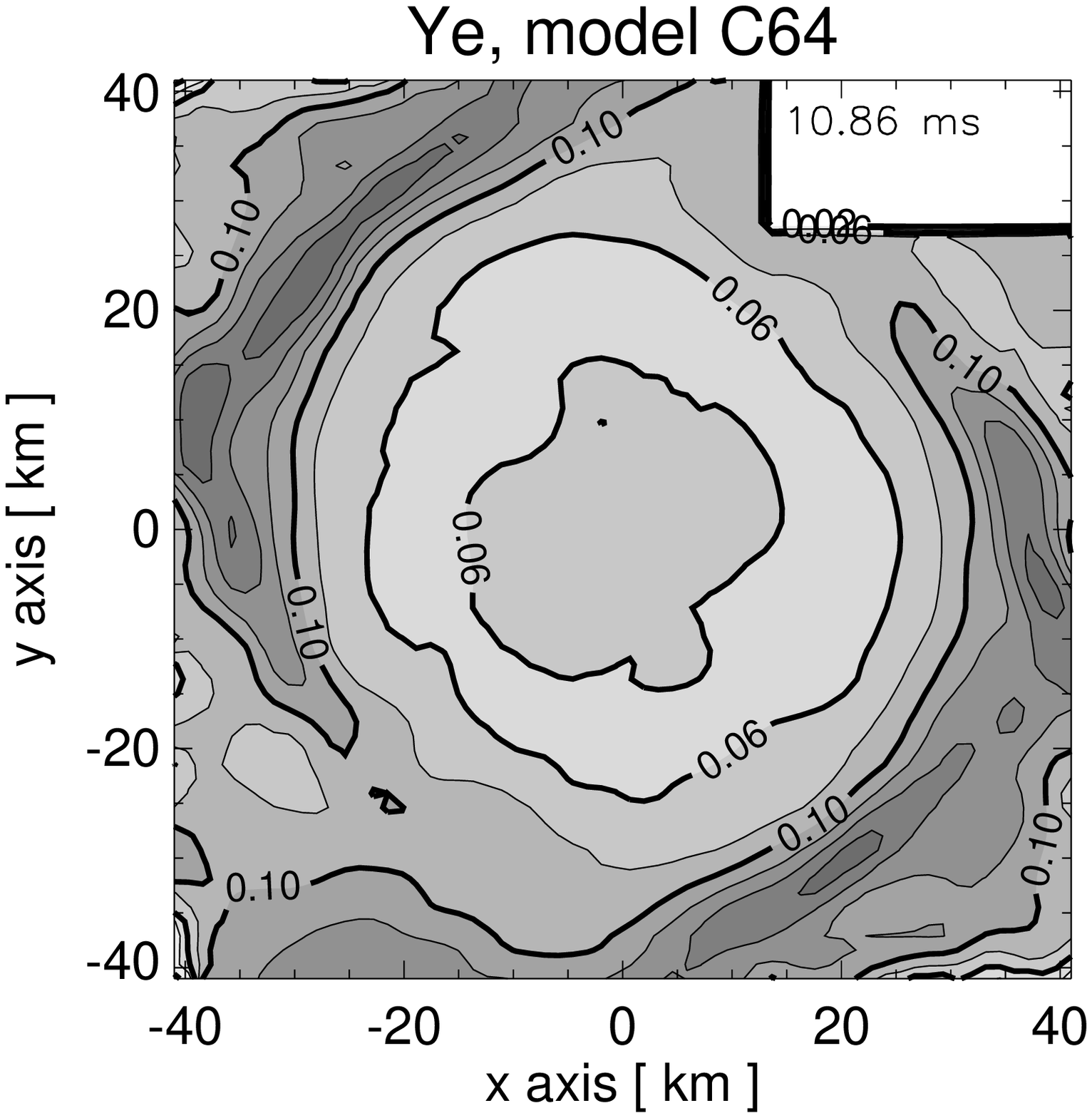}
  \put(-1.0,6.8){{\Large \bf c}} &
\raisebox{8cm}{\parbox[t]{8.8cm}{
\caption[]{\label{fig:Yes}
The spatial distributions of the electron fraction $Y_e$
in the orbital plane for models A64, B64, and C64 at the end of
the calculations (times given in the top right corners).
The contours are linearly spaced with intervalls 
of 0.02 and the bold contours are labeled with their respective 
values. The grey shading emphasizes the levels with darker shading
indicating higher values of $Y_e$. The plots have to be compared
with panel {\bf f} of Fig.~\protect\ref{fig:A128Ye} which shows
the same information for model A128 at time $t = 8.80$~ms
}}}
 \end{tabular}
\end{figure*}

\begin{figure*}
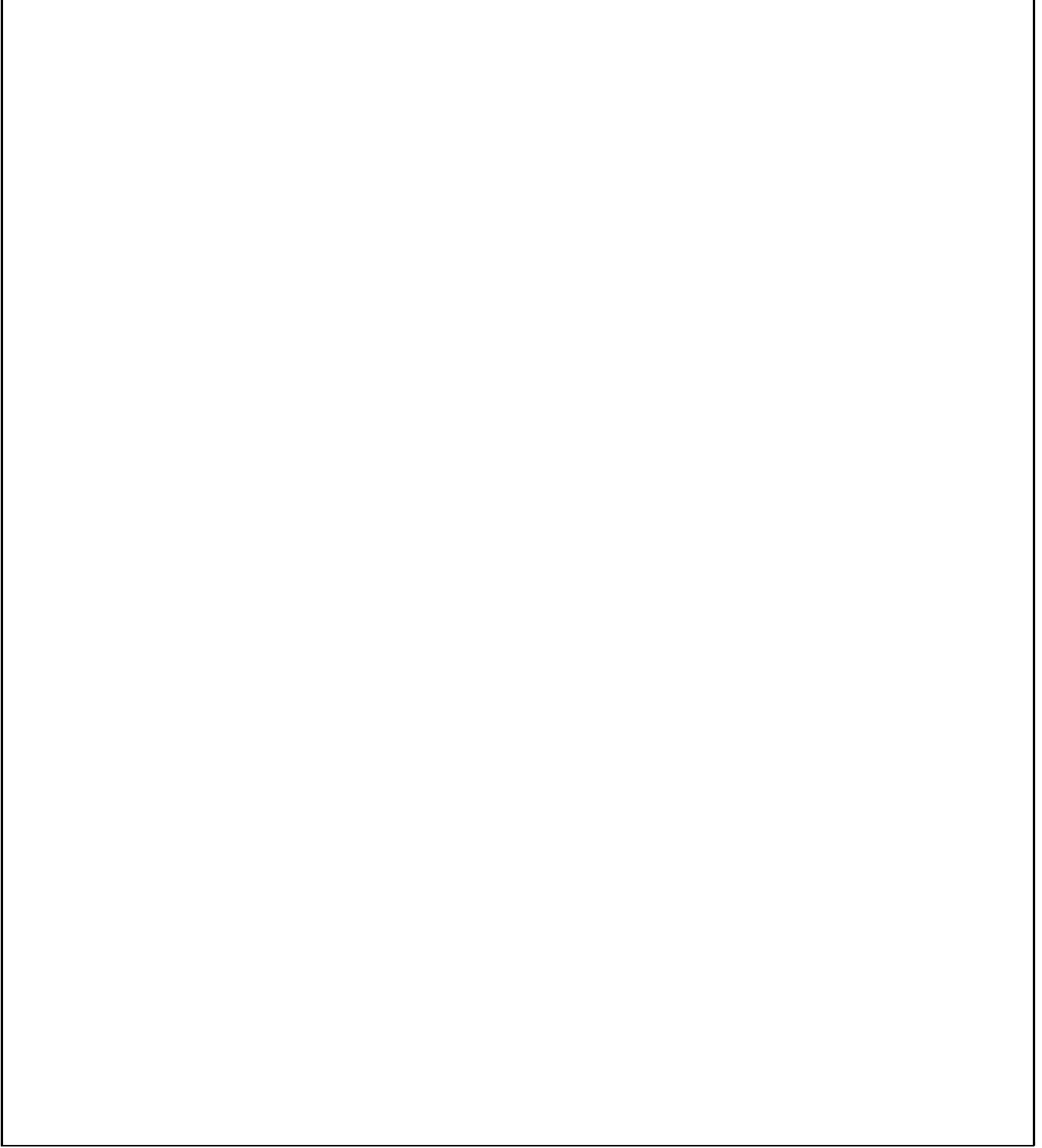

\caption[]{\label{fig:A128collect}
Cuts in the orbital plane of model A128 showing different 
thermodynamical and composition parameters at the end of the
simulated evolution (time $t = 8.80$~ms). Panel {\bf a} gives the
electron degeneracy parameter $\eta_e$ (contours linearly spaced
with steps of one unit), panel {\bf b} shows the electron neutrino
degeneracy parameter $\eta_{\nu_e}$ (contours linearly spaced with 
steps of 0.5 units), panel {\bf c} displays the temperature 
distribution (contours linearly spaced with steps of 2~MeV),
panel {\bf d} is a plot of the entropy per nucleon (contours linearly
spaced with steps of 1~$k_{\rm B}$/nucleon), and panel {\bf e}
informs about the mass fraction of $\alpha$ particles in the medium
(contours logarithmically spaced in steps of one dex)
}
\picplace{20cm}
\end{figure*}

\subsection{Thermodynamics and composition\label{sec:therm}}

Figure~\ref{fig:munuYe} displays contour levels in the
temperature-density plane of the electron neutrino chemical
potential (measured in MeV) for an
electron fraction of $Y_e = 0.046$. In Fig.~\ref{fig:munu}
the contours corresponding to vanishing electron neutrino
chemical potential $\mu_{\nu_e} = 0$ are plotted for different
values of $Y_e$. Figures~\ref{fig:munuYe} and \ref{fig:munu} 
provide information about how chemical equilibrium is shifted
according to the equation of state of Lattimer \& Swesty (1991)
when neutron star matter changes its temperature-density state.

Cold neutron star matter at neutrinoless $\beta$-equilibrium is in 
a state with $\mu_{\nu_e} = 0$. When such gas of the outermost
$\sim 0.1$--$0.2\,M_{\odot}$ of the neutron star with density 
$\rho\la 10^{14}\,{\rm g/cm}^3$ and electron fraction $Y_e\la 0.025$ 
(see Fig.~2 in Paper~I) is expanded and heated while the lepton 
fraction $Y_{\rm lep} = Y_e + Y_{\nu_e} \approx Y_e$ stays 
roughly constant as it is the case for a fast change where
neutrino losses are too slow to compete, the $\beta$-equilibrium
is shifted into the region of negative $\mu_{\nu_e}$ values.
This usually implies that the electron degeneracy of the matter 
is drastically decreased,
too, because the electron chemical potential $\mu_e = \mu_{\nu_e}
+ \mu_n - \mu_p$ also drops when $\mu_{\nu_e}$ 
attains negative values. For hot gas (i.e., gas at conditions
lying above the nose-like feature of the curves in 
Figs.~\ref{fig:munuYe} and \ref{fig:munu}) the state with 
$\mu_{\nu_e} = 0$ corresponds to a higher value of $Y_e$ which is seen 
in Fig.~\ref{fig:munu} by moving along lines parallel to the ordinate.
On the lepton-number loss time scale 
associated with neutrino emission, the gas will tend
to evolve again towards the $\beta$-equilibrium with $\mu_{\nu_e} = 0$
by an enhanced production and emission of $\bar\nu_e$
relative to $\nu_e$. 
Notice that from Figs.~\ref{fig:munuYe} and \ref{fig:munu} 
one infers that the same arguments are true for the case that neutron 
star matter with densities above $\rho \approx 10^{14}\,{\rm g/cm}^3$ 
is strongly compressed. From the properties of the
high-density equation of state we therefore deduce that 
during the coalescence of neutron star binaries the hot gas
in the compact, compressed core region of the merger as well as 
the heated, decompressed disk matter will radiate $\bar\nu_e$
more copiously than $\nu_e$. 

This explains the relative sizes of electron neutrino and antineutrino 
luminosities as discussed in Sect.~\ref{sec:nuem}.
Driven by this imbalance of the emission of $\nu_e$ and $\bar\nu_e$,
the initially very neutron-rich matter ($Y_e \la 0.095$ everywhere in
the neutron star, see Fig.~2 in Paper~I) gains electron lepton number
and becomes more proton-rich again. Figure~\ref{fig:A128Ye} shows
this evolution for model A128 from the start of the simulation until
its end at 8.80~ms. The snapshots of the $Y_e$ distribution in the
orbital plane visualize how, as a consequence of the rapid neutrino
loss from the disk region, $Y_e$ in this region climbs from initial
values of 0.02--0.06 to values of more than 0.18 in some parts.
In the core region the neutrino emission proceeds much more slowly 
so that 
$Y_e$ changes only slightly during the simulated time. If we continued
our computations for a long enough time to see the matter cooling 
again by neutrino losses (provided the configuration is stable for
a sufficiently long period), this process of $Y_e$ increase would
again be inverted and the gas would evolve towards the cold, 
deleptonized, very neutron-rich state again. 

Figure~\ref{fig:Yes} displays the final situations 
($t \approx 10$--11~ms) in the models
A64, B64, and C64 to be compared with panel~{\bf f} of 
Fig.~\ref{fig:A128Ye}. As in case of the neutrino emission,
one notices very similar properties of all four models. In model A64
the peak $Y_e$ values in the disk region are as high as 0.22.
Note that in all four models very neutron-rich matter is swept
off the grid, a tiny fraction of which might potentially become
unbound (see Sect.~\ref{sec:hydevol} and Paper~I). It is also
interesting to see the still very elongated and deformed 
neutron-rich inner region of models A64 and B64 which indicates
ongoing strong dynamical and pulsational activity of the massive 
core. This is not so pronounced in model C64 where
the anti-spin setup of the initial model has caused the
dissipation of a large fraction of the rotational energy 
during the coalescence of the
neutron stars. Model A128 has also a much more circular core
region because, as described in Paper~I, the better resolution 
allows for a much more fine-granular flow pattern which contains
a large fraction of the initial vorticity and kinetic energy 
in small vortex structures.

Figure~\ref{fig:A128collect} presents a collection of plots of 
parameters that give information about the thermodynamical 
state in the orbital plane of model A128 and about the 
nuclear composition of the gas at the end of the simulation. 
Panels~{\bf a} and {\bf b} show
the electron degeneracy parameter $\eta_e = \mu_e/(k_{\rm B}T)$ 
($k_{\rm B}$ is the Boltzmann constant, $\mu_e$ the electron chemical
potential) and the electron neutrino degeneracy parameter
$\eta_{\nu_e} = \mu_{\nu_e}/(k_{\rm B}T)$, respectively. 
Concordant with the discussion above, the $\beta$-equilibrium 
conditions in the whole star are characterized by 
$\eta_{\nu_e} \la 0$ (panel~{\bf b}). In the core values between
$-3$ and $-6$ can be found, while in the disk moderately negative
values are present (around $-1$) and in some regions the medium
has evolved back to a state close to $\eta_{\nu_e} \approx 0$.
Comparison with panels~{\bf a} and {\bf b} of Fig.~\ref{fig:neutA128}
shows that in these regions the emission rates of $\nu_e$ and 
$\bar\nu_e$ are already very similar again whereas the production  
of electron antineutrinos is clearly dominant in those parts of
the disk with the most negative values of $\eta_{\nu_e}$.
The electron degeneracy is moderate ($\eta_e \approx 2$--3)
in the disk but climbs to numbers around $\eta_e \approx 25$
near the center.

The temperature $k_{\rm B}T$ and entropy are displayed in 
panels~{\bf c}
and {\bf d}, respectively, of Fig.~\ref{fig:A128collect}. 
Detailed information
about the evolution of the temperature in all models was given
in Paper~I (Figs.~4--7, 14--17, and 20, 21). At the end of the 
simulation model A128 has the highest temperatures of 
$k_{\rm B}T\approx 30$~MeV in hot spots located in a shell
around the central high-density core where $k_{\rm B}T\approx 10$~MeV.
The disk has been heated up to $k_{\rm B}T \approx 10$~MeV in regions
of density $\rho\approx 10^{12}$--$10^{13}\,{\rm g/cm}^3$, 
$k_{\rm B}T \ga 6$~MeV where $\rho \ga 10^{11}\,{\rm g/cm}^3$, and
$k_{\rm B}T \ga 1$--2~MeV for $\rho \ga 10^{10}\,{\rm g/cm}^3$. 
The corresponding entropies are less than 1~$k_{\rm B}$/nucleon in the
core region and between 3 and slightly more than 7~$k_{\rm B}$/nucleon 
in the disk (panel~{\bf d}). In model C64 similar disk entropies
are found while in model A64 specific entropies up to about 
9~$k_{\rm B}$/nucleon and in model B64 up to even 
10~$k_{\rm B}$/nucleon develop towards the end of the simulated 
evolution.

Such high entropies allow only minor contributions of nuclei
to be present in the gas in nuclear statistical equilibrium
at the densities found for the disk matter on our computational 
grid. Most of the nuclei are completely disintegrated into free
nucleons (the mass fraction of heavy nuclei is below the lower limit
of $\sim 10^{-8}$ returned from the equation of state of Lattimer
\& Swesty (1991)), and only small admixtures of $\alpha$ particles
are possible. Panel~{\bf e} of Fig.~\ref{fig:A128collect} shows
that the mass fraction $X_{\alpha}$ of $\alpha$ particles is 
typically less than about $10^{-3}$. Only in the outermost
parts of the disk $X_{\alpha}\sim 10^{-2}$ because there
the temperatures are low enough, $k_{\rm B}T\sim 1$--2~MeV, 
that some of the free nucleons can recombine.

%
%

\section{Neutrino-antineutrino annihilation\label{sec:annihil}}

Neutrino-antineutrino annihilation in the surroundings of the 
merger has been proposed to create a sufficiently energetic 
fireball of $e^+e^-$-pairs and photons 
to explain gamma-ray bursts at cosmological distances.
We attempt to put this idea to a quantitative test.
With the given information about the fluxes and spectra of the
neutrino emission of all grid cells (see Paper~I for technical
details), it is possible to evaluate our hydrodynamical models 
for the energy deposition by $\nu\bar\nu$-annihilation in a
post-processing step. Since the neutrino luminosities become
large only after the merging of the two neutron stars and in 
particular
after the gas torus around the compact central body has formed,
we consider the late stages of our simulated merger evolutions 
as the most interesting ones to perform the analyses. In the
phase when quasi-stationary conditions have been established, the
neutrino luminosities have reached their saturation levels and
the annihilation rates have become maximal.

\subsection{Numerical evaluation\label{sec:anninum}}

Neglecting phase space blocking effects in the phase spaces of
$e^-$ and $e^+$, the local energy deposition rate
(energy ${\rm cm}^{-3}{\rm s}^{-1}$) at a position ${\bf r}$ by
annihilation of $\nu_i$ and $\bar\nu_i$ into $e^+e^-$-pairs (which 
is the dominant reaction between neutrinos and antineutrinos) can be
written in terms of the neutrino and antineutrino phase space
distribution functions
$f_{\nu_i} = f_{\nu_i}(\epsilon,{\bf n},{\bf r},t)$
and $f_{\bar\nu_i} = f_{\bar\nu_i}(\epsilon',{\bf n'},{\bf r},t)$
as (Goodman et al.~1987, Cooperstein et al.~1987, Janka 1991)
\begin{eqnarray}\label{8}
\lefteqn{ Q^+_{\nu\bar\nu}(\nu_i\bar\nu_i) =  {1\over 4}\,
{\sigma_0c\over \rund{m_e c^2}^2\rund{hc}^6} } \nonumber\\
& \Biggl\lbrace & \,
{\rund{C_1 + C_2}_{\nu_i\bar\nu_i}\over 3}\,\,\cdot\,\int_0^\infty {\rm d}
\epsilon\int_0^\infty {\rm d}\epsilon'\,\rund{\epsilon + \epsilon'}\,
\epsilon^3\epsilon'^3 \nonumber\\
& & \phantom{{\rund{C_1 + C_2}_{\nu_i\bar\nu_i}\over 3}}
\oint_{4\pi}{\rm d}\Omega
\oint_{4\pi}{\rm d}\Omega'\,f_{\nu_i}f_{\bar\nu_i}\,\rund{1-\cos\theta}^2
\,+
\nonumber\\
& + &  C_{3,\nu_i\bar\nu_i}\rund{m_e c^2}^2\cdot
\int_0^\infty {\rm d}\epsilon \int_0^\infty {\rm d}\epsilon'\,
\rund{\epsilon + \epsilon'}\,\epsilon^2\epsilon'^2 \nonumber\\
& & \phantom{{\rund{C_1 + C_2}_{\nu_i\bar\nu_i}\over 3}}
\oint_{4\pi}{\rm d}
\Omega\oint_{4\pi}{\rm d}\Omega'\,f_{\nu_i}f_{\bar\nu_i}\,\rund{1-\cos\theta}
\,\, \Biggr\rbrace \,\, .
\end{eqnarray}
When the energy integrations are absorbed into
(energy-integrated) neutrino intensities
$I_{\nu_i}$ and $I_{\bar\nu_i}$,
\begin{equation}\label{9}
I_{\nu}\ =\ I_{\nu}({\bf n},{\bf r},t)\ \equiv\
{c \over (hc)^3}\,\int_0^\infty{\rm d}\epsilon\,\,
\epsilon^3\,f_{\nu}(\epsilon,{\bf n},{\bf r},t) \ ,
\end{equation}
\Eq{8} can be rewritten as
\begin{eqnarray}\label{10}
\lefteqn{Q^+_{\nu\bar\nu}(\nu_i\bar\nu_i) =  {1\over 4}\,
{\sigma_0\over c\,\rund{m_e c^2}^2}
 \, \Biggl\lbrace \,
{\rund{C_1 + C_2}_{\nu_i\bar\nu_i}\over 3}\,\cdot } \nonumber\\
 & & \oint_{4\pi}{\rm d}
\Omega\,I_{\nu_i}\,\oint_{4\pi}{\rm d}\Omega'\,
I_{\bar\nu_i}\,
\eck{\ave{\epsilon}_{\nu_i} + \ave{\epsilon}_{\bar\nu_i}}\,
\rund{1 - \cos\theta}^2\,+
\nonumber\\
& + & C_{3,\nu_i\bar\nu_i}\rund{m_e c^2}^2\cdot \nonumber \\
& & \oint_{4\pi}{\rm d}\Omega\,I_{\nu_i}
\,\oint_{4\pi}{\rm d}\Omega'\,I_{\bar\nu_i}\,
{\ave{\epsilon}_{\nu_i} + \ave{\epsilon}_{\bar\nu_i}\over
\ave{\epsilon}_{\nu_i}\,\ave{\epsilon}_{\bar\nu_i}}\,
\rund{1 - \cos\theta} \,\, \Biggr\rbrace \,\, .
\end{eqnarray}
The integrals over $\Omega$ and $\Omega'$ sum up
neutrino and antineutrino radiation incident
from all directions. $\theta$ is the angle between neutrino and
antineutrino beams and $\ave{\epsilon}_{\nu_i}$ and
$\ave{\epsilon}_{\bar\nu_i}$ are suitably
defined average spectral energies of
neutrinos and antineutrinos, respectively. The weak interaction
cross section is $\sigma_0 = 1.76\times 10^{-44}\,{\rm cm}^2$,
$m_ec^2 = 0.511\,{\rm MeV}$ is the electron rest-mass energy,
$c$ the speed of light, and the weak coupling constants are
$\rund{C_1 + C_2}_{\nu_e\bar\nu_e} = \rund{C_V - C_A}^2 +
\rund{C_V + C_A}^2 \approx 2.34$,
$C_{3,\nu_e\bar\nu_e} = {2\over 3}\rund{2C_V^2 - C_A^2} \approx
1.06$ and
$\rund{C_1 + C_2}_{\nu_x\bar\nu_x} = \rund{C_V - C_A}^2 +
\rund{C_V + C_A - 2}^2 \approx 0.50$,
$C_{3,\nu_x\bar\nu_x} = {2\over 3}\eck{2(C_V - 1)^2 -
(C_A - 1)^2} \approx -0.16$ for $\nu_x = \nu_\mu,\,\nu_\tau$ and
$C_A = {1\over 2}$, $C_V = {1\over 2} + 2\sin^2\theta_{\rm W}$
with $\sin^2\theta_{\rm W} = 0.23$. The total energy deposition
rate at the position ${\bf r}$ is given as the sum of 
the contributions from
annihilation of $\nu_e$ and $\bar\nu_e$, $\nu_\mu$ and $\bar\nu_\mu$,
and $\nu_\tau$ and $\bar\nu_\tau$:
\begin{equation}\label{11}
Q^+_{\nu\bar\nu}({\bf r})\ =\ Q^+_{\nu\bar\nu}(\nu_e\bar\nu_e)
+ Q^+_{\nu\bar\nu}(\nu_\mu\bar\nu_\mu)
+ Q^+_{\nu\bar\nu}(\nu_\tau\bar\nu_\tau) \ .
\end{equation}

When working with a discrete grid the integrals in \Eq{10} are
replaced by sums over all cells $k$,
\begin{equation}\label{12}
\oint_{4\pi} {\rm d}\Omega\,I_{\nu}\ \longrightarrow\ \
\sum_k \Delta\Omega_k\cdot I_{\nu,k} \ .
\end{equation}
$\Delta\Omega_k$ is the solid angle with which cell $k$ is seen
from a position ${\bf r}$ at distance $d_k \equiv |{\bf r} - {\bf r}_k|$
when ${\bf r}_k$ is the location of the center of cell $k$.
In order to avoid the need to take into account projection effects, 
we define an effective radius $D$ associated with
the cells of the cartesian grid used
for the hydrodynamical modelling by setting the cell volume
$V_k = \Delta x \Delta y \Delta z = (\Delta x)^3$ equal to the volume
of a sphere $V = 4\pi D^3/3$:
\begin{equation}\label{13}
D\ =\ \rund{{3 \over 4\pi}}^{1/3}\,\Delta x \ .
\end{equation}
With the projected area $A = \pi D^2$ we obtain
\begin{equation}\label{14}
\Delta \Omega_k \ =\ \pi\,\rund{{3 \over 4\pi}}^{2/3}\,\rund{{\Delta x
\over d_k}}^2 \ .
\end{equation}
Using the simplifying assumption
that a grid cell radiates neutrinos with isotropic intensity
into the half space around the outward direction defined by the
local density gradient ${\bf n}_\rho = \nabla\rho/|\nabla\rho|$,
the flux $j_{\nu_i,k}$ is related to
the neutrino radiation intensity $I_{\nu_i,k}$ by $j_{\nu_i,k}
= \pi I_{\nu_i,k}$. With an effective emissivity $Q_k^{\rm eff}(\nu_i)$
(see Paper I) which represents the energy emission of cell $k$
per ${\rm cm}^{-3}{\rm s}^{-1}$ in a single neutrino species
$\nu_i = \nu_e,\,\bar\nu_e,\,\nu_\mu,\,\bar\nu_\mu,\,\nu_\tau$ or
$\bar\nu_\tau$, the intensity $I_{\nu_i,k}$ is therefore given by
\begin{equation}\label{15}
I_{\nu_i,k} = {j_{\nu_i,k}\over \pi} = {1\over \pi}\,
Q_k^{\rm eff}(\nu_i)\,{V\over A} = Q_k^{\rm eff}(\nu_i)\,
{4\over 3\pi}\rund{{3 \over 4\pi}}^{1/3}\!\!\!\Delta x \,\,\, .
\end{equation}
\Eq{14} and \Eq{15} allow us to evaluate the sum of \Eq{12}
sufficiently accurately by
\begin{equation}\label{16}
\sum_k \Delta\Omega_k\cdot I_{\nu_i,k}\ \approx\ {1\over \pi}\,
\rund{\Delta x}^3\,\sum_k {Q_k^{\rm eff}(\nu_i) \over d_k^2}\ .
\end{equation}
With the average energy of neutrinos emitted from cell $k$,
$\ave{\epsilon}_{\nu_i,k}$, (see Paper I) and the angle enclosed by
the radiation from cells $k$ and $k'$ at the position ${\bf r}$,
$\cos\theta_{k k'} = \rund{{\bf r} - {\bf r}_k}\rund{{\bf r}
- {\bf r}_{k'}}/\rund{d_k d_{k'}}$, the integrals of \Eq{10}
finally become sums over all combinations of grid cells $k$
with all cells $k'$. This double sum has to be evaluated at all
positions ${\bf r}$ where the energy deposition rate by
$\nu\bar\nu$-annihilation is to be determined. The corresponding
computational load is appreciable but
can be significantly reduced by taking into account only those
grid cells which emit towards ${\bf r}$,
i.e.~whose emission into the outward directed half space around
${\bf n}_\rho$ is also pointing to position ${\bf r}$. The
criterion for this being fulfilled is
${\bf n}_\rho\cdot \rund{{\bf r} - {\bf r}_k} > 0$.

\begin{figure*}
 \begin{tabular}{cc}
  \epsfxsize=8.8cm \epsfclipon \epsffile{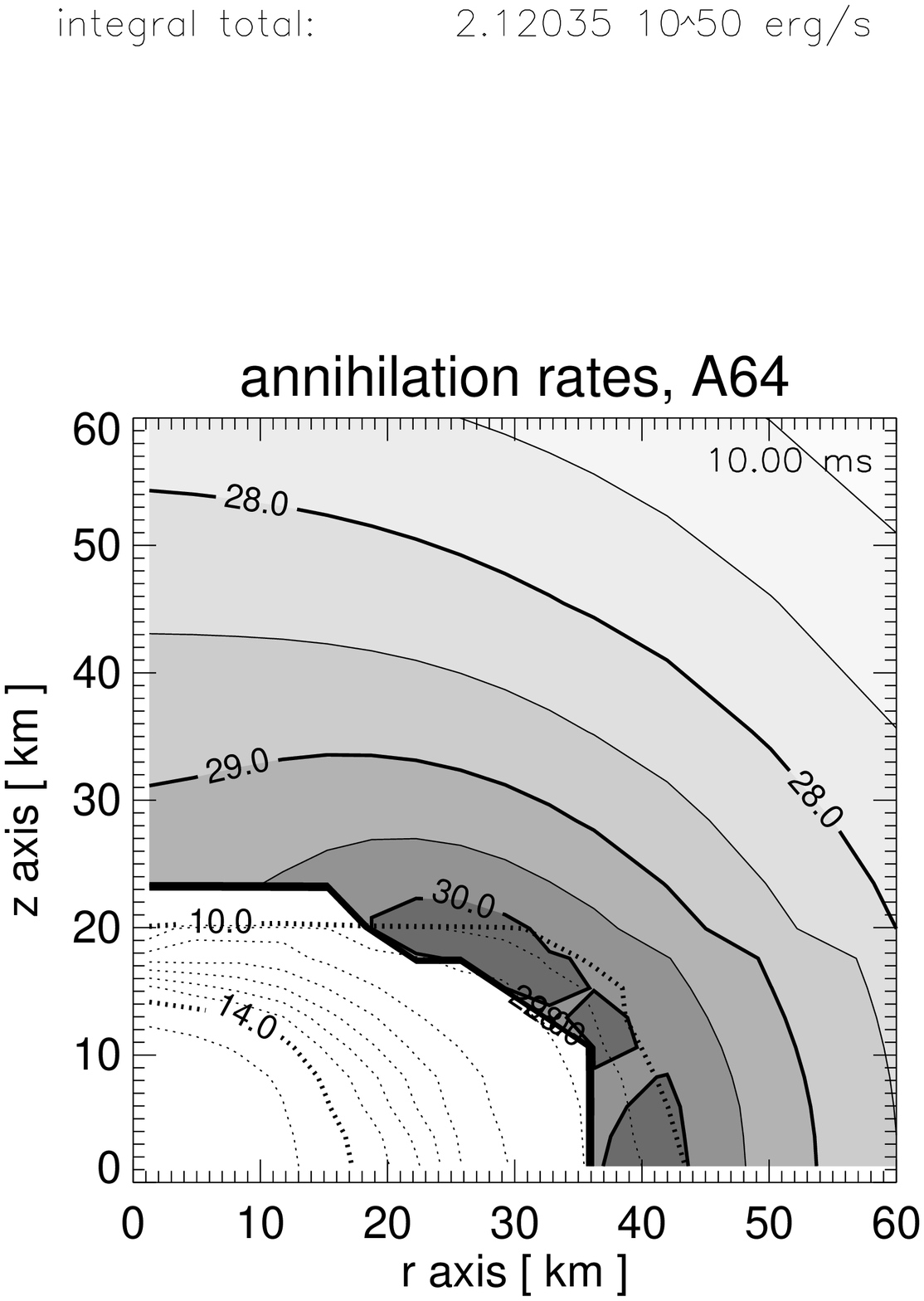} &
  \epsfxsize=8.8cm \epsfclipon \epsffile{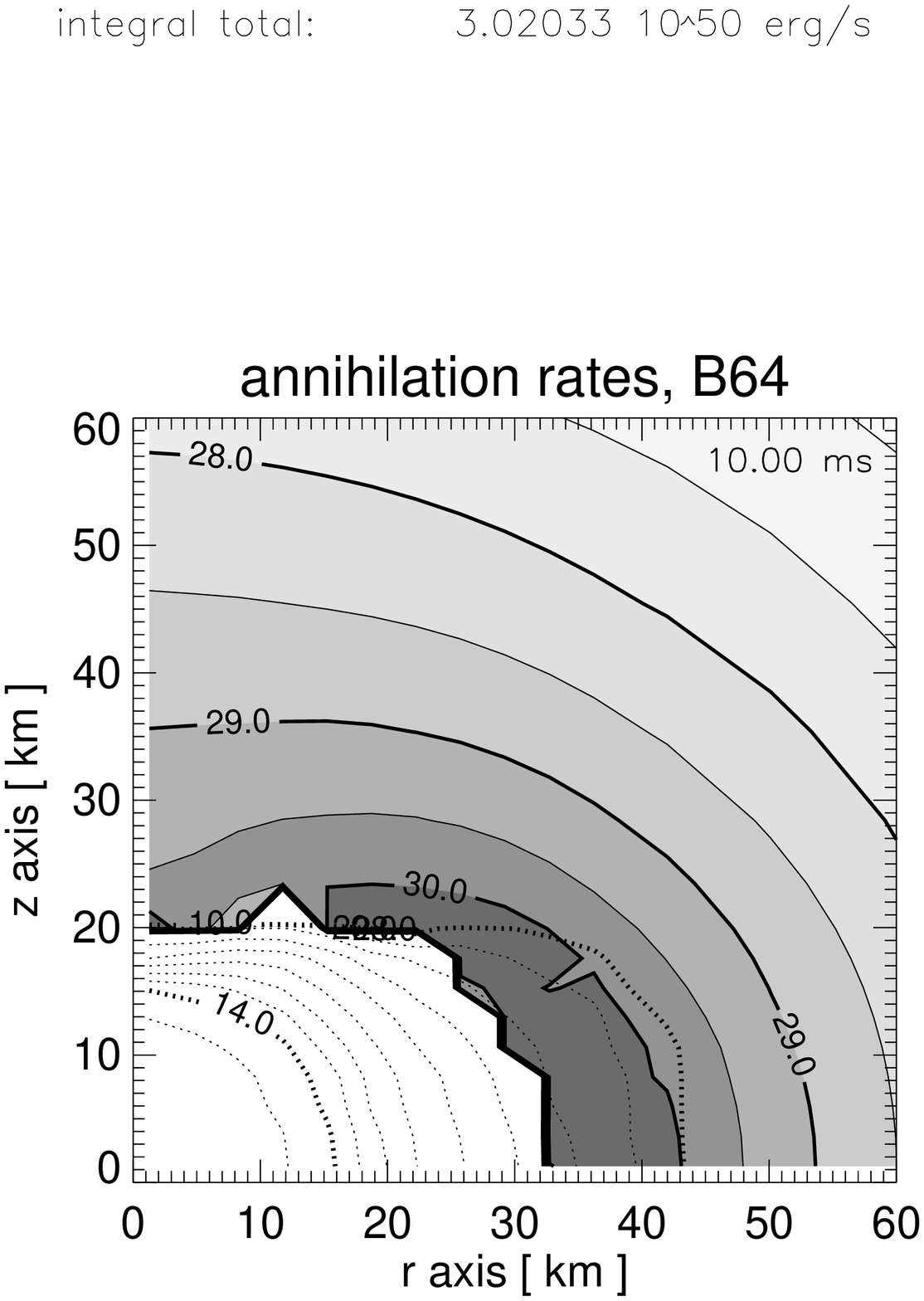} \\
  \epsfxsize=8.8cm \epsfclipon \epsffile{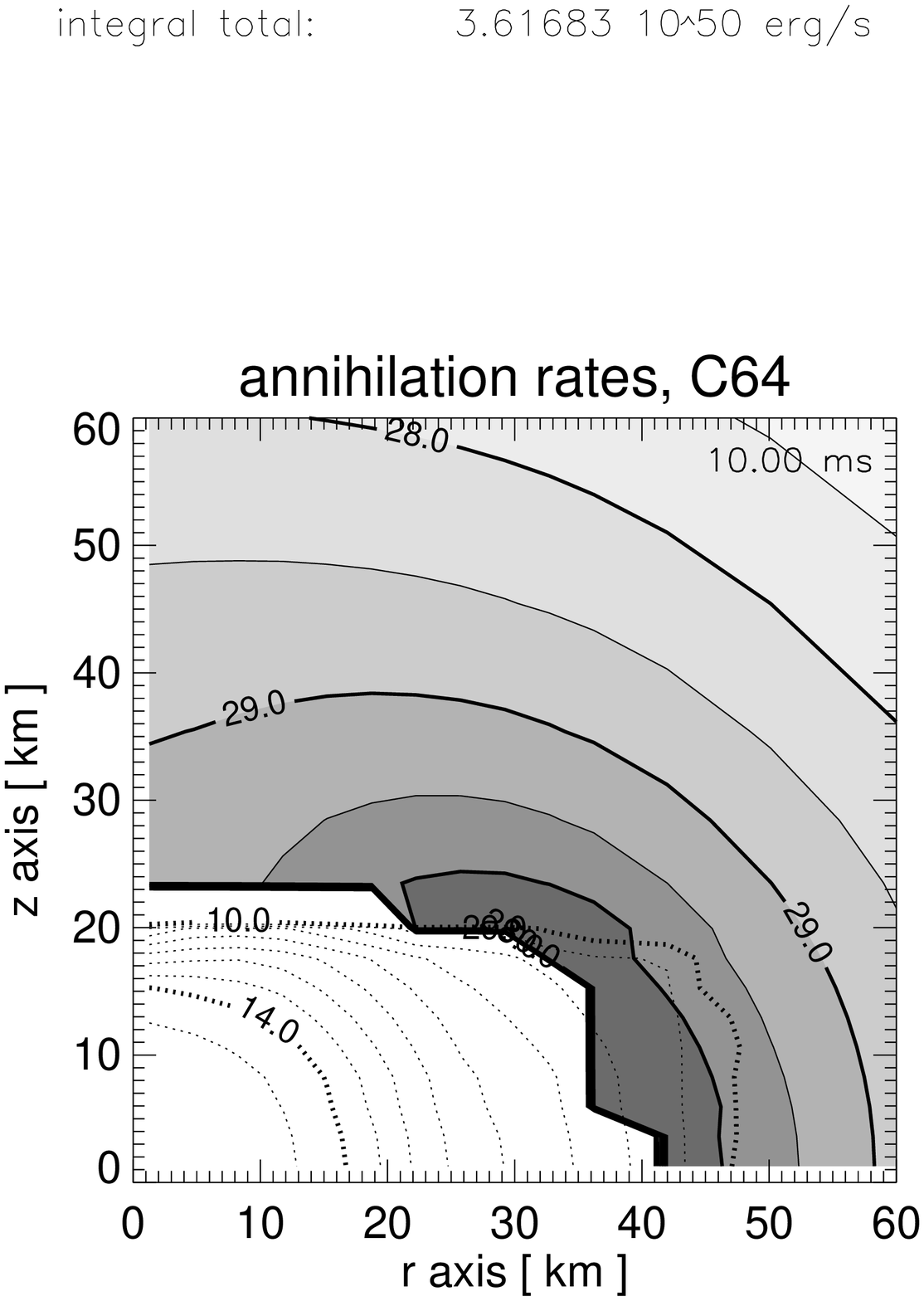} &
\raisebox{8cm}{\parbox[t]{8.8cm}{
\caption[]{\label{fig:annmap}
Maps of the total local energy deposition rates (in erg/cm$^3$/s)
by neutrino-antineutrino annihilation into electron-positron
pairs in the vicinity of the merger for the three models 
A64, B64, and C64 at time $t = 10$~ms after the start of the
simulations. In one quadrant of  
the $r$-$z$ plane orthogonal to the orbital
plane (at $z = 0$), the plots show values obtained as averages 
of the energy deposition rate over azimuthal angles. 
The corresponding solid contour lines are
logarithmically spaced in steps of 0.5 dex, the grey shading
emphasizes the levels with dark grey meaning high energy
deposition rate. The dashed contours indicate levels of the 
azimuthally averaged density, also logarithmically spaced with
intervals of 0.5 dex. The energy deposition rate was
evaluated only in that region around the merged object, where
the mass density is below $10^{11}$~g/cm$^3$
{\it and} the energy loss rate by neutrino emission is 
smaller than $10^{30}$~erg/cm$^3$/s. One can see that due to
the closeness to the main neutrino radiating disk
region, most annihilation energy is deposited in the ``outer''
(in $r$-direction) and ``upper'' (in $z$-direction) regions 
of the disk (dark grey areas)
}}}
 \end{tabular}

\end{figure*}

\begin{figure}
 \epsfxsize=8.8cm \epsfclipon \epsffile{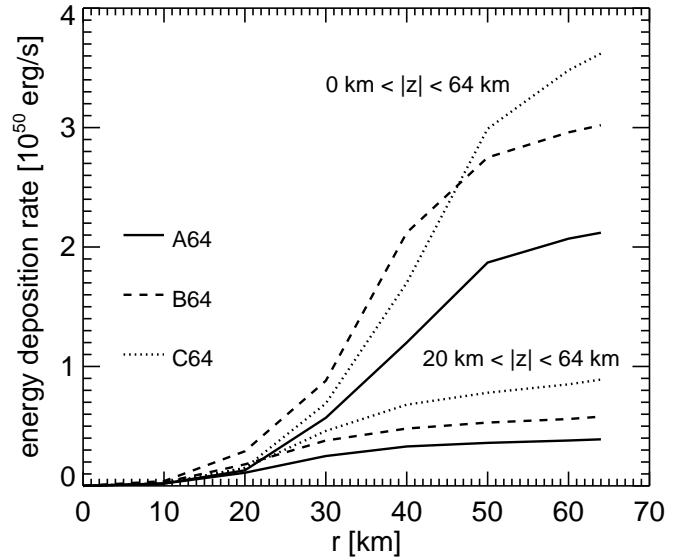}
\caption[]{
Cumulative total energy deposition rates by neutrino-antineutrino
annihilation into electron-positron pairs for the three models
A64, B64, and C64 as functions of the equatorial distance $r$ from 
the center of the merger once for the case that the integration is 
done for polar distances $20\,{\rm km}\le |z|\le 64\,{\rm km}$
and another time for $0\,{\rm km}\le |z|\le 64\,{\rm km}$.
The integrals were performed over the values of the local energy 
deposition rates plotted in the three panels of 
Fig.~\protect\ref{fig:annmap}
}
\label{fig:anndis}
\end{figure}

\subsection{Numerical results\label{sec:annires}}

The numerical post-processing procedure described in the previous
section yields the total energy deposition rate 
$Q_{\nu\bar\nu}^+$ by neutrino-antineutrino annihilation as a 
function of the position~${\bf r}$. Figure~\ref{fig:annmap}
shows the results, averaged over azimuthal angles, in a
quadrant of the $r$-$z$-plane perpendicular to the orbital
plane for the three models A64, B64, and C64. The evaluation
was performed only in that region around the central part of
the merger where the local energy loss rate by neutrino emission
is less than $10^{30}\,{\rm erg/cm^3/s}$ and the density is 
below $10^{11}\,{\rm g/cm}^3$. Density levels are indicated by
dashed contour lines in the plots.

One can see that in all three models the highest rate of energy
deposition ($> 10^{30}\,{\rm erg/cm^3/s}$) occurs in the outer
regions of the disk ($35\,{\rm km}\le r \le 45\,{\rm km}$ in the
orbital plane) within about 25~km above and below the 
orbital plane. Here the energy deposition rate is much larger
than the energy loss rate. Because this deposition transfers
energy into gas layers with densities of still more than 
$10^{10}\,{\rm g/cm}^3$, one must suspect that a baryonic wind
will be created very similar to the neutrino-driven wind caused by 
neutrino energy deposition in the surface layers of the nascent
neutron star in a type-II supernova (for information about the 
neutrino-driven wind from forming neutron stars, see 
Duncan et al.~1986,
Woosley \& Baron 1992, Witti et al.~1994, Woosley et al.~1994;
see also Hernanz et al.~1994). Instead of creating a fireball
of a nearly pure relativistic electron-positron-photon plasma which
might lead to an energetic gamma-ray burst, this energy is used up
to lift baryons in the strong gravitational field of the massive
central body. Consequently, the expansion velocities of this 
matter are nonrelativistic. If too much of this wind material
gets mixed into the pair-photon plasma, the baryonic load  
$M_{\rm w}$ can become too high to allow for Lorentz factors
$\Gamma$ in the required range of $\Gamma = 
1/\sqrt{1-(v/c)^2} \sim E_{\nu\bar\nu}/(M_{\rm w}c^2) \ga 100$.

The large energy deposition rates at radii between 30~km 
and 50~km in the equatorial plane 
and at moderate heights above and below the orbital 
plane can be explained by the closeness to the main neutrino 
emitting ring in the disk between 20~km and 30~km (see 
Figs.~\ref{fig:neutA64}--\ref{fig:nepeC64}). According to 
Eqs.~(\ref{10}), (\ref{12}), and (\ref{16}) the annihilation rate 
decreases
at least with the fourth power of the distance to the neutrino
radiating grid cells $k$ and $k'$: $Q_{\nu\bar\nu}^+\propto
1/(d_k^2d_{k'}^2)$. Because of the influence of the geometrical
factor $(1-\cos\theta)^2$ in Eq.~(\ref{10}) (this factor accounts
for the dependence of the annihilation probability on the
relative velocity of the interacting neutrinos and the  
kinematically allowed phase space for the reaction), 
$Q_{\nu\bar\nu}^+$ decreases even more steeply when one moves
away from the source and large-$\theta$ collisions between
neutrinos become less and less frequent. At distances from the 
merged object $d \equiv \sqrt{r^2 + z^2} \ga 50$--60~km
we find that along each radial beam the local energy deposition rate
as a function of $d$ becomes a power law with power law index
between $-6$ and $-8$. Towards the rotation
($z$-) axis, the growing distance to the neutrino producing 
disk region cannot be compensated by the higher chance of head-on
collisions between neutrinos. This is the reason why the 
contour levels in the plots of Fig.~\ref{fig:annmap} bend
towards the polar region of the merger near the $z$-axis.

The integral values of the energy deposition rate 
by $\nu\bar\nu$-annihilation in the surroundings of the
merger out to equatorial distances $r$ are
shown in Fig.~\ref{fig:anndis}. For each of the models A64,
B64, and C64 we have evaluated the spatial integral once for
vertical heights $20~{\rm km}\le |z|\le 64~{\rm km}$ and another
time for $0~{\rm km}\le |z|\le 64~{\rm km}$. The maximum upper 
integration limit is determined by the largest distances where the 
local rates $Q_{\nu\bar\nu}^+$ were calculated. However, one can
see from Fig.~\ref{fig:anndis} that the curves tend to approach
a saturation level for $r\to 64$~km from which we conclude that
extending the integrations into the region $d > 64$~km would 
not change the results significantly. From a comparison of both
cases one can recognize that only a minor fraction of 
about $1/5$--$1/4$ of the annihilation 
energy is converted into $e^+e^-$-pairs in the region above
and below the disk. Only at heights $|z| \ga 20~{\rm km}$ is
the baryon density low enough that most of the 
converted energy might end up in a relativistic fireball.
However, in the whole region $0~{\rm km}\le |z|\le 64~{\rm km}$
the energy deposition rate is rather small, only about 
2--$4\cdot 10^{50}\,{\rm erg/s}$; the ``useful'' fraction is probably
less than $1/4$ of that. The models were evaluated at times
when the neutrino emission of the models had already achieved
a maximum value and a quasi-stationary state. At earlier times the
neutrino luminosities are much lower and therefore the
integral values of the energy deposition rate are even smaller
than those displayed in Fig.~\ref{fig:anndis}.

For the disk or torus geometry of our models
the annihilation efficiency, defined as 
$e_{\nu\bar\nu} \equiv ({\rm d}E_{\nu\bar\nu}/{\rm d}t)/
(L_{\nu_e} + L_{\bar\nu_e} + 4L_{\nu_{\mu}})$ where 
${\rm d}E_{\nu\bar\nu}/{\rm d}t$ is the total energy deposition
rate by $\nu\bar\nu$-annihilation, can be found to be of the
order of a few tenths of a per cent:
\begin{equation}
e_{\nu\bar\nu}\,=\, (2\,...\,3)\cdot 10^{-3}\,
\frac{L_{\nu_e}}{1.5\cdot 10^{52}{\rm erg/s}}\,
\frac{\ave{\epsilon_{\nu_e}}}{13\,{\rm MeV}}\,
\frac{20\,{\rm km}}{R_{\rm d}}\ .
\label{eq:eff}
\end{equation}
The factors in this equation can be deduced by simple 
analytical transformations and dimensional considerations of the 
volume integral of the annihilation rate (Eq.~(\ref{eq:eff})). 
$R_{\rm d}$ is the inner radius of the disk, the radius which roughly
separates the compact inner core from the more dilute gaseous
cloud of matter around. The inserted numbers
are representative values taken from Figs.~\ref{fig:neutrad},
\ref{fig:ergne}, and \ref{fig:mene}. The efficiency of 
Eq.~(\ref{eq:eff}) for
converting emitted neutrino energy into electron-positron pairs
by $\nu\bar\nu$-annihilation is in good agreement with what is
obtained in supernova simulations or analytical estimates
assuming emission in a spherical geometry instead 
of the disk-dominated emission of our numerical models.

From the results discussed here we 
conclude that there is no chance to obtain the energies needed
for a cosmological gamma-ray burst by $\nu\bar\nu$-annihilation
during the instant of the merging of binary neutron stars. 
Before the hot cloud of gas around the central, dense body has
formed, the neutrino luminosities stay rather low and only very
little energy is deposited by $\nu\bar\nu$-annihilation. Even
later when a disk around the compact, merged body has formed
and the neutrino emission has reached a high level, the 
energy deposition rates of 2--$4\cdot 10^{50}\,{\rm erg/s}$
lead to a total deposited energy of only a few
$10^{48}\,{\rm erg}$ within the computed time of 10~ms.
One would need the strong neutrino emission to be maintained
for periods of about 10~s to pump an energy of more than 
$10^{51}\,{\rm erg}$ into a fireball of $e^+e^-$-pairs and 
photons. Note that these statements are {\it not} changed 
if the observed gamma-ray bursts are beamed events. If one
relies on the emission geometry of our models, a constrained
solid angle $\Delta\Omega$ into which the fireball expands 
and becomes visible in a jet-like outburst would 
also (and by the same factor) reduce the spatial volume where 
$\nu\bar\nu$-annihilation in the surroundings of the merger 
could deposit the useful energy. Correspondingly, the numbers 
given in Fig.~\ref{fig:anndis} would have to be scaled down
by a factor $\Delta\Omega/(4\pi)$. 
In view of the fact discussed in Sect.~\ref{sec:hydevol} that
the central, compact core of the merger has a mass that can
neither be stabilized by internal pressure nor by rotation and
should therefore collapse into a black hole 
on a time scale of milliseconds,
we conclude that the energy available by pair annihilation
of neutrinos is lower than the desired and required ``canonical'' 
value of $10^{51}/(4\pi)\,{\rm erg/steradian}$
by more than a factor of 1000.

So far we have considered the phase of the merging
of binary neutron stars and the evolution that follows immediately 
afterwards. This, however, leaves the question unanswered
whether a disk around the most
likely forming black hole could produce and emit neutrinos on
a much longer time scale and could thus power a gamma-ray burst
by the discussed mechanism of neutrino-pair annihilation?
Since simulations of this scenario are beyond the present
capabilities of the employed numerical code and, in addition, 
the results depend on the unknown viscosity in the disk,
we attempt to
develop a simple model of the behavior and properties of such
a disk with respect to its neutrino emission and the strength of 
$\nu\bar\nu$-annihilation.

\subsection{Simple model for the post-merging 
emission from the disk\label{sec:simod}} 

Our simulations suggest that
some material, possibly about $0.1\,M_{\odot}$, could remain in a 
disk around the central black hole. This disk will be heated by 
viscous dissipation and will emit neutrinos and antineutrinos
until its matter is accreted into the central black hole. 
The efficiency of $\nu\bar\nu$-annihilation increases linearily
with the $\nu$ luminosity (Eq.~(\ref{eq:eff})) and thus a short,
very luminous neutrino burst is more efficient to create an
$e^+e^-$-pair fireball than the same energy emitted on a longer
time scale with smaller neutrino fluxes. It has to be investigated
whether enough energy can be provided in the pair-photon fireball
by the neutrino emission
from the disk to explain a $\gamma$-ray burst at cosmological
distances. Viscosity effects have a crucial influence on the  
disk evolution and on the neutrino emission. Viscous forces, on the 
one hand, transfer angular momentum between adjacent fluid elements
and determine the accretion time scale and accretion rate.
Viscous dissipation of rotational energy, on the other hand, heats
the disk and is thus essential for the neutrino emission.
Disk size, disk temperature, disk viscosity, and neutrino emission
properties can therefore not be chosen independent of the accretor
mass and disk mass. In the following we shall attempt to relate 
these quantities by simple considerations and conservation arguments.

The lifetime of the disk will decrease with larger dynamic 
viscosity $\eta$ because the viscous force that generates a
torque carrying angular momentum outward is increased. 
For a (Newtonian) Keplerian disk the viscous force 
(per unit area) in the 
angular ($\phi$) direction, $f_{\phi}$, is simply expressed
by the component $t_{r\phi}$ of the viscous stress tensor
(see, e.g., Shapiro \& Teukolsky 1983): 
\begin{equation}
f_{\phi} \,=\,-t_{r\phi} \,=\,{3\over 2}\,\eta\,\Omega_{\rm k}\ ,
\label{eq:visf}
\end{equation}
where $\Omega_{\rm k} = \sqrt{GM/r^3}$ is the Keplerian 
angular velocity. The torque $T$ exerted by the viscous stress is
given by $T = f_{\phi}\,r\,(2\pi r\,2h)$
when $2h\sim 2R_{\rm s}$ is taken as the vertical diameter of the
thick disk around the black hole with mass $M$ and Schwarzschild radius
$R_{\rm s} = 2GM/c^2 \approx 9(M/3\,M_{\odot})\,{\rm km}$. 
The accretion rate
${\rm d}M/{\rm d}t$ can be estimated (roughly) by setting the 
viscous torque equal to the rate 
${\rm d}J/{\rm d}t\sim ({\rm d}M/{\rm d}t)\Omega_{\rm k}r^2$ at which
angular momentum is consumed by the black hole due to the accretion 
of matter from the disk:
\begin{equation}
{{\rm d}M\over {\rm d}t} \,\approx\, 6\pi\eta R_{\rm s} \ .
\label{eq:dmdt}
\end{equation}
From that, the accretion time scale of a disk of mass $\Delta M_{\rm d}$
is estimated to be
\begin{equation}
t_{\rm acc}\,\approx\,{\Delta M_{\rm d}\over 6\pi\eta R_{\rm s}} \ .
\label{eq:tacc1}
\end{equation}
Thus, the lifetime of the disk is determined by the outward
transport of angular momentum through the viscous torque.
Equation~(\ref{eq:tacc1}) shows that it decreases
with the value of the dynamic viscosity as $1/\eta$.

Viscous dissipation generates heat in the disk at a rate 
per unit volume of (see Shapiro \& Teukolsky 1983)
\begin{equation}
{{\rm d}Q\over {\rm d}t}\,=\, {-f_{\phi}t_{r\phi}\over \eta}\,=\,
{9\over 4}\,\eta\,\Omega_{\rm k}^2\ .
\label{eq:dqdt1}
\end{equation}
At steady-state conditions the maximum dissipation rate occurs at 
a radius $r \approx 1.36\,R_{\rm d}\approx {4\over 3}\,R_{\rm d}$ 
when $R_{\rm d}$ is
the inner radius of the disk which is taken to be the innermost
stable circular orbit around the central accreting black hole,
$R_{\rm d} \approx 3R_{\rm s} \approx 6GM/c^2$. Using this
in Eq.~(\ref{eq:dqdt1}) one obtains for the maximum
rate at which frictional heat is liberated,
\begin{equation}
{{\rm d}Q\over {\rm d}t}\,\approx\, {1\over 28}\,\eta\,
{GM\over R_{\rm s}^3} \ .
\label{eq:dqdt2}
\end{equation}
Thus, the viscous heating rate increases linearly with $\eta$.

For small viscosity $\eta$ the viscous heating time scale is long
and the disk remains rather cool, also because cool matter is 
comparatively transparent for neutrinos and therefore the neutrino
cooling time scale is short. In that case the neutrino luminosity
for a disk with volume $V_{\rm d}$ is
$L_{\nu}\sim V_{\rm d}\, {\rm d}Q/{\rm d}t \propto \eta$ and the 
total energy radiated in neutrinos, 
$E_{\nu}\sim L_{\nu}t_{\rm acc}$, becomes independent of $\eta$
because of $t_{\rm acc}\propto 1/\eta$. With Eq.~(\ref{eq:eff}) 
one finds that the
energy converted into $e^+e^-$ by $\nu\bar\nu$-annihilation
increases proportional to $\eta$:
$E_{\nu\bar\nu} = e_{\nu\bar\nu}E_{\nu}\propto
L_{\nu}^2t_{\rm acc}\propto \eta$. In the optically thin case the
mean energy of emitted neutrinos, $\ave{\epsilon_{\nu}}$, which enters
the calculation of $e_{\nu\bar\nu}$ will also increase with $\eta$
and cause a slightly steeper than linear dependence of $E_{\nu\bar\nu}$
on $\eta$. 

If $\eta$ is large,
the disk is heated rapidly and strongly and thus becomes opaque
for neutrinos. With a neutrino diffusion time scale $t_{\rm diff}$
that is much longer than the heating time scale the 
neutrino luminosity is
$L_{\nu}\sim V_{\rm d}\, ({\rm d}Q/{\rm d}t) t_{\rm acc}/t_{\rm diff}$
which is only indirectly dependent on $\eta$ through $t_{\rm diff}$ 
and thus the (viscosity dependent) gas temperature $T$.
In that case $E_{\nu}\propto 1/(\eta t_{\rm diff})$ and
$E_{\nu\bar\nu}\propto 1/(\eta t_{\rm diff}^2)$.  
Note that the average energy of
emitted neutrinos, $\ave{\epsilon_{\nu}}$, which also determines
$e_{\nu\bar\nu}$, is only very weakly dependent on the viscosity 
of the disk in the optically thick case 
because it reflects the conditions at the neutrino decoupling 
sphere (see Eq.~(\ref{eq:tsurf}) below). 
The diffusion time scale increases with the disk temperature
and thus with the disk viscosity
due to the energy dependence of the weak interaction cross sections.
This leads to a decrease of $E_{\nu\bar\nu}$ with $\eta$
that is steeper than $1/\eta$. 

The considerations above suggest that
the annihilation energy $E_{\nu\bar\nu}$ has a pronounced maximum
at a particular value $\eta^\ast$ of the dynamic viscosity. 
Because of $E_{\nu\bar\nu}\propto L_{\nu}^2$ the annihilation of
neutrinos and antineutrinos is more efficient when a certain energy
is emitted in a short time with a high luminosity rather than over
a long period with a moderate flux.
If $\eta$ is small, $L_{\nu}$ stays low. If $\eta$ is very large and
the interior of the disk very hot and thus neutrino-opaque, the 
neutrino luminosity $L_{\nu}$ scales with the inverse of the 
neutrino diffusion time scale and with the total energy $E_{\nu}$
that can be emitted in neutrinos during the lifetime $t_{\rm acc}$
of the disk. This energy $E_{\nu}$ decreases in case of very large $\eta$
because $t_{\rm acc}$ becomes shorter and
the internal energy cannot be completely radiated away in neutrinos
before the neutrino-opaque matter is accreted into the black hole. The 
kinetic energy that is converted into internal energy by viscous friction
is entirely transported away by neutrinos and the fluxes are largest,
if the diffusion time scale is similar to the accretion time scale but
not much longer. The optimum value $\eta^\ast$ is therefore
determined by the condition $t_{\rm acc}\approx t_{\rm diff}$.

Let us assume that the part of the disk where 
most of the neutrinos are emitted has a mass $\Delta M_{\rm d}$
and is a homogeneous torus with center at $4R_{\rm s}$
and radius $R_{\rm s}$ (inner radius $3R_{\rm s}$, outer radius 
$5R_{\rm s}$) (Mochkovitch et al.~1993; Jaroszy\'nski 1993).
This is a fairly good picture in view of the shape and structure
of the disk that we obtained in our numerical simulations\footnote{In
the simple model considered here, the structure and geometry of the
disk is assumed to be given. It is not self-consistently determined
in dependence of the gas temperature and thus in dependence of 
the competing effects of viscous heating and neutrino cooling.}.
In terms of $R_{\rm s}$ the volume of the disk torus is
$V_{\rm d} = 8\pi^2 R_{\rm s}^3$ and its surface 
$S_{\rm d} = 16\pi^2 R_{\rm s}^2$. With the neutrino mean
free path $\lambda = \rho\sigma_{\rm eff}/m_u$ and 
$\Delta M_{\rm d} = \rho V_{\rm d}$
the diffusion time scale is approximately given by
\begin{equation}
t_{\rm diff}\,\sim\,{3R_{\rm s}^2\over c\lambda}\,\sim\,
{3(\Delta M_{\rm d})\sigma_{\rm eff}\over
8\pi^2R_{\rm s}m_{\rm u} c}\ .
\label{eq:tdiff}
\end{equation}
Here $m_{\rm u}$ is the atomic mass unit and the thermally averaged
effective neutrino interaction cross section 
$\sigma_{\rm eff} \equiv \sum_i Y_i\sigma_i$ is defined as the
sum of the cross sections $\sigma_i$ times the number fractions 
$Y_i = n_i/n_B$ of the corresponding reaction targets for all 
neutrino processes in the medium, i.e., neutrino scattering off $n$,
$p$, $e^-$, $e^+$, charged-current absorptions of $\nu_e$ and 
$\bar\nu_e$ by $n$ and $p$, respectively, and 
$\nu$-pair interactions. We find 
\begin{equation}
\sigma_{\rm eff}\,\sim\,
(1\,...\,4)\cdot 10^{-41}\,\rund{{k_{\rm B}T\over 5\,{\rm MeV}}}^{\! 2}
\ {\rm cm}^2 \ ,
\label{eq:sigma}
\end{equation}
the exact value depending on the neutrino type, the neutrino 
degeneracy and neutrino spectra, and the
detailed composition of the medium. For the entropies, densities, 
and temperatures obtained in our simulations the gas in the disk 
is completely disintegrated into free nucleons; nucleon as well
as lepton degeneracy plays a negligible role 
(see Sect.~\ref{sec:therm}). Therefore fermion
phase space blocking effects are unimportant. The thermal 
average of the neutrino cross section was evaluated by using 
a Fermi-Dirac distribution function with a vanishing neutrino chemical
potential, $\mu_{\nu} = 0$. In case of incomplete dissociation 
of the nuclei the neutrino opacity should still be within the 
uncertainty range associated with the cross section variation of
Eq.~(\ref{eq:sigma}). Setting $t_{\rm acc}$
(Eq.~(\ref{eq:tacc1})) with $\Delta M_{\rm d} = \rho V_{\rm d}$
equal to $t_{\rm diff}$ (Eq.~(\ref{eq:tdiff})), one determines 
the value of the shear viscosity in the disk,
where $\nu\bar\nu$-annihilation yields the largest energy,
as 
\begin{equation}
\eta^{\ast}\,\approx\,{4\pi\over 9}\,{m_{\rm u}c\over\sigma_{\rm eff}}
\,\sim \,(1.7\,...\,6.9)\cdot 10^{27}\,\rund{{k_{\rm B}T\over
5\,{\rm MeV}}}^{\!-2}\ {\rm {g\over cm\,s}}\ .
\label{eq:opvi1}
\end{equation}
The range of values accounts for the uncertainty in the 
effective neutrino interaction cross section. For the typical 
composition of the disk material the cross section is more likely
near the upper limit of the given interval, in which case the lower 
value of $\eta^{\ast}$ is favored.

Let us now consider a disk with this optimum value $\eta^{\ast}$.
Making use of $\eta^{\ast}$, the
interior temperature of the disk can be estimated by setting
the integral rate of viscous energy generation in the disk, 
$L_{\rm visc}\sim V_{\rm d}({\rm d}Q/{\rm d}t)$,
equal to the luminosity due to neutrino diffusion,
$L_{\nu}\sim V_{\rm d}\varepsilon_{\nu}/t_{\rm diff}\sim
({1\over 3}c\lambda)S_{\rm d}(\varepsilon_{\nu}/R_{\rm s})$.
Here $\varepsilon_{\nu}/R_{\rm s}$ is an approximation to the
gradient of the neutrino energy density in the disk and
$\varepsilon_{\nu}\sim 3\cdot {7\over 8}a_{\rm rad}
T_{\rm int}^4$ is the sum of the energy densities of all three 
kinds of non-degenerate $\nu\bar\nu$-pairs with $a_{\rm rad}$ being
the radiation constant. One finds
\begin{equation}
{k_{\rm B}T_{\rm int}\over 5\,{\rm MeV}}\,\approx \,5.6\,
(\Delta M_{{\rm d},01})^{1/4}R_{{\rm s},9}^{-3/4} 
\label{eq:tint}
\end{equation}
where
$\Delta M_{{\rm d},01}\equiv \Delta M_{\rm d}/(0.1\,M_{\odot})$ and
$R_{{\rm s},9} \equiv
R_{\rm s}/(9\,{\rm km})$ is normalized to the
Schwarzschild radius of a $3\,M_{\odot}$ black hole. Note that
due to the dependence of $t_{\rm diff}$ and of $\eta^{\ast}$
on $\sigma_{\rm eff}$ (according to Eqs.~(\ref{eq:tdiff}) and
(\ref{eq:opvi1}), respectively) $k_{\rm B}T_{\rm int}$ does not depend
on the neutrino interaction cross section and is therefore 
insensitive to its uncertainty. 

Plugging the result of Eq.~(\ref{eq:tint}) for the interior disk
temperature into Eq.~(\ref{eq:opvi1})  
yields for the optimum disk viscosity
\begin{equation}
\eta^{\ast}\,\sim\,(5.5\,...\,22)\cdot 10^{25}\, 
(\Delta M_{{\rm d},01})^{-1/2}\,R_{{\rm s},9}^{3/2}\ 
\ {\rm {g\over cm\,s}}\ ,
\label{eq:opvi2}
\end{equation}
the interval of values again corresponding to the range of
possible values of the effective neutrino interaction cross section. 
$\eta^{\ast}$ from Eq.~(\ref{eq:opvi2}) can now be used in
Eq.~(\ref{eq:dqdt2}) to calculate 
$L_{\rm visc}\sim V_{\rm d}({\rm d}Q/{\rm d}t)$, which, when
set equal to the neutrino luminosity expressed in terms of
temperature and surface area $S_{\rm d}$ of the neutrinosphere, 
$L_{\nu} \sim S_{\rm d}(3{7\over 8}a_{\rm rad}T_{\rm surf}^4)c/4$,
leads to an estimate of the neutrinospheric temperature 
\begin{equation}
{k_{\rm B}T_{\rm surf}\over 5\,{\rm MeV}}\,\approx\,(0.7\,...\,1.0)\cdot
(\Delta M_{{\rm d},01})^{-1/8}\,R_{{\rm s},9}^{1/8}\ .
\label{eq:tsurf}
\end{equation}
The temperature of the neutrino emitting disk surface 
is around $5\,{\rm MeV}$ and rather insensitive to the exact 
value of the effective neutrino interaction cross section 
(slightly larger result
for smaller cross section), to the disk mass $M_{\rm d}$, and
to the inner disk radius $R_{\rm d}\sim 3R_{\rm s}$. 

The optimum value $\eta^{\ast}$ for the dynamic viscosity as
given in Eq.~(\ref{eq:opvi2}) corresponds to an effective 
$\alpha$-parameter of
$\alpha\approx \eta^{\ast}/(\rho c_{\rm s}R_{\rm s})\sim
(1.7\,...\,7.0)\cdot 10^{-3}$ when $R_{\rm s} = 9\,{\rm km}$,
$\Delta M_{\rm d} = 0.1\,M_{\odot}$, and 
$\rho = \Delta M_{\rm d}/(8\pi^2R_{\rm s}^3) = 
3.44\cdot 10^{12}\,{\rm g/cm}^3$ are used, and 
the sound speed $c_{\rm s}$ is evaluated with 
$k_{\rm B}T = 28\,{\rm MeV}$.
For these values of density and temperature the gas pressure
is dominated by relativistic particles, i.e., photons, electrons,
positrons, and neutrinos. Neutrino shear viscosity does not 
contribute significantly to $\eta^{\ast}$. In the neutrino-opaque 
case it is estimated to be
\begin{eqnarray}
\eta_{\nu}\, & = & \,{1\over 3}\,{\varepsilon_{\nu}\lambda\over c}
\,=\,{\varepsilon_{\nu}m_{\rm u}\over 3\rho\sigma_{\rm eff}c} 
\nonumber \\
\, & \sim & \,
(1\,...\,4)\cdot 10^{23}\,\rund{{k_{\rm B}T\over 5\,{\rm MeV}}}^{\! 2}
\rund{{10^{12}\,{\rm g/cm^3}\over \rho}}
\,\,{\rm {g\over cm\,s}}\ ,
\label{eq:nuvis}
\end{eqnarray}
where the interval of the numerical value is again associated with
the uncertainty of the effective cross section $\sigma_{\rm eff}$.
For $\rho = 3.44\cdot 10^{12}\,{\rm g/cm}^3$ and
$k_{\rm B}T = 28\,{\rm MeV}$ one finds $\eta_{\nu}\sim 
(0.9\,...\,3.6)\cdot 10^{24}\,{\rm g\,cm^{-1}s^{-1}}$.
A temperature as high as $k_{\rm B}T\sim 70$--$80\,{\rm MeV}$ is
required for the neutrino viscosity to become large enough to
account for $\eta^{\ast}$.

For the diffusion and accretion time scales one obtains by 
inserting Eq.~(\ref{eq:opvi2}) into Eq.~(\ref{eq:tacc1})
\begin{equation}
t_{\rm acc}\,\approx\, t_{\rm diff}\,\sim\,(53\,...\,212)
\,(\Delta M_{{\rm d},01})^{3/2}\,R_{{\rm s},9}^{-5/2}\ \,{\rm ms}\ .
\label{eq:tacc2}
\end{equation}
This time is much longer than the dynamical time scale 
(${\cal O}(1\,{\rm ms}$)) and the neutrino equilibration time scale
($\la 1\,{\rm ms}$). Therefore our assumptions that neutrinos
diffuse in the disk and are in equilibrium with the matter are
confirmed a posteriori. The total neutrino luminosity is
\begin{equation}
L_{\nu}\,\sim\, (0.62\,...\,2.48)\cdot 10^{53}
(\Delta M_{{\rm d},01})^{-1/2}
R_{{\rm s},9}^{5/2}\ {\rm erg/s}\ .
\label{eq:nulum}
\end{equation}
In Eq.~(\ref{eq:tacc2}) the smaller values and in 
Eq.~(\ref{eq:nulum}) the larger ones correspond to the case
of larger viscosity $\eta^{\ast}$ and thus smaller neutrino
cross section $\sigma_{\rm eff}$ according to the postulated
equality of Eqs.~(\ref{eq:tacc1}) and (\ref{eq:tdiff}).
The total energy $E_{\nu} = L_{\nu}t_{\rm acc}$ radiated away 
over the time $t_{\rm acc}$ is independent of both and becomes
\begin{eqnarray}
E_{\nu}\, & \approx & \,{\pi\over 7}\,{G M(\Delta M_{\rm d})
\over 3R_{\rm s}} \,=\,{\pi\over 42}\,(\Delta M_{\rm d})c^2 
\nonumber \\
& \approx & \,1.3\cdot 10^{52}\Delta M_{{\rm d},01}\ \,\,{\rm erg}
\label{eq:nuerg}
\end{eqnarray}
which is (approximately) equal to the Newtonian gravitational binding
energy $E_{\rm bind} = {1\over 2} GM(\Delta M_{\rm d})/R_{\rm d} =
|E_{\rm grav} + E_{\rm rot}|$ of mass $\Delta M_{\rm d}$
at the inner disk radius 
$R_{\rm d} = 3R_{\rm s}$ where the matter is swallowed by the black
hole ($E_{\rm grav}$ is the gravitational
potential energy, $E_{\rm rot}$ the rotational energy).
Here it is assumed that no rotational 
kinetic energy is extracted from the black hole which is equivalent
to a zero stress boundary condition at $R_{\rm d}$. This requires
that within $R_{\rm d}$ the gas spirals into the black hole 
rapidly without radiating, an idealization which is probably 
justified (see, e.g., Shapiro \& Teukolsky 1983). The small 
discrepancy between the factors ${1\over 2}$ and ${\pi\over 7}$
of our calculation results from the fact that we consider a 
simple one-zone model of a homogeneous disk. We find that
the radiation efficiency of the disk in our simplified treatment
is $E_{\nu}/(\Delta M_{\rm d}c^2) = {\pi\over 42} \approx 7.5\%$
(exact value for a thin, Newtonian accretion disk: 
${1\over 12}\approx 8.3\%$). This result has to be compared
with the radiation efficiency of about 5.7\% for relativistic disk 
accretion onto a nonrotating black hole and with the radiation 
efficiency of 42.3\% for a maximally rotating black hole with
a prograde accretion disk
(see Shapiro \& Teukolsky 1983). Since our numerical
simulations suggest the formation of a central black hole with
a relativistic rotation parameter $a = Jc/(GM)$ that is clearly
less than 1 (Ruffert et al.~1996),
the reference value for the radiation efficiency 
for disk accretion onto a nonrotating black hole is relevant
and our Newtonian disk evolution model most likely overestimates 
the amount of energy that can be carried away by neutrinos
before the accreted mass finally plunges rapidly from $R_{\rm d}$ to
the event horizon.

Using the results of Eqs.~(\ref{eq:tsurf}) and (\ref{eq:nulum})
to compute the $\nu\bar\nu$-annihilation
efficiency $e_{\nu\bar\nu}$ according to 
Eq.~(\ref{eq:eff}) and employing the integral energy $E_{\nu}$ 
emitted in
neutrinos as given from Eq.~(\ref{eq:nuerg}), one can obtain a result
for the energy $E_{\nu\bar\nu} = e_{\nu\bar\nu} E_{\nu}$
deposited in an $e^+e^-$-pair-photon fireball by the annihilation
of $\nu$ and $\bar\nu$ radiated from the disk.
With $R_{\rm d} = 3R_{\rm s} = 27\,{\rm km}$, 
$\ave{\epsilon_{\nu_e}}\approx 3k_{\rm B}T_{\rm surf}\sim 
(10.5\,...\,15.0)\,{\rm MeV}$, and 
$L_{\nu_e}\sim \rund{{1\over 8}\,...\,{1\over 6}}L_{\nu}$
(see Sect.~\ref{sec:nuem}), i.e, 
$L_{\nu_e} \la (1\,...\,4)\cdot 10^{52}\,{\rm erg/s}$, we get
\begin{equation}
E_{\nu\bar\nu}\,\sim\,(1.1\,...\,9.4)\cdot 10^{49}\,
(\Delta M_{{\rm d},01})^{3/8}\,R_{{\rm s},9}^{13/8}\ \,{\rm erg}\ .
\label{eq:ann}
\end{equation}
The upper and lower bounds of 
the interval for $E_{\nu\bar\nu}$ correspond to the most extreme
(maximum and minimum, respectively) choices for 
$e_{\nu\bar\nu}$, $L_{\nu_e}$, and $\ave{\epsilon_{\nu_e}}$.
Notice that the analytical estimates of the neutrino luminosity
(Eq.~(\ref{eq:nulum})) and the mean energy of neutrinos emitted
from the disk, $\ave{\epsilon_{\nu}}\approx 3k_{\rm B}T_{\rm surf}
\approx 11...15\,{\rm MeV}$ (Eq.~(\ref{eq:tsurf})), agree well 
with our numerical results for the phase shortly after the 
merging. While the dynamical time scale of the
merging is of the order of 1~ms and the post-merging evolution
was followed by our numerical simulations for a period of about
10~ms, the disk emits neutrinos with similar luminosities for
a much longer time of a few hundred milliseconds
(cf.~Eq.~(\ref{eq:tacc2})). Therefore, Eq.~(\ref{eq:ann}) gives a 
number for the energy
deposition by $\nu\bar\nu$-annihilation that is a factor of 
10--100 larger than the $\sim 10^{48}\,{\rm erg}$ calculated in
Sect.~\ref{sec:annires}. 

%
%

\section{Discussion\label{sec:end}}

In this paper we have reported about hydrodynamical calculations 
of the merging of equal-mass binary neutron stars with different
initial spins. We have analysed the models for
their neutrino emission, for neutrino-antineutrino annihilation
in the surroundings of the merger, and for the thermodynamical 
conditions in the merged object. 

\subsection{Mass loss and nucleosynthesis\label{sec:sumalo}}

The dynamical merging proceeds within a few milliseconds after
the simulations were started from an initial center-to-center
distance of 42~km of the two 1.6~$M_{\odot}$ neutron stars. 
Shortly after the two stars have fused into one
compact object with a mass of about 3~$M_{\odot}$
and an average density of more than $10^{14}\,{\rm g/cm}^3$,
spun-off matter forms a less dense toroidal cloud
($\rho\approx 10^{12}\,{\rm g/cm}^3$) that is heated to
temperatures of 5--10~MeV by friction in shock waves and strong 
pressure waves sent into the surrounding gas by the oscillations
and periodic pulsations of the central high-density body.
Roughly 0.1~$M_{\odot}$ of
material receive a large momentum and are pushed beyond the
grid boundaries. However, only a small fraction of at most
$10^{-5}$...$10^{-3}\,M_{\odot}$ of the matter has
a total energy (internal plus kinetic plus gravitational) 
large enough to allow the gas to become unbound. In model A128
which has the best numerical resolution, in fact none of the matter
can escape from the gravitational potential of the merger, in
contrast to model A64 where it is several $10^{-5}\,M_{\odot}$.
Notice, however, that energy released by the recombination of
free nucleons into nuclei --- an effect which is taken into
account in our simulations by the use of the ``realistic''
equation of state of Lattimer \& Swesty (1991) --- and energy
production by nuclear reactions proceeding in the cooling and 
expanding gas (Davies et al.~1994) can aid the mass ejection 
and could increase the unbound mass relative to our estimates.

The ejection of matter is also very sensitive to the amount of 
angular momentum that is present in the merging binary. The 
largest mass ejection was found in model B64 immediately after
the merging,
because the initial configuration of this model had a solid-body 
type rotation and thus the largest specific angular momentum in
regions far away from the system axis. In contrast, model C64 
showed the largest mass loss a few milliseconds later. The initial 
anti-spin setup of this model led to vigorous vibrations of the
central body and to the outward acceleration of material some time
after the two neutron stars had formed a single object. 
Whether and how much mass can be dynamically lost during the 
post-merging evolution, 
however, will also depend on the stability of the merged
object. The central, compact core of the merger 
is so massive that it can be stabilized neither
by internal pressure for the currently favored
supranuclear equations of state, nor by its rapid rotation
(for details, see Sect.~4.1.3 of Paper~I).
One therefore has to expect its collapse into a black hole 
within a few milliseconds. In case of model C64 not much matter
would be expelled if the gravitational instability
sets in before the large-amplitude post-merging oscillations
have taken place. 

The less dense cloud of gas that surrounds the massive, very 
dense central
body is stabilized by internal pressure because its rotational
velocities are significantly less than the Kepler velocity 
(see Fig.~12 of Paper~I). When the massive core collapses into
a black hole, the ambient matter will therefore be swallowed 
up by the black hole on a dynamical time scale of 
$10^{-4}$...$10^{-3}$~s. Only gas with a sufficiently large
angular momentum will have a chance to remain in a disk 
or extended torus around
the black hole. From there it will spiral into the black hole 
on the much longer time scale of viscous angular momentum transport.
One can estimate that with typical orbital velocities of about
$0.25c$ as found in our models, only gas at radii beyond about
44--107~km has an angular momentum that is large enough (for more 
information, see Sect.~4.1.3 of Paper~I). This minimum orbital 
radius is outside of the grid boundaries of
our simulations. We therefore conclude that essentially all the
mass on the computational grid will disappear in a forming black
hole more or less immediately, and only the
$\sim 0.015$--$0.15\,M_{\odot}$ of material that have been swept
off the grid might be able to end up in a toroidal ``disk'' 
around the central
black hole. A disk mass of about $0.1\,M_{\odot}$ should be taken
as an extreme upper limit for the considered scenario. Like the
dynamically ejected mass, the amount of gas that ends up in a disk
is sensitive to the initial angular momentum of the neutron star 
binary and to the relative times of black hole formation and mass
spin-off. Last but not least, a quantitative
answer seems to depend also on the numerical resolution, with the
trend that better resolved models yield smaller estimates for the
possible disk mass and mass ejection.

\begin{figure}
 \epsfxsize=9.25cm \epsfclipon \epsffile{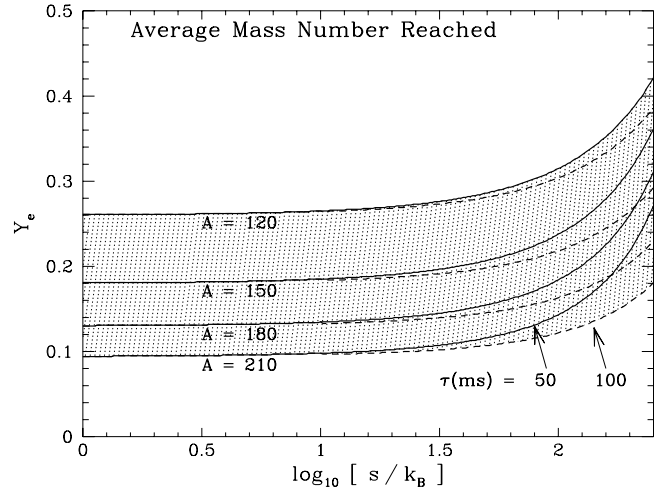}
\caption[]{
Expected average mass numbers of nuclei formed by an r-process
starting with NSE conditions for low entropies and with
the $\alpha$-process for high entropies. The
plot shows contours of constant average mass number in the
$Y_e$-$s$ plane ($s$ in $k_{\rm B}$ per nucleon). $Y_e$ and $s$
define the thermodynamical conditions in the expanding gas
at a temperature of about $5\cdot 10^9$~K, i.e., before 
the r-processing takes place. The dynamical evolution is 
characterized by the time scale $\tau$ of the adiabatic expansion 
between $T\sim 7\cdot 10^9$~K and $T\sim 3\cdot 10^9$~K.
Solid lines correspond to a time scale of 50~ms, dashed lines to 
100~ms. Between the contours
for mass numbers $A = 120$ and $A = 210$ the shaded area marks the
region where a suitably mass-weighted combination of 
the r-process yields for different conditions will produce a
solar-system like abundance pattern. For larger values
of $Y_e$, no significant r-processing can take place, for lower
values of $Y_e$ very strong r-processing will primarily lead to
nuclei in the region of the actinides
}
\label{fig:rprocess}
\end{figure}

A mass of $10^{-4}$...$10^{-3}\,M_{\odot}$ that is dynamically
ejected during the merging of binary neutron stars might have 
important implications for nucleosynthesis 
(Lattimer \& Schramm 1974, 1976; Eichler et al.~1989).
Dependent on the phase when the mass loss occurs, the expelled 
gas will start its expansion from different initial conditions 
of entropy and composition. The ejection of initially very cool, 
low-entropy material might be caused by the tidal interaction
during the last stages of the inspiral and during the mass transfer 
phase of
very close non-equal mass binaries. If the two components have
``nearly equal'' initial masses, the mass transfer is unstable
and within a few orbital periods the lighter star can be 
completely dissipated into a thick, axially symmetric disk around 
the primary. Some fraction of the surface material might
escape the system (Lattimer \& Schramm 1974, 1976).
If the initial mass ratio of the two stars is 
large, the binary is stable against dynamical-time scale mass
transfer and there is the interesting possibility that the 
secondary (the less massive component) 
is stripped to the minimum mass of stable neutron 
stars, at which stage it will explode
(Page 1982; Blinnikov et al.~1984, 1990; Eichler et al.~1989;
Colpi et al.~1989, 1991, 1993; Colpi \& Rasio 1994). However,
recent investigations suggest that stable mass transfer is unlikely
because the initial mass of the secondary must already 
be very small (below $\sim 0.4\,M_{\odot}$; Bildsten \& Cutler 1992,
Kochanek 1992, Rasio \& Shapiro 1994, Lai et al.~1994). In 
addition, an unreasonably high value of the neutron star viscosity
is needed to enforce corotation and to maintain tidal locking,
because stable mass transfer requires a dynamically stable Roche
limit configuration which can only exist in synchronized systems
with extreme mass ratios. Such systems are essentially ruled out     
for neutron stars (Lai et al.~1994). If mass shedding in the
discussed situations occurs, it would lead to the ejection of
initially cold, very low-entropy and very neutron-rich material. 
Subsequent radioactive $\beta$-decays of unstable, neutron-rich
heavy nuclei that are present in the decompressed matter will 
heat the expanding gas to temperatures around 0.1~MeV, which will
give rise to r-process conditions (Lattimer et al.~1977,
Meyer 1989, Eichler et al.~1989).

Gas spun off the exterior parts of the dilute toroidal cloud 
that surrounds the compact core of the merged binary, instead,
has been heated to temperatures of $\ga 1$~MeV by
friction during the merging and post-merging evolution. This heating
has produced entropies of a few $k_{\rm B}$ per nucleon. Neutrino
emission has already raised $Y_e$ from initial values 
$Y_e\approx 0.02$ to slightly less neutron-rich conditions with
$Y_e\approx 0.05$--0.2. The electron degeneracy is only moderate,
$\eta_e \approx 2$.
All these parameters are very similar to the conditions found in
the shocked outer layers of the collapsed stellar core in a type-II
supernova where the site of the classical r-process has been
suggested (see, e.g., Hillebrandt 1978). Compared with the supernova
case, the range of $Y_e$-values in the potentially ejected 
merger material is on the low side. An r-processing occurring under 
such neutron-rich conditions would be very efficient and should 
preferentially produce r-process nuclei with very high mass numbers
(Fig.~\ref{fig:rprocess}).
The predominant production of high-mass r-process elements
would be in concordance with the fact that one cannot expect
the formation of all Galactic r-process material 
in the considered low-entropy ejecta. With our numerical
estimates of a few $10^{-4}\,M_{\odot}$ for the mass loss per
merger event and with the possible event rate of NS-NS
and NS-BH mergers of $10^{-6}$...$10^{-4}$ per year per 
galaxy (Narayan et al.~1991, Phinney 1991, Tutukov et al.~1992
Tutukov \& Yungelson 1993),
which corresponds to about $10^4$...$10^6$ events
during the lifetime of the Galaxy, only 1--$100\,M_{\odot}$ of the
$\sim 10^5\,M_{\odot}$ of Galactic r-process elements could be
produced. But if the r-process were strong enough, all the 
Galactic actinides (e.g., about 40--50~$M_{\odot}$ of Th) might be
accounted for by the material shed during neutron star merging!

However, the neutron-rich wind that is driven by neutrino 
energy deposition in the outer disk regions can also contribute 
to the nucleosynthetic input into the interstellar medium. With
a total neutrino luminosity of about $10^{53}\,{\rm erg/s}$ the 
wind will have a mass outflow rate of the order of 
0.001--$0.01\,M_{\odot}{\rm s}^{-1}$ 
(Qian \& Woosley 1996, Woosley 1993b), depending on the 
gravitating mass of the merger, the disk mass and 
geometry, and the neutrino emission as a function of time.
For a duration of the outflow between some fractions of
a second and a few seconds, one might therefore have another 
$10^{-4}\,M_{\odot}$ up to several
$10^{-2}\,M_{\odot}$ of material that are expelled with very
interesting thermodynamical properties. 

Like the neutrino-driven 
wind from new-born neutron stars, the neutrino-heated material 
should have significantly higher entropies than the matter
that is dynamically ejected from the
disk or torus by momentum transfer during core pulsations. 
Owing to the particularities of geometry, gravitational potential,
and neutrino emission in the merger situation, the
expansion time scales as well as the degree of neutronization 
might be significantly different from the supernova case. 
Since the neutrino emission from the core and the disk of the merger 
is dominated by the $\bar\nu_e$ fluxes, absorptions of $\bar\nu_e$ 
in the wind material ($\bar\nu_e + p \to n + e^+$) will be more
frequent than the absorption of less abundant 
$\nu_e$ ($\nu_e + n \to p + e^-$) and will keep the expanding 
material neutron-rich. Because of a larger luminosity ratio
$L_{\bar\nu_e}/L_{\nu_e}\approx 3$--4.5 but similar mean $\nu_e$
and $\bar\nu_e$ energies with
$\ave{\epsilon_{\bar\nu_e}}/\ave{\epsilon_{\nu_e}}\approx 1.5$--1.8, 
the expanding wind will have a lower electron fraction
than in the supernova case (see Qian \& Woosley 1996 
for a discussion of the electron fraction in neutrino-driven winds).
Values as low as 
$Y_e\approx \eck{1+L_{\bar\nu_e}\ave{\epsilon_{\bar\nu_e}}/
(L_{\nu_e}\ave{\epsilon_{\nu_e}})}^{-1}\approx 0.1$--0.2 seem 
possible in the neutrino wind from the merger. For such low values
of $Y_e$ a strong r-processing can happen even at modest entropies 
of $s\sim 50$--$100\,k_{\rm B}/{\rm nucleon}$ 
(see Sect.~\ref{sec:gamburst} for an estimate of the wind entropies)
which are too low to
allow for the formation of r-process nuclei in the neutrino-driven
winds from protoneutron stars for the typical electron fractions
of $Y_e \ga 0.35$--0.4 found there (Witti et al.~1994, 
Takahashi et al.~1994, Woosley et al.~1994, Qian \& Woosley 1996). 
Figure~\ref{fig:rprocess} visualizes this and
shows that even for rather slow expansions with expansion time scales
of more than 100~ms nuclei with mass numbers $A$ between 150 and
210 can be formed.

Unfortunately our current models allow only rough estimates
of the mass loss and the conditions in the ejected matter. Our
simulations could not directly follow the ejection of mass 
from the merger because they suffered from the
limitations due to the use of the computational grid and due to
an insufficient numerical resolution, 
especially of matter at low densities.
A detailed and meaningful analysis of the very interesting aspects 
of a possible r-processing in the dynamically ejected 
low-entropy material and in the high-entropy
neutrino-driven winds from NS-NS or NS-BH mergers
has to be postponed until models are available which 
yield more quantitative information about the long-time evolution
of the merged object, the torus geometry, and
the neutrino-matter interactions in the outer parts of the torus.
Only such models can give evidence about the duration of the mass
loss and the amount of material that is ejected with different 
entropies, different expansion time scales, and different degrees 
of neutronization, all of which determine the nucleosynthetic
processes (see Fig.~\ref{fig:rprocess} and also 
Woosley \& Hoffman 1992, Witti et al.~1994).

\subsection{Neutrino emission and gamma-ray bursts\label{sec:gamburst}}

The luminosities and mean energies of the neutrinos
emitted from merging neutron stars are very similar to those
calculated for supernovae and protoneutron stars. After the 
two neutron stars have merged, luminosities up to several
$10^{52}$~erg/s are reached for every neutrino species and
the average energies of $\nu_e$ leaking out of the merger are
10--13~MeV, of $\bar\nu_e$ they are 19--21~MeV, 
and of heavy-lepton neutrinos
around 26--28~MeV. However, the neutrino emission exhibits
characteristic differences from the supernova case, too.

The total neutrino luminosity from merging neutron stars does not 
increase to a value above $10^{52}$~erg/s before the
hot, toroidal gas cloud around the dense and compact core of the
merger starts to form. More than 90\% of the peak neutrino emission 
of about $10^{53}$~erg/s stems from the ``disk'' region where
the optical depths and thus the neutrino diffusion time scales are 
significantly smaller than in the core. Another pecularity is the
fact that the very neutron-rich, decompressed and heated neutron
star matter predominantly emits electron antineutrinos. This should
hold on until the electron fraction in the medium has grown to 
a level where the increase of the lepton number by $\bar\nu_e$ 
emission is compensated by the $\nu_e$ losses. If the
merged configuration remained stable for a sufficiently long time,
the deneutronization phase will be superseded by an extended period
where the heated gas deleptonizes and cools again and thus evolves
back to the state of cold neutron star matter. 
However, it is very likely that the merged object, which 
contains essentially the baryonic mass of two typical neutron 
stars (about $3\,M_{\odot}$), does not remain gravitationally stable.
For all currently favored nuclear equations of state it should
collapse to a black hole long before the cooling is finished.

If the gravitational instability of the massive core of the 
merged object sets in before the hot gaseous
torus has formed and if all the surrounding gas falls into the
black hole immediately, the neutrino emission from
the merger will stay fairly low with a total luminosity of less
than $\sim 10^{52}$~erg/s. This is much too low to get sufficient
energy for a cosmological gamma-ray burst from 
$\nu\bar\nu$-annihilation during the final stages of the inspiral
of the two neutron stars and during the first 1--4~ms right after
the merging. The annhilation efficiency of neutrinos and
antineutrinos,        
$e_{\nu\bar\nu} = ({\rm d}E_{\nu\bar\nu}/{\rm d}t)/L_{\nu}$,
increases proportional to the neutrino luminosity 
(Eq.~(\ref{eq:eff})) and the neutrino energy deposition rate 
$E_{\nu\bar\nu}$ with the product of neutrino and antineutrino
luminosities. At the time of maximum neutrino emission some
6--8~ms after the stars have merged, we calculate an 
annihilation efficiency of $(2$--$3)\cdot 10^{-3}$ and 
a neutrino energy deposition rate of 2--$4\cdot 10^{50}$~erg/s
in the whole space outside the high-density regions of the compact
core and the surrounding ``disk''. The integrated energy deposition 
during the simulated evolution of about 10~ms is therefore 
less than $4\cdot 10^{48}$~erg (assuming maximum neutrino
fluxes during the whole considered times). Even with the 
unrealistic assumption that all the $\nu\bar\nu$-annihilation 
energy could be useful to power a relativistic pair-photon
fireball, this energy would fail to account for the canonical
$\sim 10^{51}/(4\pi)$~erg/steradian of a typical gamma-ray 
burst at cosmological distances (e.g., Woods \& Loeb 1994;
Quashnock 1996) by nearly three orders of magnitude.
Of course, these arguments are not conclusive if there is 
strong focussing of the expanding fireball into a 
narrow solid angle $\delta\Omega$. In this case an observer would
deduce a largely overestimated value (by a factor 
$4\pi/\delta\Omega$) for the energy in the fireball
and in the gamma-ray burst if he assumed
isotropy of the emission. However, for the considered merger
scenario and the geometry of the post-merging configurations 
in our simulations, it is very hard to imagine how the required
strong beaming of the fireball into a jet-like outflow could be 
achieved.

It is interesting to note that if the neutrino emission
calculated for our merger models would continue for a few seconds,
which is a typical duration of observed gamma-ray bursts 
(e.g., Norris et al.~1994, Kouveliotou 1995), a burst energy of
about $10^{51}$~erg could well be accounted for by the 
annihilation of neutrinos and antineutrinos. An accretion
disk or torus around the central black hole could provide
a luminous neutrino source for the required period of time. 
This time span is much longer than the times
covered by our hydrodynamical modelling. Since the current
numerical simulations were neither able to mimic the effects 
of a central black hole nor to track the evolution of the merger
for a sufficiently long time, we attempted to develop a simple
analytical model in Sect.~\ref{sec:simod} to give us insight 
into the principal dependences of the energy deposition by
the annihilation of neutrinos emitted from the disk or torus. 

This torus
model was based on Newtonian physics and did not determine the
torus structure self-consistently. However, the analytic 
treatment took into
account the effects of viscous angular momentum transport, 
viscous heating, neutrino cooling, and partial neutrino 
opaqueness. To first order and on a qualitative level, our 
considerations should also be valid for accretion disks around 
black holes in general relativity (see, e.g., 
Shapiro \& Teukolsky 1983) and for tori around 
Schwarzschild black holes (see, e.g., Chakrabarti 1996). 
In fact, the employed assumptions 
about the torus geometry are supported by general relativistic
investigations of neutron tori
(Witt et al.~1994; Jaroszy\'nski 1993, 1996)
and the neutrino luminosities from our
simple torus model are compatible with those obtained from the 
relativistic analyses. Moreover, the neutrino flux and 
neutrinospheric temperature calculated analytically are also
in good agreement with the results of our numerical models
for the post-merging phase when the neutrino emission has 
reached its saturation level.

Equation~(\ref{eq:ann}) gives the estimate of the 
$\nu\bar\nu$-annihilation energy $E_{\nu\bar\nu}$ 
for our analytical disk model. We find that $E_{\nu\bar\nu}$ 
could at best lie between $1.1\cdot 10^{49}\,{\rm erg}$
and $9.4\cdot 10^{49}\,{\rm erg}$ for a disk of $0.1\,M_{\odot}$ 
around a $3\,M_{\odot}$ black hole. The lifetime of such a disk
is determined by the time scale of the outward transport of
angular momentum and
is estimated to be several ten up to a few hundred milliseconds.
The ranges of values account
for the uncertainties in the neutrino opacity (which depends on
the composition of the medium and on the neutrino spectra)
and for the corresponding variation of the neutrino
luminosity and mean energy of the emitted neutrinos. Further
uncertainties due to the unknown disk viscosity are circumvented
by deriving the result for a value 
$\eta^\ast$ of the dynamic shear viscosity (Eq.~(\ref{eq:opvi2}))
which assures a maximum result for $E_{\nu\bar\nu}$. 

Unless focussing or beaming of the expanding pair-photon
fireball towards the observer
plays an important role, the annihilation energy of Eq.~(\ref{eq:ann})
is too low by more than a factor of 10 to explain powerful
cosmological gamma-ray bursts. Weak bursts with an energy of
about $10^{50}\,{\rm erg}$ and durations of less than or around
a second, however, do not seem to be completely excluded on grounds
of Eqs.~(\ref{eq:ann}) and (\ref{eq:tacc2}). 
The result of Eq.~(\ref{eq:ann}) and in
particular the upper value of $\sim 10^{50}\,{\rm erg}$
should be considered as a very optimistic maximum estimate for
the energy deposited by $\nu\bar\nu$-annihilation 
in the surroundings of the disk. Equation~(\ref{eq:ann}) 
was derived for the
most favorable conditions and by making a whole sequence of
most extreme assumptions, the combination of all of which 
appears rather unlikely.

In the first place it was assumed that the dynamic viscosity
of the disk adopts the optimum value $\eta^{\ast}$ of 
Eq.~(\ref{eq:opvi1}). For much smaller viscosities the toroidal disk 
should stay rather cool and the neutrino fluxes correspondingly
low. For much larger values of the viscosity the lifetime of the
disk will decrease because the mass accretion rate into the 
black hole increases with the rate of the outward transport of
angular momentum mediated by viscous forces. In this case 
the neutrino luminosities will be bounded by the fact
that the viscous friction will heat up the disk to very high
temperatures and thus the neutrino absorption and scattering
cross sections, which scale roughly with the square of the neutrino
energy, will increase. Therefore the neutrino diffusion 
time scale will increase, too, and the neutrino cooling will become
inefficient. As a consequence, most of the dissipated gravitational
and rotational energy could be advected into the black hole when the
matter, after having lost part of its angular momentum, 
spirals in through the innermost stable circular orbit
(advection-dominated regime).

Equation~(\ref{eq:ann}) also represents an optimistically high
value of the energy deposition by $\nu\bar\nu$-annihilation 
because the radiation efficiency of about 8\%
obtained for our Newtonian model of a Keplerian
accretion disk is an upper bound to the radiation efficiency of a
relativistic disk around a nonrotating black hole where it is less 
than 6\%. Moreover, it should be remembered that the annihilation 
efficiency
$e_{\nu\bar\nu}$ of Eq.~(\ref{eq:eff}) most likely overestimates
the useful fraction of the annihilation energy by a considerable
factor. Our hydrodynamical simulations show that the fraction of
the neutrino energy deposited in the possibly baryon-poor region above
and below the disk but not in the plane of the disk (where it
will serve to drive a baryonic, nonrelativistic wind instead of 
creating a relativistically expanding pair-photon fireball) could be
as small as some 20--25\% of the number given in Eq.~(\ref{eq:ann})
(cf.~Fig.~\ref{fig:anndis}).
Even more, general relativistic effects were neglected in the 
numerical simulations and analytical
considerations presented here. Jaroszy\'nski (1993) showed 
that they lower the energy that can be transported to infinity
significantly (by about 80\%). 

For the whole uncertainty range of the neutrino
interaction cross section (Eq.~(\ref{eq:sigma})) and for the 
corresponding range of neutrino luminosities and mean energies of 
emitted neutrinos, we find that the result of Eq.~(\ref{eq:ann})
falls short of the desired value $E_{\nu\bar\nu}/(4\pi) \sim
10^{51}/(4\pi)$~erg/steradian by at least an order of
magnitude if the disk mass is of the order of $0.1\,M_{\odot}$
and the central black hole has a mass of about $3\,M_{\odot}$.
From our numerical models of NS-NS mergers, one concludes
that a disk with a mass close to or even larger than 
$0.1\,M_{\odot}$ might be formed only under very special conditions.
Even if there
is a high angular momentum in the system due to neutron star spins
as in our model B64, the amount of
material that has a chance to form a disk is hardly
as much as $0.1\,M_{\odot}$ (see Paper~I).
In models A64 and C64 the lower angular momentum
allows only little material to possibly remain in a disk. 
Only matter outside of the boundary of our computational grid
has enough angular momentum to be rotationally stabilized. Before
some matter in models A64 and C64 has acquired a sufficiently 
large angular momentum to be lost off the computational
grid --- a process that is partly
aided by pressure waves created by the wobbling and ringing
of the central, compact object --- the merger, however, 
has probably already collapsed to a black hole which
swallows up most of the surrounding, pressure supported 
matter on a dynamical time scale.

Massive, self-gravitating tori around black holes may be subject
to general relativistic global instabilities that lead to 
catastrophic runaway mass loss and may provide by far the
shortest evolutionary time scale of such tori as recently 
argued for stationary polytropic tori by Nishida et al.~(1996) and
for stationary neutron tori by Nishida \& Eriguchi (1996). Such an
instability would have the same implications as the 
case of extremely large disk viscosity discussed above
where the accretion time scale of the torus due to the rapid 
viscous angular momentum transport could be much smaller than
the neutrino diffusion time scale. As a consequence of the 
rapid accretion of the torus material, most of the internal 
energy of the gas would be carried into the black hole along
with the gas instead of being radiated away by neutrinos.
Because the duration of the neutrino emission would be very
short without a compensating increase of the neutrino 
luminosity, the total energy emitted in neutrinos would be
much smaller than in the extreme and optimum situation considered 
in the derivation of Eq.~(\ref{eq:ann})
in Sect.~\ref{sec:simod}. Therefore, global disk instabilities might
be another threat to neutrino-powered gamma-ray bursts.
However, it has still to be demonstrated whether global runaway 
instabilities develop in the non-stationary situation and how
they behave and evolve in the presence of changes of the angular
momentum distribution.

Even if the lifetime of the merger and of the neutrino radiating
accretion torus is long enough and
neutrino annihilation could provide a powerful ``engine''
for creating a fireball, yet another major concern for the viability 
of the considered $\gamma$-ray burster scenario comes from the 
baryonic wind that is blown off the surface of the merger and
accretion torus by neutrino heating. This neutrino-driven 
wind is unavoidable when large neutrino fluxes are emitted and a
small fraction of these neutrinos
annihilate or react with nucleons in the low-density gas in the
outer layers of the merger. In order to obtain bulk Lorentz factors
$\Gamma \sim \dot{E}_{\nu\bar\nu}/(\dot{M}_{\rm w}c^2)\ga 100$ 
one can only allow for mass loss rates 
$\dot{M}_{\rm w} \la 2\cdot 10^{-6}\,M_{\odot}{\rm s}^{-1}$ if the 
rate at which the pair-photon fireball is supplied with energy is
$\dot{E}_{\nu\bar\nu}\approx 3\cdot 10^{50}$~erg/s. With about $2/3$
or $\sim 2\cdot 10^{50}$~erg/s of this energy being transferred to the
dilute outer regions of the accretion torus in our models 
(Sect.~\ref{sec:annires}) by
$\nu\bar\nu$-annihilation (neglecting additional heating by
neutrino-electron scattering and neutrino absorption), 
we compute a mass loss rate of at least
$\dot{M}_{\rm w}\ga {2\over 3}\dot{E}_{\nu\bar\nu}\cdot 
\eck{GM/(3R_{\rm s})}^{-1}
\ga 7\cdot 10^{-4}\,M_{\odot}{\rm s}^{-1}$ for a black hole with mass
$M = 3\,M_{\odot}$ and Schwarzschild radius $R_{\rm s}\approx 27$~km.
From this we get $\dot{E}_{\nu\bar\nu}/(\dot{M}_{\rm w}c^2)\la 0.25$ 
and estimate an entropy 
$s\approx (\varepsilon + P)/(k_{\rm B}T\rho/m_{\rm u})$ ($\varepsilon$
internal energy density, $P$ pressure) of 
$s\approx \eck{4 m_{\rm u}c^2/(3 k_{\rm B}T)}
\eck{(\dot{E}_{\nu\bar\nu}/3)/
(\dot{M}_{\rm w}c^2)}\la 100$--200~$k_{\rm B}/{\rm nucleon}$ when
$k_{\rm B}T\sim 0.5\,...\,1$~MeV. Therefore, unless the
neutrino-driven wind can be hindered to penetrate into the 
pair-photon fireball, e.g., in a region along the system axis 
by centrifugal forces, there is no chance to obtain highly
relativistic fireballs with $\Gamma \ga 100$.

These issues might also be critical when a neutron star merges with
a black hole or when two non-equal mass neutron stars coalesce.
Simulations indicate (Lee \& Klu\'zniak 1995) that already after the
dynamical interaction of a neutron star and a black hole a 
cloud of baryonic material might ``pollute'' the surroundings 
even near the system axis. Moreover, instead of an accretion torus,
a stable binary system might form with
a more massive black hole circulated by a low-mass neutron star 
companion which is possibly unstable to explosion (see also
Sect.~\ref{sec:sumalo} and
Blinnikov et al.~1984, Eichler et al.~1989). Future,
better resolved computations covering a wider range of 
equations of state and masses of the interacting stars
will have to show the conclusiveness of their
results. 

It is interesting to speculate whether 
there is a chance to get tori more massive than in the merging of 
two (nearly) equal-mass neutron stars when a neutron star merges
with a black hole or when a small neutron star is tidally disrupted
before merging with a companion neutron star which has a 
significantly larger mass (e.g., Eichler et al.~1989,
Narayan et al.~1992, Mochkovitch et al.~1993, and references therein).
On the one hand, a more massive disk can
radiate neutrinos for a longer time than the accretion time scale
of the innermost, most strongly neutrino radiating part near the 
last stable orbit around the black hole. With much more matter being
at larger radii, the neutrino emitting torus region considered
and normalized to a mass of $0.1\,M_{\odot}$ in the derivation of
Eq.~(\ref{eq:ann}) will be continuously refed by material advected
inward. On the other hand, Eq.~(\ref{eq:ann}) suggests that
the energy that can be provided in the pair-photon fireball by
$\nu\bar\nu$-annihilation increases steeply with the mass $M$
of the central black hole, $E_{\nu\bar\nu}\propto R_{\rm s}^{13/8}
\propto M^{13/8}$. Larger gravitating masses at the center, e.g., 
a massive stellar black hole, might therefore allow
for more powerful $\gamma$-ray bursts, at least, if one relies on
the simplified picture developed in the derivation of 
Eq.~(\ref{eq:ann}) which does not take into account the dynamics
of the tidal interaction between neutron star and black hole and
its implications for the disk formation and the disk mass.

For the reasons outlined above it appears to be extremely hard
to account for the energies of cosmological $\gamma$-ray bursts by
$\nu\bar\nu$-annihilation, at least in the case of merging binary
neutron stars if beaming or focussing of the fireball towards the
observer does not play a crucial role. The same conclusion was 
arrived at by Jaroszy\'nski (1993, 1996) who investigated models of
relativistic tori around rotating stellar mass 
black holes and tested different
values of specific angular momentum, viscosity, and entropy.
He found that the 
neutrino emission and annihilation energy from these tori is
insufficient to explain the energies of cosmological $\gamma$-ray
bursts except for tori around Kerr black holes with very
high angular momentum, i.e., with relativistic rotation
parameters $a\sim 1$. This might indeed be realized in the 
collapsed cores of rapidly rotating Wolf-Rayet stars in the
``failed supernova'' or ``collapsar'' scenario 
(Woosley 1993a). However, models for the stellar evolution of
Wolf-Rayet stars seem to indicate that the angular momentum loss
during the mass-loss phases is too large to allow for the formation
of very rapidly spinning black holes (Langer, personal communication).
Moreover, with $\nu\bar\nu$-annihilation as the source of the energy
of the pair-photon fireball,
these models might come into additional trouble if,
as claimed by Quashnock (1996), the homogeneity of the distribution 
of the bursts in the BATSE 3B Catalog and the non-association 
of the bursts with
large-scale structures of luminous matter in our extragalactic 
neighbourhood implies such large distances of the $\gamma$-ray
burst sources that the required total energy output in 
$\gamma$-rays is larger than $3\cdot 10^{51}\,{\rm erg}$.

Our investigations do not include
effects due to possible convective overturn and instabilities in the
disk that might occur as a result of specific entropy (or composition)
inversions caused by neutrino effects or viscous heating. Such dynamical
processes were recently found to be present in multi-dimensional
models of type-II supernovae (Herant et al.~1994; Burrows et al.~1995;
Janka \& M\"uller 1995, 1996) and were computed for axisymmetric
advection-dominated accretion flows with two-dimensional hydrodynamical
simulations by Igumenshchev et al.~(1996) (see also Chen 1996). 
Although the disk remains globally stable, shorter-wavelength
modes may affect the flow dynamics and effective disk viscosity 
significantly. In addition, such instabilities might
increase the neutrino fluxes and the average neutrino
energies considerably and thus might help
$\nu\bar\nu$-annihilation. Our three-dimensional 
hydrodynamical simulations were not followed for a 
sufficiently long time to see whether such overturn processes occur
in the merging scenario, and the simplified one-zone torus model
does not take into account an enhancement of the neutrino fluxes by 
possible convective transport.
Also, magnetic fields in the merging neutron stars and in the
torus were disregarded. Within a lifetime of a few tenths of a second,
initial $B$-fields of $\sim 10^{12}$--$10^{13}\,{\rm G}$
might be amplified by a factor of 100 or more in
the rapidly rotating disk around the black hole (rotation periods
$\sim 1\,{\rm ms}$) and might become energetically important
(Rees, personal communication). A relativistic 
magneto-hydrodynamical wind of extremly high luminosity, perhaps
associated with a binary neutron star merger, was suggested to generate
$\gamma$-ray bursts by Thompson (1994). In this respect NS-BH mergers 
and neutron star collisions and coalescence need further theoretical
investigation. Moreover, neutron star collisions were recently pointed
out as potential origin of short 
cosmological $\gamma$-ray bursts by Katz \& Canel (1995a) who argued
that hot matter fragments could 
be ejected and, when they become neutrino-transparent on dynamical 
time scales, could lead to the emission of neutrinos with very
high luminosities.

%
%

\section{Summary and conclusions\label{sec:summary}}

The neutrino emission and neutrino-antineutrino annihilation 
during the coalescence of binary neutron stars were investigated.
To this end the three-dimensional Newtonian equations of
hydrodynamics were integrated by the Riemann-solver based
``Piecewise Parabolic Method'' on an equidistant Cartesian grid
with a resolution of $64\times 64\times 32$ or 
$128\times 128\times 64$ zones. The properties of neutron
star matter were described by the equation of state of Lattimer
\& Swesty (1991). Energy loss and changes of the electron abundance
due to the emission of neutrinos were taken into account by an 
elaborate ``neutrino leakage scheme''. We have simulated the
coalescence of two identical, cool (initially 
$k_{\rm B}T_{\rm c}\approx 7\,{\rm MeV}$) neutron stars with a 
baryonic mass of about $1.6\,M_\odot$, a radius of 15~km, and an 
initial center-to-center distance of 42~km for three different
cases of initial neutron star spins.

The total neutrino luminosity prior to and during the dynamical 
phase of the coalescence is very 
small ($L_{\nu}\la 10^{51}$~erg/s), becomes about 
1--$2\cdot 10^{52}$~erg/s when the stars have merged into one 
rapidly spinning massive body, and climbs to 
1--$1.5\cdot 10^{53}$~erg/s after spun off material has
formed a hot toroidal cloud with a mass of 0.1--$0.2\,M_{\odot}$ 
around the wobbling and pulsating central object. The neutrino
fluxes are clearly dominated ($\sim$90--95\%) by the emission 
from this ``disk''.
Since the disk matter is neutron-rich, $\bar\nu_e$
are radiated with a luminosity that is a factor 3--6 higher than 
the (individual) luminosities
of $\nu_e$ and $\nu_x$ ($\equiv \nu_{\mu},\,\bar\nu_{\mu},\, 
\nu_{\tau},\,\bar\nu_{\tau}$). The mean energies of the emitted
neutrinos are very similar to those of supernova neutrinos,
$\ave{\epsilon_{\nu_e}}\approx 12\,{\rm MeV}$, 
$\ave{\epsilon_{\bar\nu_e}}\approx 20\,{\rm MeV}$,
and $\ave{\epsilon_{\nu_x}}\approx 27\,{\rm MeV}$.

When the neutrino luminosities are highest, only
about 0.2--0.3\% of the energy emitted in neutrinos is deposited
in the immediate neighborhood of the merger by 
$\nu\bar\nu$-annihilation, and the maximum integral energy deposition
rate is found to be about 3--$4\cdot 10^{50}$~erg/s. Thus,
to pump an energy of the order of
$10^{51}/(4\pi)$~erg/steradian into a fireball of $e^+e^-$-pairs 
and photons, the strong neutrino emission would have to continue for
several seconds. Since a collapse of the central core of the merger
with a mass of $\ga 3\,M_{\odot}$ into a black hole  
within milliseconds seems unavoidable,
we conclude that the available energy is insufficient by a factor of
about 1000 to explain gamma-ray bursts at cosmological distances.
However, it appears possible that an accretion torus with a mass
of $\sim 0.1$--$0.2\,M_{\odot}$ remains around the central black 
hole and is accreted on the time scale of viscous angular momentum
transport. Analytical estimates suggest that even under the most
favorable conditions in this torus and with an optimum value of the 
disk viscosity, annihilation of $\nu\bar\nu$ pairs emitted from
this torus provides an energy that is still more than a factor of
10 too small to account for powerful cosmological gamma-ray bursts,
unless focussing of the fireball expansion plays an important
role.

A few $10^{-4}\,M_{\odot}$ of very neutron-rich, low-entropy
matter may be dynamically ejected shortly after the neutron stars
have merged, and another $10^{-4}$ up to a few $10^{-2}\,M_{\odot}$
of strongly neutronized, high-entropy material might be carried
away from the accretion torus
in a neutrino-driven wind on a time scale between a fraction of
a second and a few seconds. The contamination with these baryons
is a severe threat to a relativistic fireball.
Aspects of nucleosynthesis in these ejecta were discussed.
Because of the neutron-richness of the ejected material and the
dominance of the $\bar\nu_e$ luminosity from the merged object
and its accretion torus, conditions suitable for
the formation of r-process elements might be realized more easily 
than in the neutrino wind from newly formed neutron stars.

It seems to be very difficult to fulfill the energetic
requirements of cosmological gamma-ray bursts with the annihilation
of $\nu\bar\nu$ pairs emitted from an accretion disk or torus
around a stellar mass black hole. 
If $\nu\bar\nu$-annihilation is nevertheless
to be saved as energy source for relativistic 
pair-photon fireballs --- despite of the problems exposed by our
numerical and analytical results and the critical issues addressed
in the discussion of Sect.~\ref{sec:end} --- then one is forced to
consider the following possibilities. 

The neutrino luminosities from the accretion torus could be
considerably higher than obtained in our models, but the
mechanism to achieve this has yet to be identified, e.g.,
it is possible that the neutrino transport in the torus is enhanced
by convective instabilities. Because of the quadratic dependence
on the neutrino luminosities, an increase of the neutrino fluxes
would affect the $\nu\bar\nu$-annihilation sensitively. 
Alternatively, still relying on the simple picture described in
Sects.~\ref{sec:annires} and \ref{sec:simod}, one might feel tempted
to interpret the estimates of the annihilation energy 
towards the optimistic side, although they were
derived by employing a number of very generous and favorable 
assumptions. In this case some interesting implications for the
hypothesis of stellar mass accretion disks around black holes
as sources of gamma-ray bursts (Paczy\'nski 1991; 
Narayan et al.~1992; Woosley 1993a, 1996) can be inferred from
combining the theoretical results with information about measured
burst time scales (Meegan et al. 1995a, Kouveliotou 1995) and
energies of cosmological bursts (e.g., Woods \& Loeb 1994;
Quashnock 1996). 

Short gamma-ray bursts have a typical duration $t_{\gamma}$
of several tenths of a second and a typical total energy that
is a factor of $\sim 20$ smaller than that of long bursts
(Mao et al.~1994). They require the release of neutrino energy
from accretion tori with masses $M_{\rm d}\sim
0.1\,M_{\odot}\eck{t_{\gamma}/(0.05\,...\,0.2\,{\rm s})}
\sim$~few~$0.1\,M_{\odot}$ and a beaming of the expanding
fireball into a solid angle $\delta\Omega\sim
4\pi\eck{(1\,...\,9)\cdot 10^{49}\,{\rm erg}/E_{\gamma}}
\eck{M_{\rm d}/0.1\,M_{\odot}}$, when $E_{\gamma}$ is
the typical energy of a cosmological gamma-ray burst if the
gamma-rays were radiated isotropically. $\delta\Omega$ becomes
smaller if the useful energy from $\nu\bar\nu$-annihilation is
less than $(1\,...\,9)\cdot 10^{49}$~erg. 
For $E_{\gamma}\sim 10^{51}$~erg the focussing of the fireball
is noticable, $\delta\Omega\sim 4\pi/10$, whereas for 
$E_{\gamma}\sim 10^{50}$~erg it is essentially absent. The required
accretion mass might suggest merging events of compact binaries
as the origin of the bursts in this case. The disk masses and a
possible focussing of the fireball expansion towards the observer
have to be explained by theoretical modelling.

For long bursts, most of which have durations $t_{\gamma}$ between 
some 10~s and about 100~s, the accretion of 
$M_{\rm d}\sim$~few~$M_{\odot}$ is needed and could provide an 
energy $E_{\gamma}\sim (1\,...\,9)\cdot 10^{50}
\eck{M_{\rm d}/M_{\odot}}$~erg which is of the order of 
$\sim 10^{51}$~erg without significant focussing of the
relativistic pair-photon plasma being necessary. For 
more energetic bursts some beaming of the fireball expansion would
have to be invoked. The large accretion mass favors the ``failed
supernova'' or ``collapsar'' scenario (Woosley 1993a). In this
model several solar masses of material surround the most strongly 
neutrino
radiating region of the accretion torus close to the innermost stable
orbit around the black hole. This should lead to mixing of baryons 
with a significant fraction of the pair-photon plasma and will 
confine the volume where $\nu\bar\nu$-annihilation might create
a relativistic fireball to a baryon-poor region along the 
system axis. The fireball will expand into a limited solid angle
which should compensate for the reduction of the useful fraction 
of the $\nu\bar\nu$-annihilation energy.

This interpretation of the bimodal distribution of burst durations
(Kouveliotou et al.~1993) employs two different kinds of 
astrophysical events. In contrast, Wang (1996) attempted
to explain the bimodality by a superposition of two distinct time
scales in the temporal structure of individual bursts corresponding
to peak widths and separations between adjacent peaks.
Katz \& Canel (1995a,b) have recently suggested the association 
of short and long bursts with two different classes of models
and have hypothesized that short bursts are produced by neutron
star collisions and long bursts originate from accretion-induced 
collapse of bare degenerate white dwarfs (Dar et al.~1992). 
In both types of models the burst energy would be provided by
the annihilation of emitted $\nu\bar\nu$ pairs.
Accretion induced collapse, however, was ruled out as a source of
gamma-ray bursts situated at cosmological distances
by Woosley \& Baron (1992) on grounds of the unacceptably large
baryonic pollution of the surroundings of the collapsed star
caused by a nonrelativistic neutrino-driven wind. The same worries
also hold for collisions of neutron stars where explosions of 
ejected low-mass fragments might create an envelope of baryonic
material around the collision site. Moreover, one has to be 
suspicious whether the neutrino emission will be luminous and
long enough that $\nu\bar\nu$-annihilation can provide an 
energy $\ga 10^{50}$~erg for a short and most likely unbeamed
gamma-ray burst. 

Neutrino emission and $\nu\bar\nu$-annihilation would be the 
energy source for the gamma-ray bursts also in the two classes of 
models that could lead to accretion tori around black holes,
i.e., the merging of binary neutron stars or of neutron star black
hole systems in case of short bursts, and collapsing, very massive
stars which do not succeed to explode as type-II supernovae in
case of long bursts. The bimodal distribution of burst durations
would reflect the two distinct mass ranges of the accretion tori
around the accreting stellar mass black holes, some 0.1~$M_{\odot}$
or a few $M_{\odot}$, respectively. The similar peak 
luminosities of both short and long bursts (Mao et al.~1994) could
be explained by the same underlying energy source for the gamma-ray
bursts. The short-time variability of the gamma-ray signal 
could be a consequence of jet precession (Hartmann \& Woosley 1995) 
or of inhomogeneities and instabilities in the
accretion torus that give rise to fluctuations of the accretion
rate. And the individual characteristics of burst events might
be associated with different masses of accretion tori 
and accreting black holes, different torus structures due to
different angular momentum distributions, and different accretion 
time scales because of variations of the angular momentum
transport, e.g., caused by magnetic fields or viscosity producing
dissipative processes in the torus.

Movies in mpeg format of the dynamical evolution of all models are
available in the WWW at
{\tt http:\nix//www.mpa-garching.mpg.de\nix/\lower0.7ex\hbox{$\!$\~~$\!$}mor\nix/nsgrb.html}

\begin{acknowledgements}
It is a pleasure for us to thank Wolfgang Keil for transforming 
Lattimer \& Swesty's FORTRAN equation of state into a usable table.
H.-T.J.~and K.T.~are very grateful to Josef~Witti for extended 
network calculations of the $\alpha$-process and for providing
comfortable tools to visualize and scan the rich data base of 
results. M.R.~would like to thank Sabine Schindler for her
patience in their common office.
Educative discussions with M.~Camenzind, W.~Hillebrandt, 
F.~Meyer, M.~Rees, and S.~Woosley are acknowledged.
K.T.~was supported by the ``Sonderforschungsbereich 375-95 f\"ur 
Astro-Teilchenphysik'' of the Deutsche Forschungsgemeinschaft,
M.R.~by the Deutsche Agentur f\"ur Raumfahrtangelegenheiten (DARA),
F\"orderungsvorhaben des Bundes, FKZ: 50 OR 92095.
The calculations were performed at the Rechenzentrum Garching on a
Cray-YMP 4/64 and a Cray-EL98~4/256.
\end{acknowledgements}

\end{document}